\def\usetikzcache{}
\def\forarxiv{} 

\ifdefined\forarxiv
\documentclass[3p]{elsarticle}
\else
\documentclass[preprint]{elsarticle}
\fi

\usepackage[utf8]{inputenc}

\usepackage[linesnumbered,ruled,vlined]{algorithm2e}
\usepackage[dvipsnames,svgnames]{xcolor}
\usepackage{graphicx}
\usepackage{amsmath}
\usepackage{hyperref}
\usepackage{cleveref}
\usepackage[caption=false]{subfig}
\usepackage{import}
\usepackage{pgfplots}
\usepackage{siunitx}
\usepackage[frozencache=true,cachedir=_minted-2018__dkdb]{minted}

\SetCommentSty{commentsty}
\SetKwComment{Comment}{$\triangleright$\ }{}

\ifdefined\usetikzcache
\usepackage{tikzexternal}
\else
\usepackage{tikz}
\usetikzlibrary{external}
\fi

\pgfplotsset{compat=newest} 
\usepgfplotslibrary{units} 
\usepgfplotslibrary{colorbrewer}

\pgfplotsset{
  discard if/.style 2 args={
    x filter/.code={
      \edef\tempa{\thisrow{#1}}
      \edef\tempb{#2}
      \ifx\tempa\tempb
      \def\pgfmathresult{inf}
      \fi
    }
  },
  discard if not/.style 2 args={
    x filter/.code={
      \edef\tempa{\thisrow{#1}}
      \edef\tempb{#2}
      \ifx\tempa\tempb
      \else
      \def\pgfmathresult{inf}
      \fi
    }
  }
}

\definecolor{Tomato2}{rgb}{0.93,0.36,0.26}
\definecolor{red3}{rgb}{0.93,0.1,0.1}
\definecolor{blue2}{rgb}{0.24,0.41,0.64}
\definecolor{blue3}{rgb}{0.34,0.31,0.84}
\definecolor{green2}{rgb}{0.25,0.6,0.43}
\definecolor{green3}{rgb}{0.33,0.66,0.16}

\newcommand{\todolater}[1]{}

\definecolor{minted_bg}{rgb}{0.9,0.9,0.95}

\setminted[SPARQL]{bgcolor=minted_bg}
\newcommand{\sparql}{\mintinline{SPARQL}}

\usepackage{colortbl}

\ifdefined\forarxiv
\journal{arXiv}
\else
\journal{Robotics and Autonomous Systems}
\fi

\newcommand{\OP}{$^\oplus$}

\begin{document}

\begin{frontmatter}
\title{Hastily Formed Knowledge Networks and Distributed Situation Awareness for Collaborative Robotics}

\author{Cyrille Berger\fnref{fn1}\corref{cor1}}
\ead{cyrille.berger@liu.se}
  \address{Department of Computer and Information Science \\
           Link\"oping University,
           SE-581 83 Link\"oping, Sweden}

\author{Patrick Doherty%
  \fnref{fn1}}
\ead{patrick.doherty@liu.se}
\address{School of Intelligent Systems and Engineering\\
Jinan University (Zhuhai Campus), Zhuhai, China}
 \address{Department of Computer and Information Science \\
           Link\"oping University,
           SE-581 83 Link\"oping, Sweden}

\author{Piotr Rudol\fnref{fn1}}
\ead{piotr.rudol@liu.se}
\author{Mariusz Wzorek\fnref{fn1}}
\ead{mariusz.wzorek@liu.se}
  \address{Department of Computer and Information Science \\
           Link\"oping University,
           SE-581 83 Link\"oping, Sweden}

\cortext[cor1]{Corresponding author.}
\fntext[fn1]{This work has been supported by the ELLIIT Network Organization for Information and Communication Technology, Sweden (Project B09); the Swedish Foundation for Strategic Research SSF (Smart Systems Project RIT15-0097); Patrick Doherty has been a (guest) professor at Jinan University (Zhuhai Campus) during the writing of this report.}

\begin{abstract}

In the context of collaborative robotics, distributed situation awareness is essential for supporting collective intelligence in teams of robots and human agents where it can be used for both individual and collective decision support. This is particularly important in applications pertaining to emergency rescue and crisis management. During operational missions, data and knowledge is gathered incrementally and in different ways by heterogeneous robots and humans. We describe this as the creation of \emph{Hastily Formed Knowledge Networks} (HFKNs). The focus of this paper is the specification and prototyping of a general distributed system architecture that supports the creation of HFKNs by teams of robots and humans. The information collected ranges from low-level sensor data to high-level semantic knowledge, the latter represented in part as RDF Graphs. The framework includes a synchronization protocol and associated algorithms that allow for the automatic distribution and sharing of data and knowledge between agents. This is done through the distributed synchronization of RDF Graphs shared between agents. High-level semantic queries specified in SPARQL can be used by robots and humans alike to acquire both knowledge and data content from team members. 
The system is empirically validated and complexity results of the proposed algorithms are provided. Additionally, a field robotics case study is described, where a 3D mapping mission has been executed using several UAVs in a collaborative emergency rescue scenario while using the full HFKN Framework.

\end{abstract}


\end{frontmatter}


\ifdefined\forarxiv
\else
\linenumbers
\fi

\section{Introduction}\label{sec:1}

The importance of effective communication and efficient data/knowledge transfer is essential for the coordination of life-saving activities in regions affected by natural or man-made disasters. The organizations and groups involved range from disaster relief, governmental and non-governmental organizations at the macro level, to actual teams of emergency rescue responders on the ground at the micro-level. Emergency rescue teams have recently been supported by heterogeneous robotic assistants. There are an increasing number of natural disasters that include wild fires, hurricanes, earthquakes and floods that require state-of-the-art emergency response to minimize loss of life and property damage.

In a seminal article, Denning \cite{denning2006hastily}, pointed out the importance of establishing \emph{Hastily Formed Networks} (HFNs) in the broader sense as "the ability to form multi-organizational networks rapidly" and as being crucial to disaster relief.  Here, Denning's focus was on effective human communication rather than efficient data/knowledge transfer, and the use of autonomous systems in emergency rescue operations was not yet prevalent. Additionally, the term \emph{conversation space} was introduced for the medium in which such communication takes place. 

Experience has shown that first response is dependent on the quality and nature of this conversation space. This space is intended to provide a medium for acquiring situation awareness and the original concept was very much focused on setting up the physical layer for communication. Proposed components of the conversational space were the physical systems, the players involved and the interaction practices. The latter include situation awareness, acquiring and sharing information, planning, making decisions, coordination, and command and control required by the players and teams involved.

Using Denning's metaphor of Hastily Formed Networks for enhanced communication and conversation spaces among human agents in emergency rescue operations as a starting point, we extend the idea in two ways using the term \emph{Hastily Formed Knowledge Networks} (HFKNs) as a guiding metaphor for this research. 
\begin{itemize}
\item Firstly, rather than focusing solely on teams of human agents, we focus on teams of human and heterogeneous autonomous robotic agents interacting in various ways among themselves and with humans to make rescue operations more efficient and to achieve collaborative goals. 
\item Secondly, rather than just focusing on communicative aspects and conversational spaces for crisis communication, we instead focus on both communicative aspects and data/knowledge aspects in \emph{data/knowledge interaction spaces} among collaborative teams consisting of both human and robotic agents. 
 \end{itemize}
 
 The goal is to automate the goal-driven creation and exchange of local and global situation awareness among team members (both human and robotic)
 in addition to improving the basis for informed decision making by providing timely data and knowledge rich contexts for doing this. These data and knowledge contexts will be both individually and collaboratively constructed on-the-fly during the unfolding of emergency rescue missions relative to the needs and requests of human and robotic team members.
 
 The work presented in this paper is part of a larger infrastructural multi-agent based framework being developed with the goal of leveraging the use of heterogeneous teams of human agents, Unmanned Aerial Vehicles (UAVs), and surface and ground vehicles. The intent is to provide situation awareness, services and supplies to emergency rescuers on the ground in collaboration with other human resources in disaster relief scenarios.\footnote{\url{https://www.ida.liu.se/divisions/aiics/projects/coopuav.en.shtml}}
 
 
 Much of the recent work with collaborative human/robotic systems~\cite{socrob:2016,18thri-Veloso,16icsr-sharing, Luo-2015-6005,DohKvaEtAl:2016:953055, Kva:2011:483527,horvitz1999mixed-initiative} has focused on the representation and generation of shared tasks and shared goals and how such shared goals can be achieved through the coordination of agents participating in such shared tasks. In other work, we have considered this and developed a delegation-based framework for task generation, allocation, and execution for collaborative robotics~\cite{DohKvaSza:2012:489882,DOHERTY-US-2013,doherty2015}. In fact, these earlier results will be used together with the work presented in this article,  for supporting collaborative data collection tasks and other types of missions. The delegation framework will be used to setup distributed data collection, distribution, and synchronization of tasks among teams of robotic and human agents. These data stores then contribute to a shared situation awareness of different aspects of the operational environment that can be used by robotic and human agents alike in other tasks associated with emergency rescue.
 
 \subsection{SymbiCloud HFKN Framework}
 
 The focus of this paper is the \emph{SymbiCloud HFKN Framework}, which includes a  data/knowledge management infrastructure that is intended to be  used to support distributed, collaborative collection of data and knowledge and its shared use in multi-agent systems. In this framework,  each agent is assumed to have a \emph{SymbiCloud module} (SCModule)  containing its local or contextual perspective of its operational environment. This module can include geographically tagged information, sensor-data abstractions, 3D maps, static and dynamic object representations, and activity recognition structures, in addition to a rich set of reasoning engines and data/knowledge management processes. An agent's \textit{SCModule} content will vary according to its capabilities, its sensors, and its ability to gather information. An agent's data/knowledge perspective can be enhanced and extended through interaction and synchronization with other agent's data/knowledge perspectives. These shared perspectives provide enhanced situation awareness for individual team members.
 

 Given a team of agents, information in \textit{SCModules} can be aggregated and merged dynamically and virtually at different abstraction levels to allow for richer perspectives to improve timely decision making and planning processes for the individual agents and teams.  The \emph{Symbi} in SymbiCloud is intended to emphasize that we would like to semantically ground as much data and knowledge as possible. The \emph{Cloud} in SymbiCloud is intended to emphasize the highly distributed and fluid use of data and knowledge across individual team members and the goal of leveraging developments in Cloud-based technology in a wider perspective, where parts of an agent's SymbiCloud module may be stored and accessed using Cloud services. Additionally, agents with SCModules can access useful and relevant information directly from the Internet such as Semantic Web content and combine it with an agent's locally stored knowledge.
 
\begin{figure}[tb]
\centering
\includegraphics[width=0.9\linewidth]{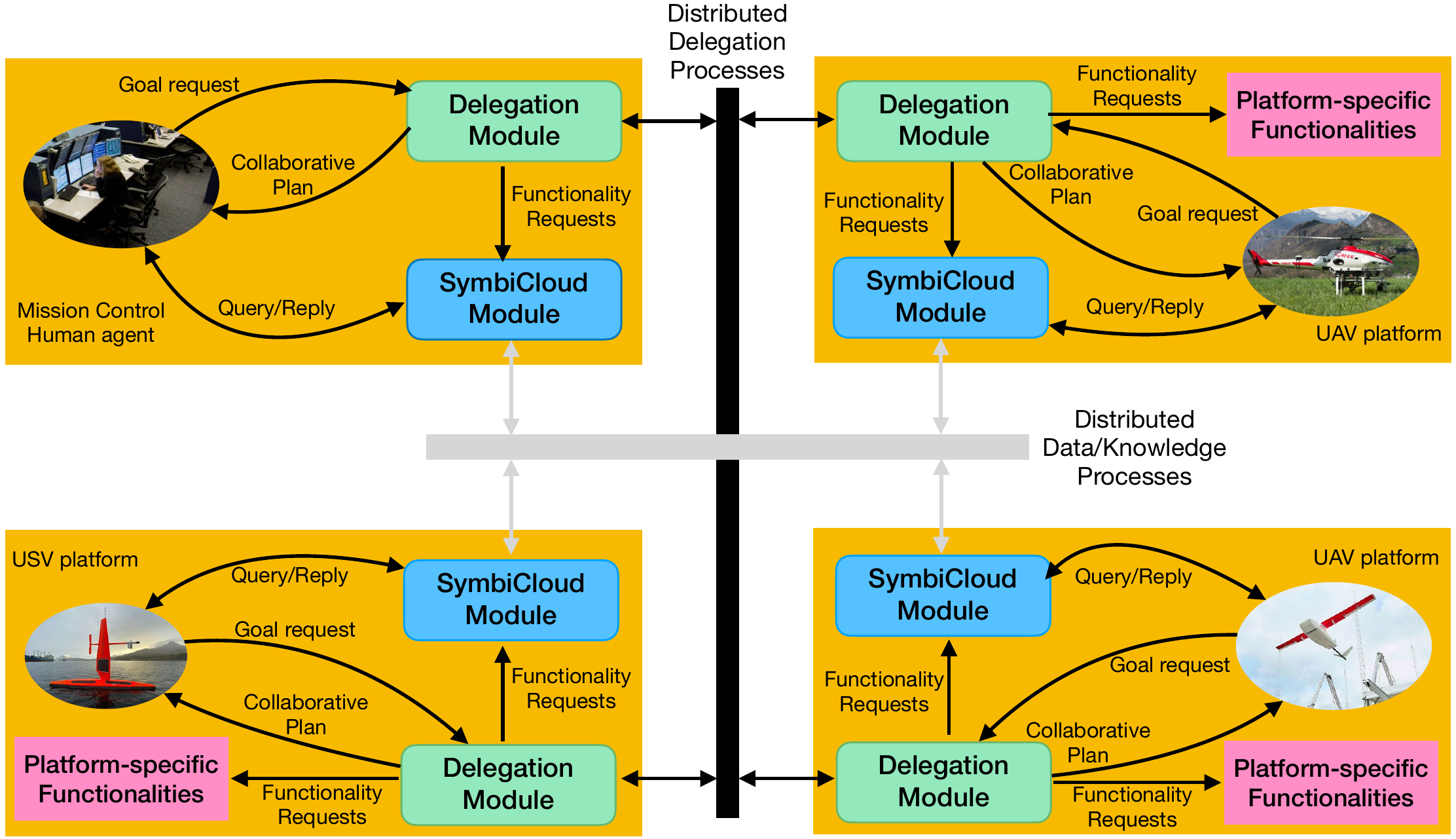}
\caption{High-level schematic - the collaborative robotics framework assumes that each participating agent has a \textit{Delegation Module} and a \textit{SymbiCloud Module}. Agents on a team generally communicate via speech acts. Each agent is assumed to be ROS compatible and leverages ROS middleware functionality while generating distributed tasks and distributed situation awareness. }
\label{fig:1}
\end{figure}

 
\Cref{fig:1}  provides a high-level schematic of the basic data/knowledge functionality we are interested in providing for  teams of human/robotic agents. In this figure, there are four agents, one human agent, one Unmanned Surface Vehicle (USV) and two Unmanned Aerial Vehicles (UAVs). Each agent has the functionality to generate and share collaborative tasks through the use of a distributed delegation framework which is part of the HFKN Framework, in addition to generating, synchronizing, querying and sharing data/knowledge through the use of the distributed HFKN Framework.

\begin{figure}[tb]
\centering
\includegraphics[width=0.8\linewidth]{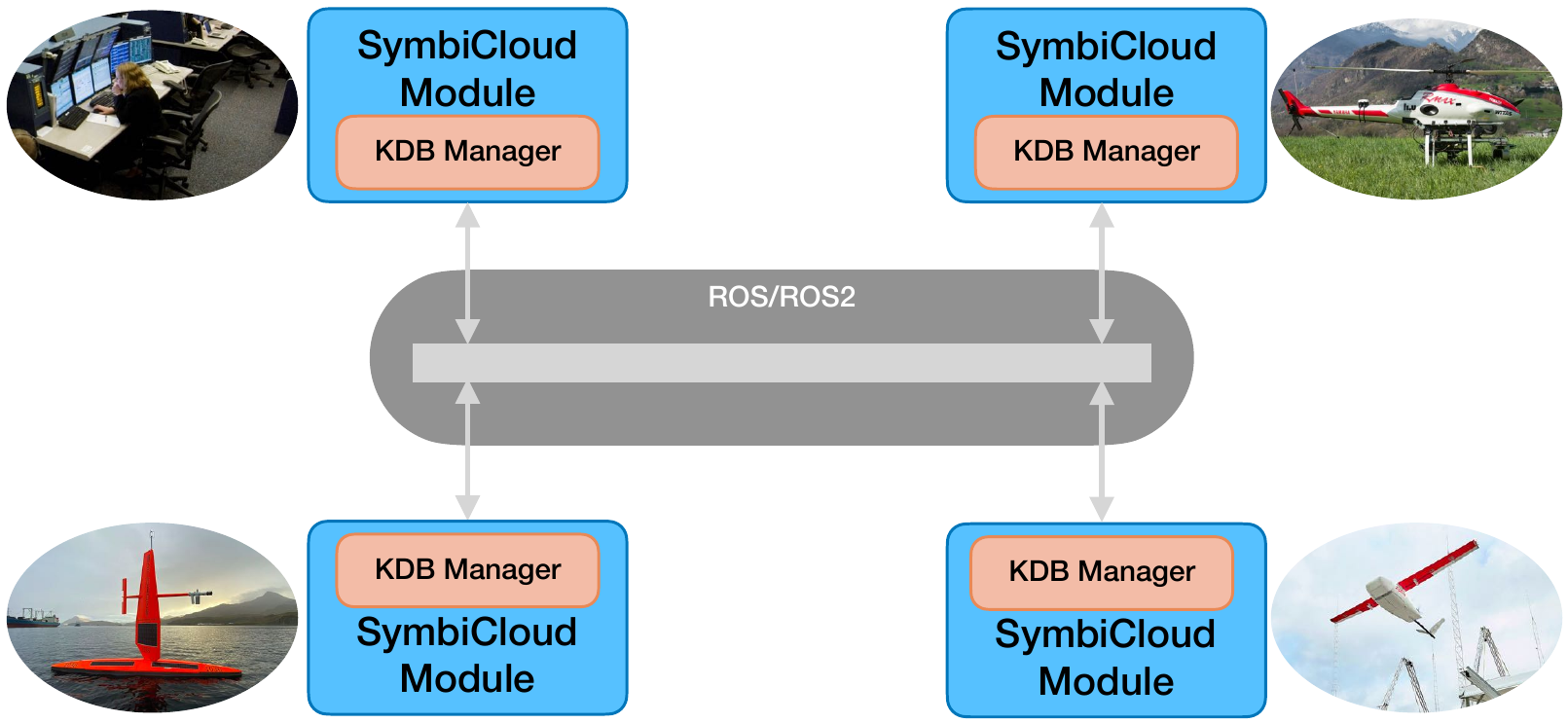}
\caption{In a multi-agent system, distributed information resources can be viewed as a dynamic graph that can grow and shrink as agent members enter and exit a particular mission in an operational environment.}
\label{fig:2}
\end{figure}

For the purposes of this paper, one can abstract the distributed knowledge network from the high-level schematic depicted in Figure~\ref{fig:1}. This aspect of the framework is shown in Figure~\ref{fig:2}.

Each agent in the system is assumed to have associated with it, a \textit{SCModule}. Additionally, it is assumed that each robotic agent is ROS~\cite{Quigley09} compatible\footnote{ROS is an acronym for the Robot Operating System, a middleware framework for robotics. For more information, see \url{https://www.ros.org}. Both ROS and ROS2 are currently used in the framework because ROS2 has better communication support for multi-agent systems. At a later date, full transition to ROS2 will be made.}. Each \textit{SCModule} stores the local or contextual situation awareness of an agent, structuring associated data and knowledge at different levels of abstraction. An \textit{SCModule} can in part, be viewed as a generalization of a layered Geographical Information System (GIS), but with a much richer variety of data and knowledge structures and more general querying mechanisms to access information in \textit{SCModules}. Additionally, SCModules include a KDB Manager for synchronizing and merging information and knowledge content across agents. 

Robotic and human agents are intended to have access to data, information and knowledge at many different levels of abstraction ranging from low-level sensor data, such as point data in single scans from LIDAR (Laser Imaging Detection And Ranging sensor), or collections of images from camera sensors. Intermediate levels of data and knowledge may contain 3D or infrared maps, that are the result of post-processing of low-level sensor data. These structures may in turn be semantically labeled with identifiable geolocated objects and additional semantic properties. Relations between such objects may then be defined and information stored about both static and dynamic activities of such objects. High-level semantic representations provide qualitative models possibly grounded in lower-levels of the knowledge and data abstractions associated with an agent.

In actual emergency rescue scenarios, there are a number of important contingencies that arise that make the design and implementation of the functionalities included in the HFKN Framework relatively complex:
\begin{itemize}
\item Unreliable communication between agent systems.
\item Out-of-range issues between agent systems.
\item Agent systems entering and exiting operational environments dynamically.
\item Agent systems turning off for such activities as recharging or refueling.
\item Queries that return no data, or only partial data in the context of data required by the querying agent.
\end{itemize}
Such contingencies present an additional level of complexity in designing the functionalities of interest. These issues must be dealt with in order to build robustness and resiliency into the framework design and use. Some of these issues can be dealt with by leveraging existing communication functionality in middleware such as ROS/ROS2, but other aspects must be taken into account in the associated algorithms as will be shown. 

\subsection{Contributions and Content}

The paper includes the following contributions:

\begin{itemize}
\item A description of a general system architecture (SymbiCloud HFKN Framework) for supporting multi-modal data/knowledge storage, in addition to the dynamic aggregation, sharing, transfer and querying of such data/knowledge in multi-agent contexts consisting of human and robotic agents. Information ranges from low-level sensor data collected by robotic sensors to high-level semantic knowledge.  
\item An integration of an existing delegation-based multi-agent framework with the HFKN Framework that together is used for automatic generation and execution of collaborative data/knowledge-collection tasks. 
 \item A synchronization algorithm and protocols which allows for the automatic distribution and sharing of information and knowledge between agents, through the synchronization of RDF (Resource Description Framework) Documents/Graphs.
 \item A data-transfer algorithm and protocols for exchanging datasets of low-level sensor data among agents based on the use of metadata about such datasets.
 \item An empirical evaluation of the HFKN Framework and associated algorithms.
 \item A field robotics emergency rescue case study based on a multi-agent data collection mission that uses all described functionalities of the HFKN Framework.
\end{itemize}

The paper is structured as follows. In~\Cref{sec:1}, the basic context for this work in addition to a conceptual description of the HFKN Framework has been introduced. In~\Cref{sec:2}, a brief summary of the Delegation Framework used by the HFKN Framework is described. In~\Cref{sec:SCModules}, the structure and content of \textit{SCModules}, that provide the data and knowledge content of agents is presented. In~\Cref{sec:dataset}, the concept of a dataset which specifies a collection of sensor data in addition to its metadata description is provided. 
In~\Cref{ssec:graph_synch_protocol}, the main algorithm and protocol for distributed data and knowledge synchronization used by the HFKN Framework is presented. In~\Cref{sec:data_exchange}, the processes as to how agents exchange low-level sensor data via a dataset transfer protocol is presented. In~\Cref{sec:validate_synch}, a number of  validation experiments related to the synchronization and dataset transfer protocols are provided. In~\Cref{sec:case_study}, a field robotics case study with collaborating UAVs and human agents that displays the power of the HFKN Framework in actual robotic scenarios is described. \Cref{sec:related_work} and \Cref{sec:conclusions} provide a description of related work and conclusions, respectively. The paper also includes two appendices that provide detailed descriptions of schemas used in \textit{SCModules} and examples of complex SPARQL queries that can be expressed by the system.

\section{Brief Summary of the Delegation Framework}\label{sec:2}

In the introduction, we provided an overview of the HFKN Framework which is the focus of this article. This component is part of a larger framework for collaborative robotics which leverages the use of a delegation framework~\cite{DohMey:2012:605381,DOHERTY-US-2013,doherty2015,HinStaEtAl:2017:1178264} for generating and executing complex, multi-agent distributed plans and tasks. An overview of the full architecture that combines the two is shown in Figure~\ref{fig:1}. In order to understand how the HFKN Framework leverages the delegation framework, it is important to have a cursory understanding of this framework. This section is intended to do that.

\begin{figure}[tb]
\centering
\includegraphics[width=0.8\linewidth]{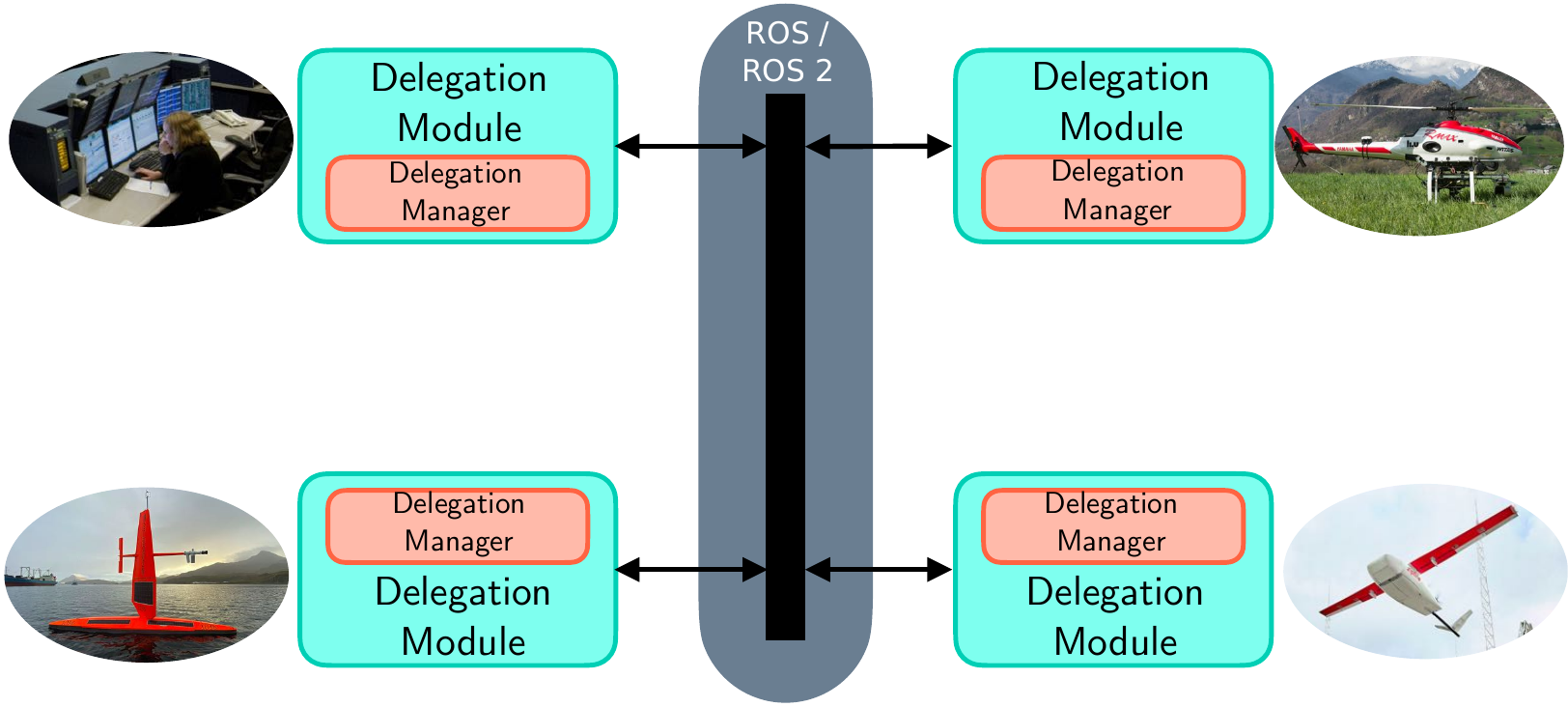}
\caption{In a multi-agent system, the distributed task generation and execution network can be viewed as a dynamic graph that can grow and shrink as agent members enter and exit a particular mission in an operational environment.}
\label{fig:5}
\end{figure}

As in the case of the distributed knowledge network, one can abstract an instance of the distributed delegation network, from the high-level schematic depicted in Figure~\ref{fig:1}. This aspect of the framework is shown in Figure~\ref{fig:5}.

\begin{figure}[tb]
\centering
\includegraphics[width=0.7\linewidth]{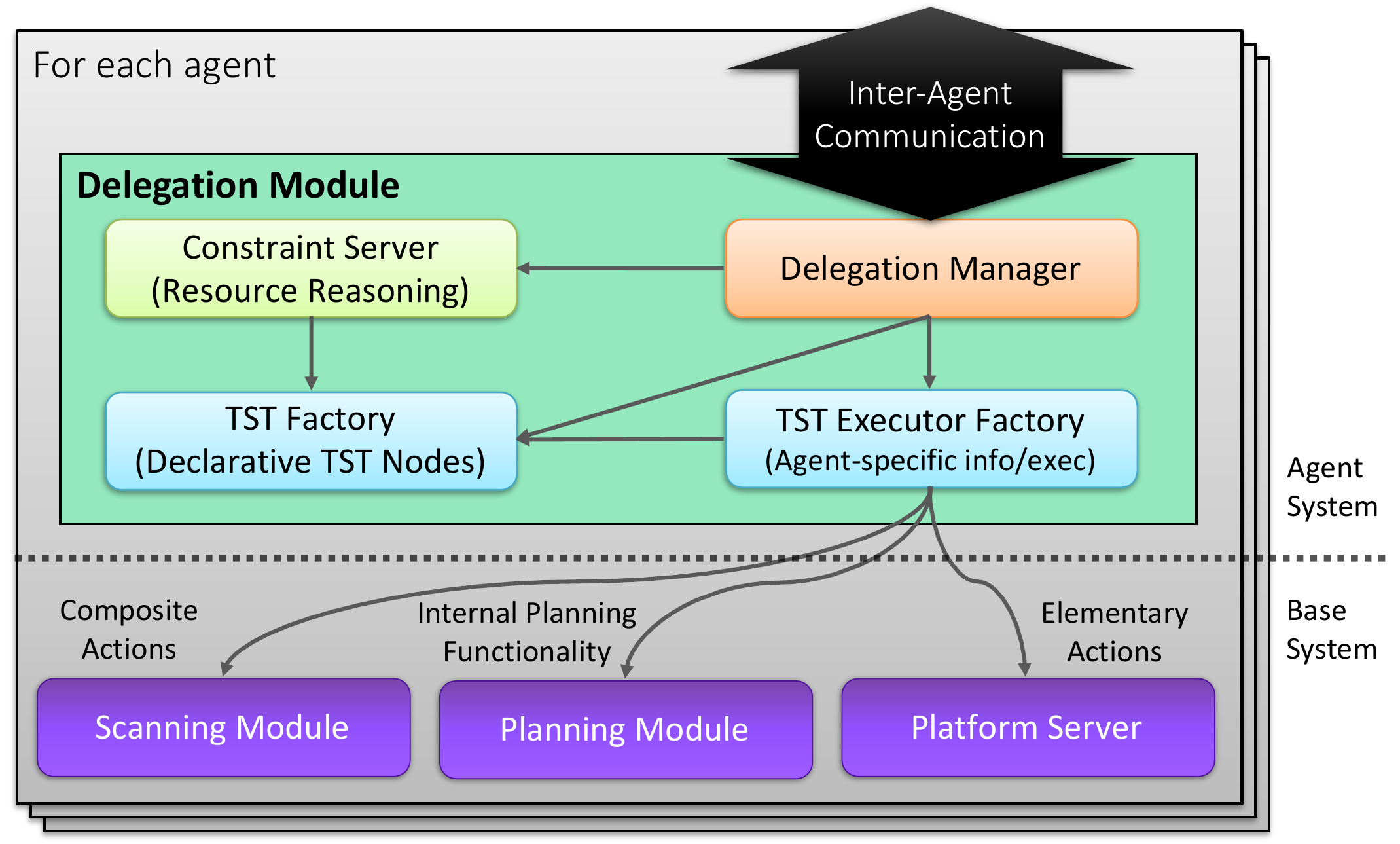}
\caption{\textit{Delegation Module} associated with each collaborative agent on a team.}
\label{fig:6}
\end{figure}

In the delegation framework, each member of a collaborative team is assumed to have a \textit{Delegation Module} associated with it. An agent's \textit{Delegation Module} contains a Delegation Manager that manages the external interactions with other agents on the team, in addition to internally managing the generation and execution of composite tasks~\cite{DohKvaSza:2012:489882}. Figure~\ref{fig:6} provides a high-level characterization of the internal architecture of a \textit{Delegation Module}.

Given a high-level mission goal specification, provided by a member of the team, the purpose of the delegation framework is to dynamically generate a task specification (often distributed among agents) to achieve the goal. This task specification often involves the use of a subset of members of the team. The task specification is generated recursively through a process where participating team members agree to do a part of the mission if they have the required resources and are able to commit to doing that part of the task specification. Each team member has the ability to broadcast for help in achieving sub-tasks associated with the larger mission specification. If successful, the net result of the process is the generation of a task specification where different parts of the specification are allocated to appropriate members of the team.

Tasks are represented as Task Specification Trees (TST)~\cite{DohHeiLan:2012:601009,DohKvaSza:2012:489882}. A TST consists of a set of control nodes and a set of elementary actions (leaf nodes) provided by participating members of the team. For simplicity, control nodes can be viewed as standard forms of control in programming languages such as sequence, concurrency, if-then, etc. During the delegation process, a tree is constructed recursively through calls to participating agents where they contribute to the overall task by providing sub-trees they are able to commit to and execute.

 It is assumed that each agent publicly commits to a set of elementary and composite actions that can be used in the collaborative delegation process. In the case of UAVs, examples of elementary actions would be such actions as \emph{FlyTo}, \emph{TakeOff}, or \emph{Land}. Composite actions might consist of more complex activities such as scanning of a region, that although internally complex for the agent, are viewed externally as elementary actions that can be used by the team to generate more complex task specifications collectively.
 
\begin{figure}[tb]
\centering
\includegraphics[width=0.8\linewidth]{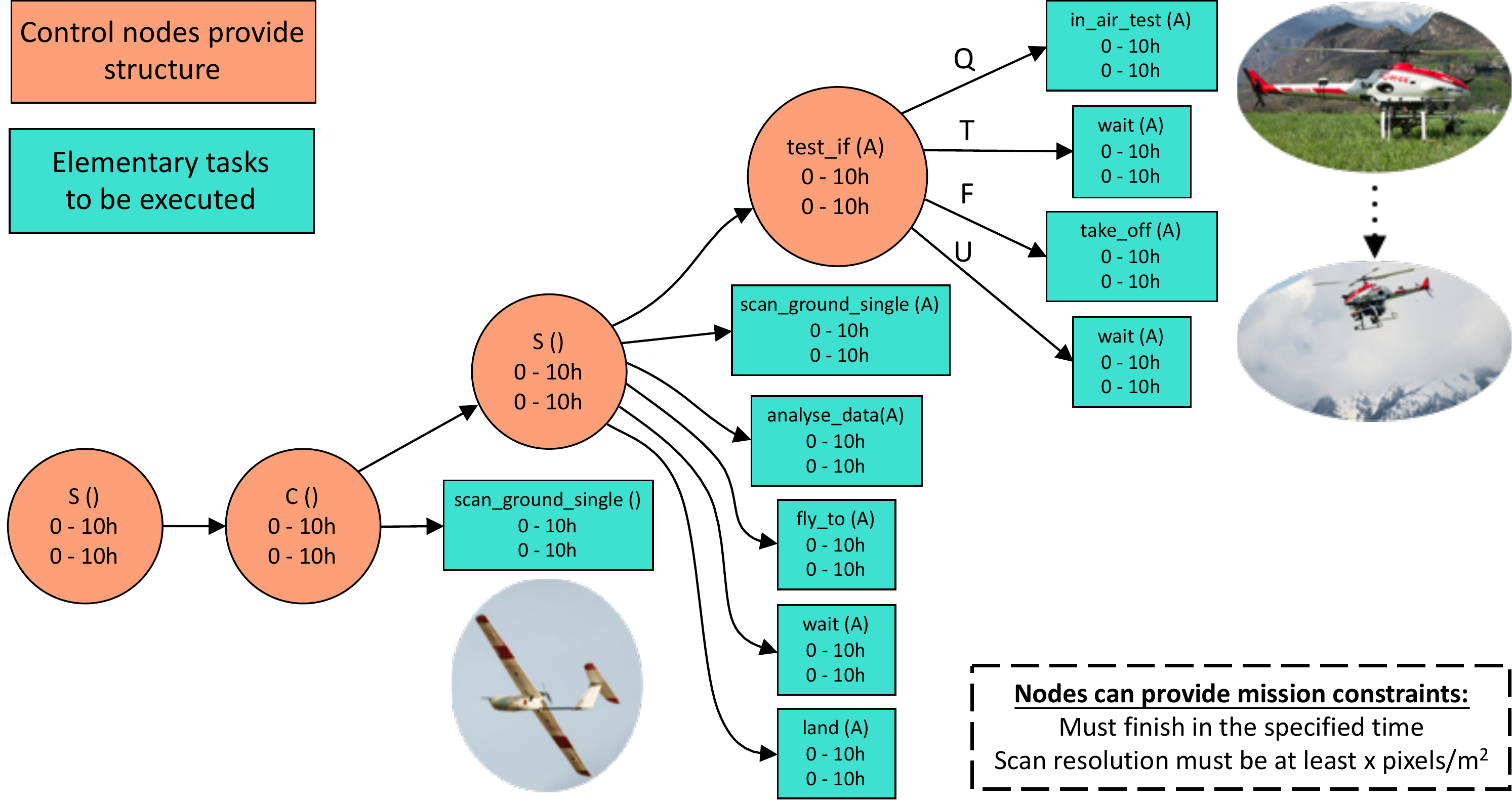}
\caption{TST generating a concurrent scan by a fixed-wing and a rotor-based system. A Test-If node first checks if the rotor-based system is on the ground or in the air. From the agent's perspective, a Scan-Ground-Single task is a composite action, while from the delegation perspective it is an elementary action.}
\label{fig:7}
\end{figure}

 Figure~\ref{fig:7} depicts an example of a TST generated for a concurrent scanning mission consisting of two UAV agents.
 
Each \textit{Delegation Module} as shown in Figure~\ref{fig:6} consists of four conceptual components:
\begin{itemize}
\item Delegation Manager - It provides inter-agent communication to other members of the team during the delegation process. Internally, it accesses the TST Factory to generate TST nodes during the TST generation phase and the TST Executor factory to execute TSTs during the execution phase.
\item TST Factory - It has the ability to generate TST nodes and TST sub-trees during the TST generation phase in the delegation process. 
\item TST Executor - Associated with each elementary or composite action publicly declared by an agent, is a platform  dependent  executor that interfaces to an agents internal functionality. The TST executor is responsible for interfacing to and managing the execution of executors associated with elementary or composite actions for a specific platform. If a TST node is a goal node type, the TST executor also has the possibility to interface with an automated planner associated with a platform to generate a sub-tree from the planner that can then be used by the TST factory.
\item Constraint Server -  TST nodes can contain constraints that are inherited as the delegation process progresses. In order for an agent  to answer the question ``\textit{can I do this?}'' when it receives a request from another agent, it autonomously sets up a constraint problem and checks the problem for consistency, possibly returning specific variable bindings. The constraint server handles this part of the generation process. For instance, constraints can be temporal, resource based, or associated with sensor capability.
\end{itemize}

In the context of the HFKN Framework, the delegation functionality will be used to generate and execute distributed information gathering and manipulation tasks for teams of collaborative humans and robots. Examples of information gathering and manipulating TSTs will be described in the field study, presented in \Cref{sec:case_study}. 

\section{SymbiCloud Modules}\label{sec:SCModules}

SCModules are associated with each agent and store combinations of data, information and knowledge. Additionally, SCModules are responsible for sharing, synchronizing and aggregating data, information and knowledge between agents.

\begin{figure}[b]
\centering
\includegraphics[width=0.8\linewidth]{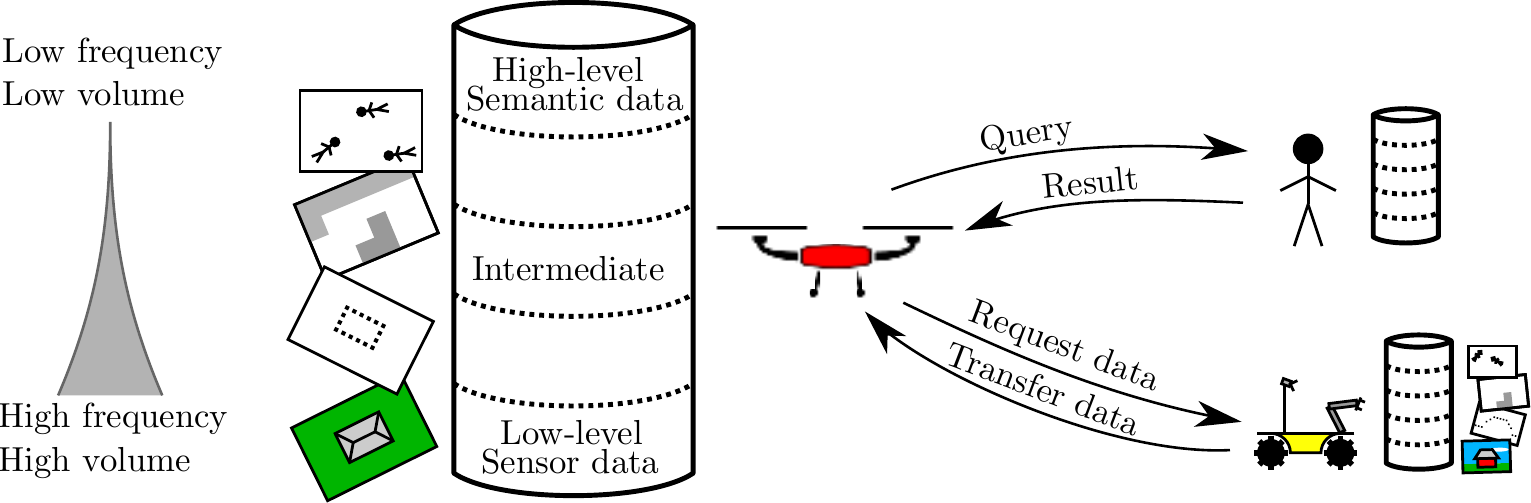}
\caption{In a multi-agent system each agent is equipped with its own \textit{SCModule} providing its particular perspective of the world. To participate in collaborative missions it is often necessary for agents to access information from other agents' \textit{SCModule} such as query for knowledge stored or request the transfer of data.}
\label{fig:4}
\end{figure}

We characterize three types of \textit{data} (in the general sense) loosely coupled to the concepts of data, information and knowledge that are acquired, stored, aggregated and shared by agents
(see \Cref{fig:4}):
\begin{itemize}
  \item \textit{Low-level sensor data} characterizes raw output from diverse sensors. Examples include data in  images from camera sensors or point data in single scans from LIDAR systems.
  \item \textit{Intermediate data} characterizes processed raw data with minimal to moderate structure. Representation of such data are often collections of feature/value pairs stored in tables but could also be the output of sensor fusion algorithms such as point clouds and 3D maps.
  \item \textit{High-level semantic data} characterizes data generally structured around objects with properties and relations between objects. Such data is normally semantically grounded. Common examples would be ontological structures, logical structures, or graph-structured knowledge.  Objects representing humans identified in search and rescue missions with feature data associated with each human is a good example. 
\end{itemize}

These different types of data pose different requirements on storage, synchronization and query mechanisms due to each having different characteristics.
Low-level sensor data is typically very well structured and normally acquired in high volumes with high frequency. It is important that efficient data structures are used for storage and retrieval. This data is typically retrieved with simple queries by requesting sets of ordered and timestamped frames. High-level semantic data, on the other hand, requires low-volume for storage and is normally generated with low frequency, but it requires more complex query mechanisms. High-level semantic knowledge  require structures and representations suitable for distributed querying, merging and synchronization. \Cref{fig:4} depicts this informatic situation for several agents. 

\begin{figure}[tb]
\centering
\includegraphics[width=0.7\linewidth]{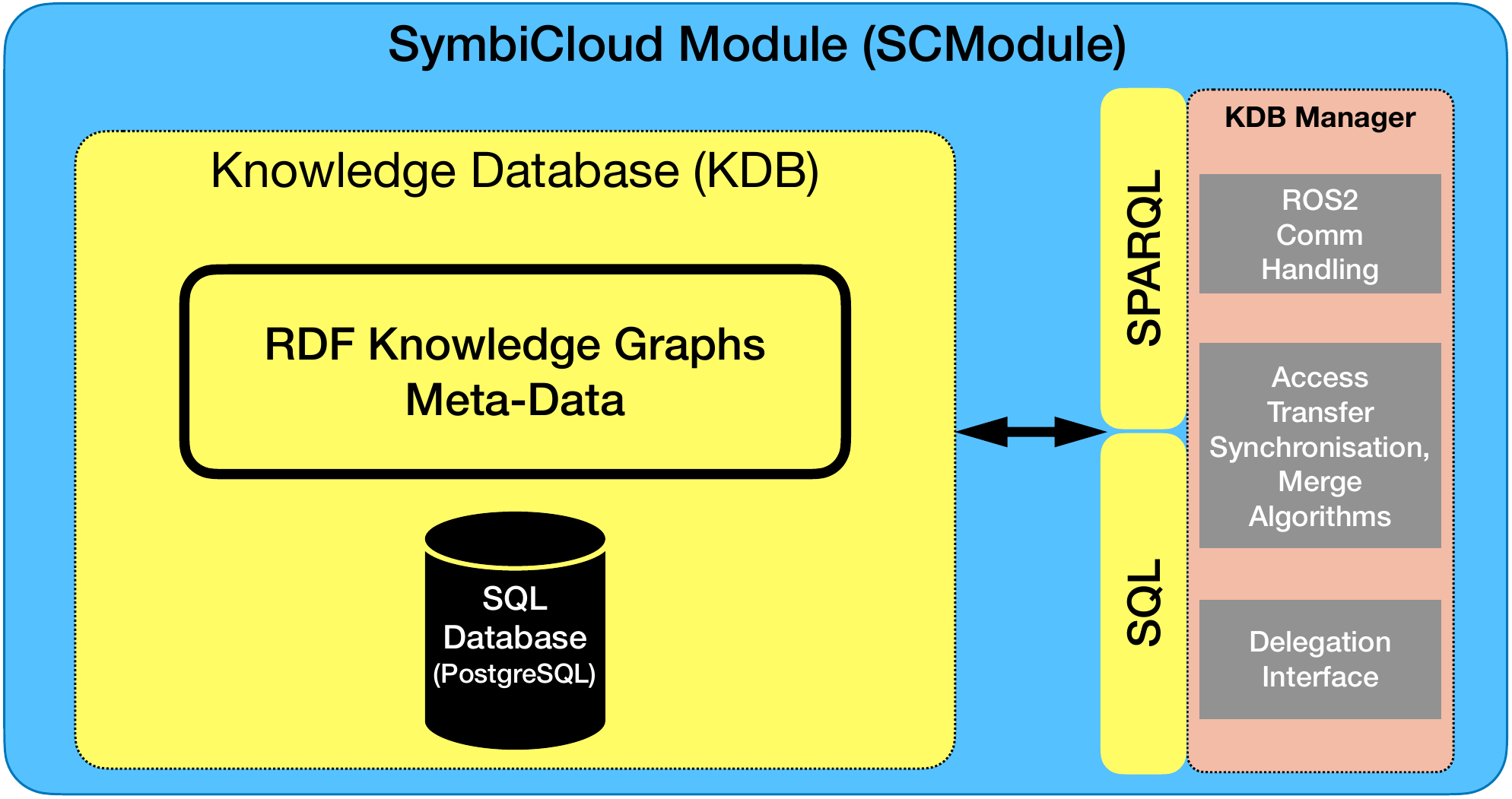}
\caption{\textit{SCModule} associated with each agent.}
\label{fig:3}
\end{figure}

 Each \textit{SCModule} as shown in \Cref{fig:3} consists of three conceptual components:
\begin{itemize}
\item A PostgreSQL database - This database generally stores low-level sensor data and intermediate data, in addition to any table-based information used by the agent.
\item A repository of  RDF Documents/Graphs - This RDF Document/Graph repository includes both object-level semantic content in addition to  metadata representations of low-level sensor  data and intermediate-level data stored in the PostgreSQL database. It is these structures that will be shared and synchronized between agents using the Knowledge Database Manager.
\item A Knowledge Database (KDB) Manager -  The KDB Manager is responsible for the management of the PostgreSQL database and the RDF Document/Graph repository. It also supports communication with other agents and handles dataset generation processes, in addition to the execution of synchronization and dataset transfer algorithms between agents. 
\end{itemize}

In order to represent high-level semantic information, \textit{SCModule}'s use the Resource Description Framework (RDF) \cite{LASSILA-TR-1999}. RDF provides a representational foundation for the Semantic Web and was developed within the World Wide Web Consortium (W3C). The RDF framework has a very simple, but powerful set of representational primitives. Most important are RDF Triples consisting of a $subject$, $predicate$ and $object$. $Subject$ and $object$ are nodes in an RDF Triple and $predicate$ is a label for the edge connecting them that expresses a relationship between the $subject$ and the $object$.
$Subject$ and $predicate$ are specified using Internationalized Resource Identifiers~\cite{RFC3987} (IRI's), that provide a unique identity to any subject or predicate,  while $object$ can have any value including an IRI.

\begin{figure}[htb]
\centering
\includegraphics[width=0.6\linewidth]{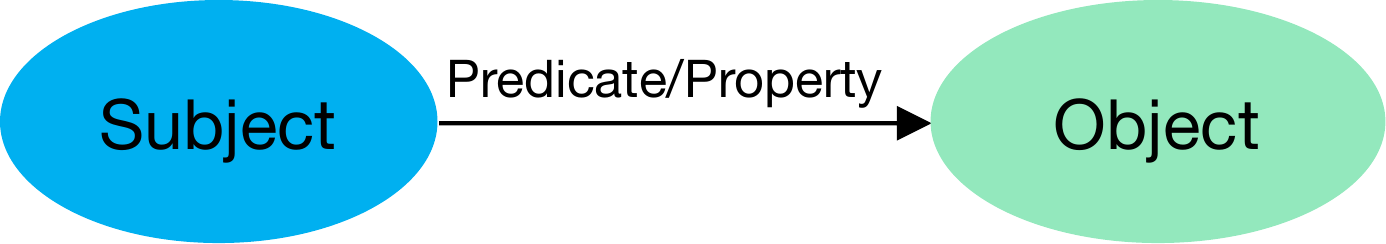}
\caption{Components in an RDF Triple.}
\label{fig:rdf_triple}
\end{figure}

It is useful to view RDF Triples graphically (see \Cref{fig:rdf_triple}), since collections of such interconnected triples can be conceptually viewed as RDF (knowledge) Graphs. Logically, an RDF Triple ($subject$, $predicate$, $object$) corresponds to an atomic formula, $predicate$($subject$, $object$). This correspondence is very powerful since it provides formal semantic interpretation of collections of RDF Triples as logical theories. There are many extensions to the RDF specification such as RDF Schema (RDFS)~\cite{brickley2014rdf} and the Web Ontology Language OWL~\cite{mcguinness2004owl} which extends RDFS and is used for specification and reasoning about ontologies.

 Although a collection of RDF Triples can be viewed as an RDF Graph, such graphs are encoded as RDF Documents. Consequently, we use the terms RDF Document and RDF Graph more or less interchangeably in the paper. Later, we will define an extended version of an RDF Document, denoted RDF$^{\oplus}$ Document, that is given as input to the RDF Graph synchronization algorithms used to synchronize shared RDF Documents between agents.
A common way to store and access content in an RDF Graph is to store the triples in a SQL database, create appropriate RDF Views for the content, and then use SPARQL (SPARQL Protocol and RDF Query Language)~\cite{SEABORNE-W3C-2008} together with these RDF Views to query RDF Graphs. 

This is the approach we take. The KDB Manager in an agent's \textit{SCModule} provides the ability for the agent, and other external agents, to access and query RDF Graphs through the use of SPARQL. From an implementational perspective, all RDF Graphs are represented in table form in the PostgreSQL database. In this case, the KDB Manager has a library of views (schemas) created using RDF View~\cite{sahoo2009survey}. RDF View provides a means of automatically converting any SQL table into a set of RDF Triples. The Sparqlify language~\cite{STADLER-LDOW-2015} is used to specify RDF Views.

Low-level sensor data and intermediate data is stored in SQL tables in order to efficiently support high volume and frequency data storage and retrieval via use of a PostgreSQL database and its associated SQL query language. Theoretically, one could translate any SQL tables into RDF Graphs, but it has been shown that SQL databases have on average between 5 to 10 times higher performance than SPARQL databases~\cite{bizer2009berlin}. This is one reason we retain a conceptual separation between low- and intermediate-level data, with high-level semantic data in the KDB. 

Although an agent can also access raw sensor data in the PostgreSQL database via the KDB Manager using SQL (or SPARQL, if the requisite RDF Views have been created), such direct querying of raw data has limited usage in the types of missions we envision. 
Instead, collections of raw data  are bundled as a \textit{dataset} and information about its contents is represented as metadata using RDF Graphs. Metadata in this context is high-level semantic information about a collection of sensor data acquired by an agent during an information gathering mission. Agents query metadata about respective datasets in order to determine what raw sensor data a team of agents has collected.  Datasets represent only a subset of information an agent can have, but this form of information and its conceptualization as a dataset, is practical and efficient in data-collection missions.

Each agent has the ability to declare RDF Documents in its KDB as public or private. Only the public information will be accessible to other agents for sharing. One could imagine more fine-grained dynamic approaches to access-control, such as mission-based or agent role-based access control, but this is left for future work.

\begin{figure}[tb]
\centering
\includegraphics[width=0.9\linewidth]{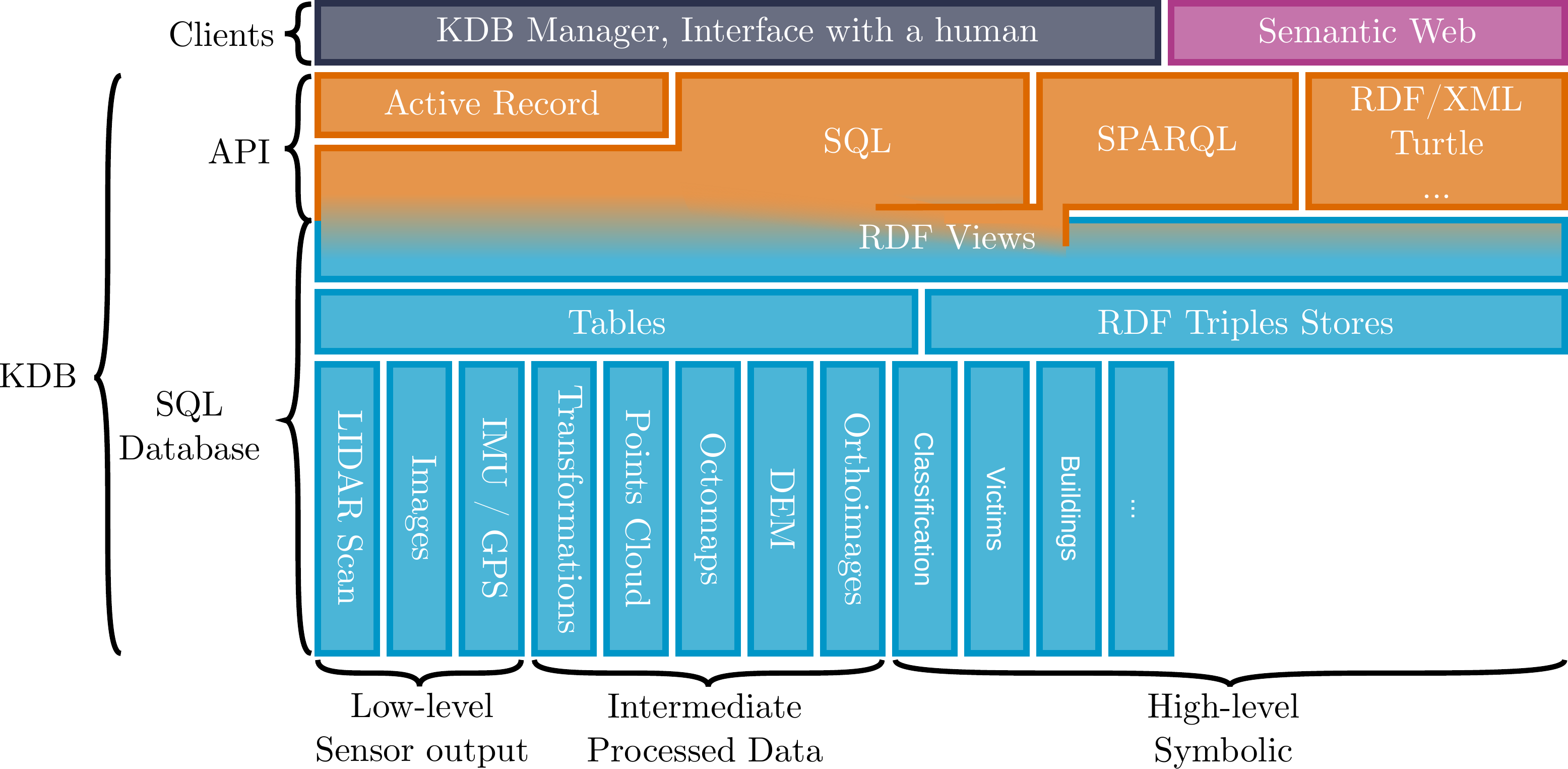}
\caption{More detailed \textit{SCModule} architecture for a single agent. All data is stored in SQL tables, and it can be queried through SQL or through semantic queries (SPARQL). The application layer handles interfaces between the KDB and a robot, a human or general sources of knowledge such as the Semantic Web.}
\label{fig:kdb_architecture}
\end{figure}

A more query-centric and implementational view of the \textit{SCModule} is shown in \Cref{fig:kdb_architecture}. All raw sensor data, intermediate- and high-level semantic data resides in the SQL Database (PostgreSQL). The low-level data is stored using tables with a schema specific to the type of data, while the high-level symbolic data is stored in tables with a schema suited for storing RDF Triples. These components are visualized with blue boxes in the figure.

The KDB provides four types of APIs to access the data stored in the SQL Database (orange boxes in the figure).
First, a high-level mapping between tables and programming language objects is provided using Active Record~\cite{fowler2012patterns}.
Second, the data in tables can be queried directly using SQL.
The two remaining APIs are compatible with Semantic Web technologies. Namely, SPARQL in combination with RDF Views and RDF/XML, Turtle (or other formats) can be used to access data represented as RDF Triples in the SQL Database. These interfaces are used by the KDB Manager, user interfaces or external Semantic Web frameworks.
In an SCModule, the application layer and the query interface for the KDB are handled by the KDB Manager.

\section{Datasets}\label{sec:dataset}

Our focus in this paper is on complex data collection missions using multiple robotic systems with human interaction. Consequently, we specify a special structure, a dataset\footnote{The sensor data based datasets considered here are not to be confused with RDF Datasets.}, that is the result of any data collection mission. Additionally, agents usually require specific subsets of raw or intermediate data specified as datasets for decision making processes during mission operation. For example, an agent may want to access part of a 3D map or an orthophoto associated with a specific region in an operational environment. Consequently, datasets should be able to be created dynamically as subsets or aggregations of already existing datasets.

The description of a dataset is characterized by high-level semantic information which describes, references and annotates low- or intermediate-level data structures. This metadata description is stored as an RDF Graph. Large data, such as point clouds, images, etc.,  cannot generally be exchanged efficiently between agents at periodic intervals, but metadata about datasets can be efficiently exchanged.

Raw sensor data collections are grouped in datasets consisting of the actual raw sensor data that is stored in a PostgreSQL database and the associated metadata  represented as an RDF Graph. In the synchronization processes managed by the KDB Manager, it is the metadata associated with datasets that is synchronised between agent platforms.  When an agent does need to access and acquire actual raw sensor data from another agent, the agent would need to make an explicit request that the sensor data should be downloaded from another agent using the Dataset Transfer Protocol presented in \Cref{sec:data_exchange}.

\subsection{Dataset Representation}

A dataset $\mathcal{D}$ consists of a set of data points $ \{ d_i \} $.
Data points can be:

\begin{itemize}
  \item Raw sensor data, such as images and image sequences, LIDAR scans, IMU frame sequences, GPS positions, robot poses, etc.
  \item Results of sensor fusion algorithms that combine different types of raw data, such as a sequence of robot poses that use both IMU and GPS data, or the locations of humans or building structures, derived from complex classification algorithms.
  \item Any other high-volume data acquired from external services, such as systems based on Semantic Web or GIS technologies.
\end{itemize}

Data points are stored in database tables according to their types. The tables contain a set of fields that are specific to the type (for instance, IMU contains fields for linear acceleration and angular velocity, as well as covariances). All data points are associated with a Uniform Resource Identifier (URI) defining the dataset they belong to. Associated with the dataset URI are additional properties about the dataset. Dataset descriptions are defined using a collection of RDF Triples:

\begin{itemize}
  \item an \textit{URI} of a dataset, for example \textsf{ex:dataset1}.
  
  \item a \textit{geometry} corresponding to the area covered by the dataset, for example:
  
        (\textsf{ex:dataset1}, \textsf{geo:hasGeometry}, "POLYGON((...))")
        
  \item a \textit{type} indicating the type of data, for example, \textit{image}, \textit{LIDAR scan}, \textit{victims} (salient points representing potential victim locations):
  
        (\textsf{ex:dataset1}, \textsf{aiics:dataset\_type}, \textsf{aiics:points\_cloud})
  
  \item a list (possibly empty) of other \textit{datasets} (URIs) included in this dataset. This is useful to allow the modular structuring of a dataset into smaller component datasets:
  
        (\textsf{ex:dataset1}, \textsf{aiics:dataset\_include}, \textsf{ex:dataset1})\\[0.3mm]
        (\textsf{ex:dataset1}, \textsf{aiics:dataset\_include}, \textsf{ex:dataset2})
        

\end{itemize}

\begin{figure}
\centering
\includegraphics[width=0.6\linewidth]{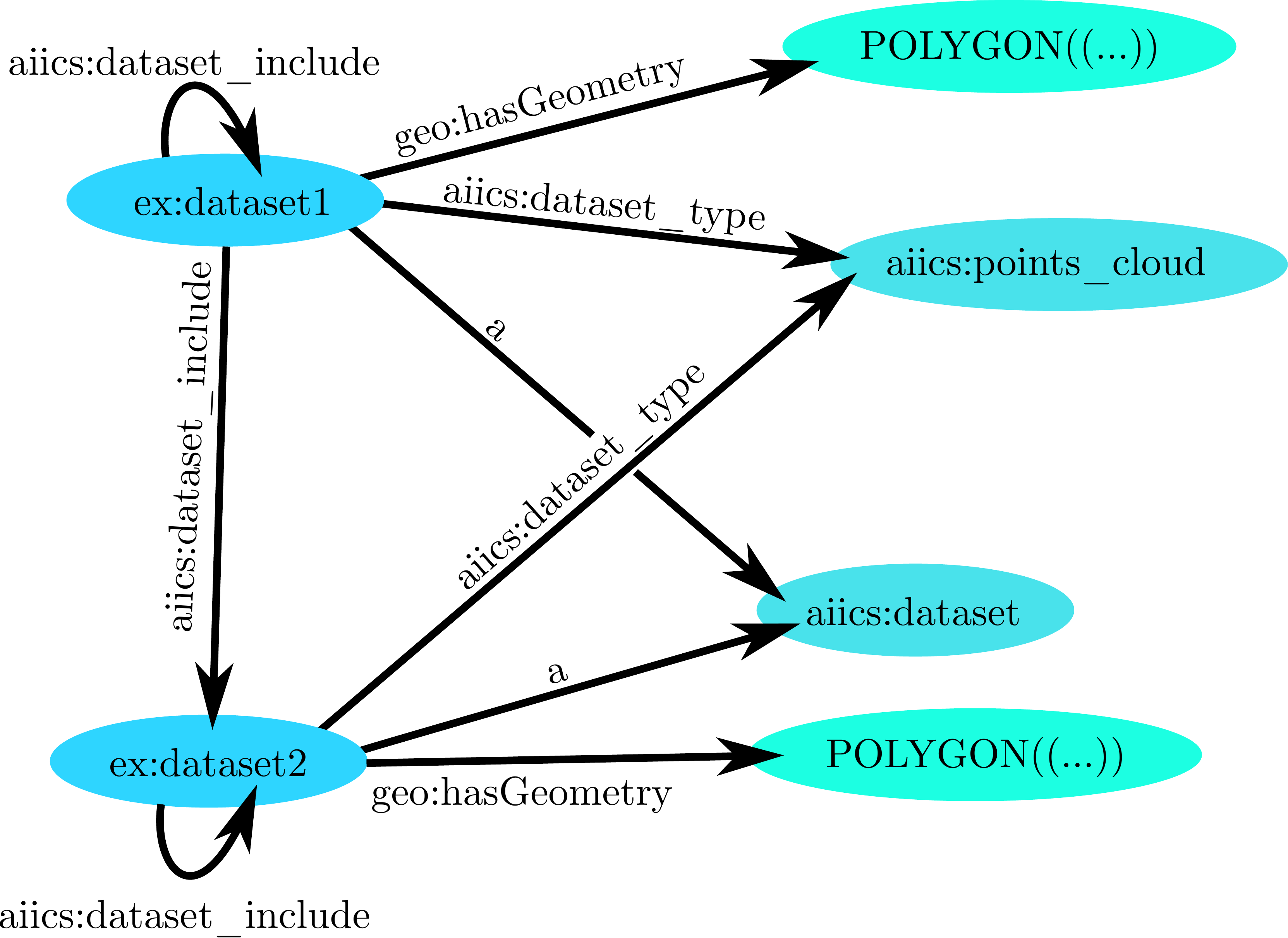}
\caption{RDF Graph representing metadata for two  point cloud datasets, where \textsf{ex:dataset2} is a subset of dataset \textsf{ex:dataset1}.}
\label{fig_example_datasets}
\end{figure}

\Cref{fig_example_datasets} shows an example RDF Graph representing metadata for two  point cloud datasets, where one is a subset of the other. For more examples of sensor data representations, see \Cref{sec:Appendix1}.

\subsection{Dataset Relation to Agents}

Given a team of agents, each agent stores a collection of RDF Graphs as part of the KDB in its SCModule. The union of collections of these RDF Graphs virtually represents the global knowledge of a team of heterogeneous agents. Since, the data/knowledge is distributed in multiple databases and owned by different agents, no agent has direct local access to the complete spectrum of information associated with a team.  In order to gain access to the required information necessary for achieving a specific mission goal (e.g. a global map of the operational environment), the agents have to collaborate to fully (or partially) synchronize their data/knowledge resources, or to eventually exchange raw sensor data. The synchronization process is intended to be fully automated. It occurs asynchronously, in particular when a team member updates a public RDF Document subscribed to by other agents.

Each agent is identified uniquely by its own URI. Datasets are coupled to specific agents by the creation of RDF Triples that relate the unique agent URI to datasets (also stored in RDF Graphs) via the following relations:

\begin{itemize}
 \item \textit{has} - which indicates that an agent has a copy of the raw data associated with the dataset in its database.
 \item \textit{created\_by} - which indicates which agent has acquired or processed raw data.
 \item \textit{created\_from} - which indicates that a dataset was created using raw data from another dataset. For instance, a point cloud can be the result of combining multiple LIDAR scans together from different agents.

\end{itemize}

\begin{figure}[tb]
  \centering
  \def\svgwidth{1.2\columnwidth}
\resizebox{0.7\linewidth}{!}{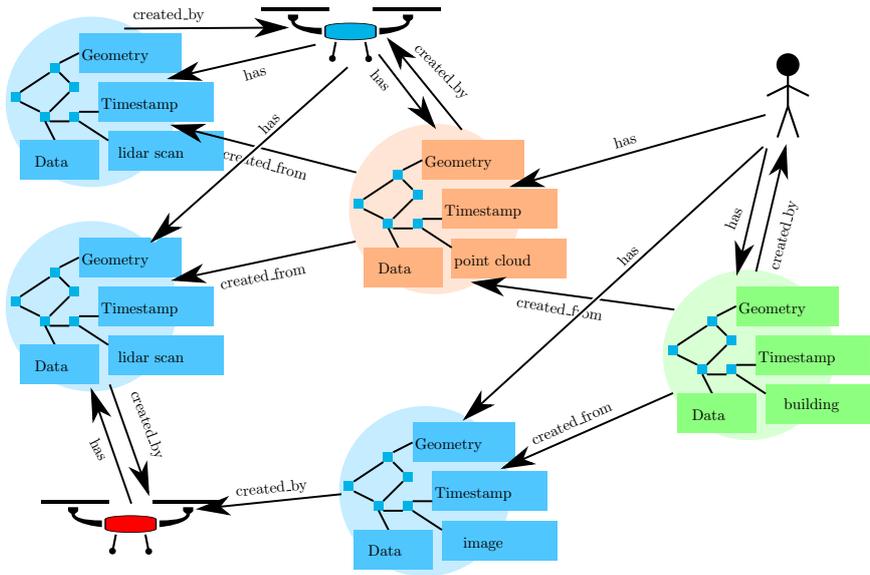}

  \caption{An example of a set of datasets and the relationship between them.
  There are three datasets (in blue) resulting from an exploration mission, created by two UAVs.
  A point cloud dataset (in orange) was created by fusing two LIDAR scan datasets. A building dataset (in green) was created by processing the point cloud and an image dataset.
  The graph shows how each dataset was created and also which agent currently holds a copy of a dataset.}
  \label{fig:multi_dataset}
\end{figure}

 \Cref{fig:multi_dataset} shows an example of a set of datasets owned or shared by two UAV platforms and one human operator.

\subsection{Dataset Discovery} 

Agents are aware of their team members via their unique URIs and by leveraging identification functionality in the framework. Additionally, as agents act in operational environments and create datasets, each has an RDF Document listing those datasets and other information publicly available. A Publish-Subscribe ROS-based strategy is used to declare publicly accessible documents that can be subscribed to by any agent. An agent's RDF Document listing public datasets is published by default. Generally, an agent subscribes to all public datasets specified by this RDF Document for each team agent, although the subscription strategy can be modified by any agent at any time. 

This team awareness implies that agents have the ability to query each other locally through SPARQL interfaces managed by their KDB Managers. For instance, if one agent is interested in a specific geographical region, it can use its synchronized shared information and discover what type of information and data associated with that geographical region is available from the team.

\begin{figure}[b]
 \centering
 \includegraphics[width=0.5\linewidth]{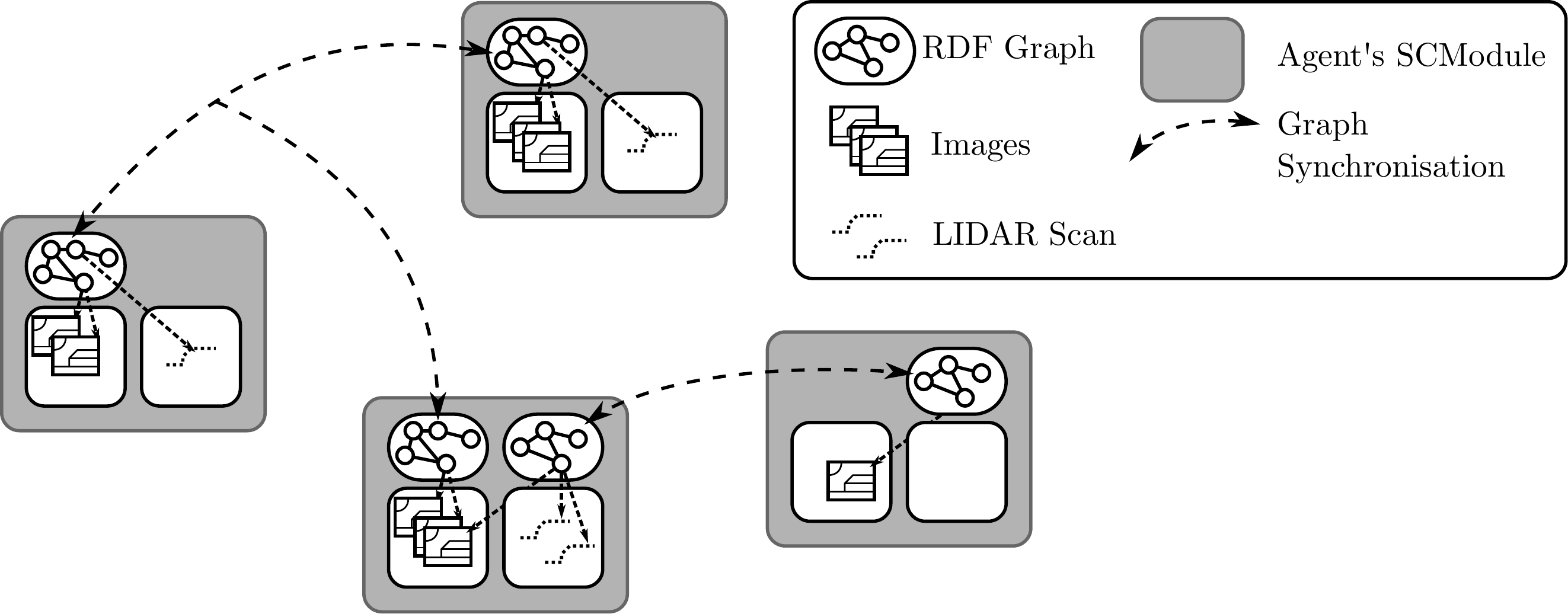}
 \caption{
 Example illustrating a team of four agents creating datasets and synchronizing selected metadata among themselves. The data stored in an agent is represented by gray box. Agents store different types of data (e.g. images or LIDAR scans) as datasets and the metadata associated with each dataset is represented as an RDF Graph. Each agent stores different datasets and it subscribes to selected documents. The dashed line shows examples where RDF Graphs are synchronised between agents.
 }
 \label{fig:kdb_sync}
\end{figure}

\Cref{fig:kdb_sync} shows an example of a set of synchronized multiple datasets created and partially shared by a team of agents after querying each other and subscribing to the relevant RDF documents of interest (datasets).

In order to exchange the  actual raw sensor data associated with a dataset as described by its metadata, the delegation framework is used with a TST (Task Specification Tree) containing two nodes for  uploading and downloading raw data between the corresponding agents. In the Delegation Framework, a TST is both declarative and executable, so such TSTs essentially drive the execution of  dataset transfer processes.
Downloading is executed on the agent receiving the data, where it first sets up the proper data storage capability. Uploading is executed on the agent sending data, where it extracts the specific data in its KDB and then sends it on the network to the receiving agent. The Dataset Transfer Protocol and processes are described in \Cref{sec:data_exchange}.




\subsection{Datasets and Multi-Agent Data Collection Missions}\label{sec:distribution_overview}

In a typical collaborative exploration scenario, each agent explores part of the environment, stores data collected during a mission locally, and shares meta-information about the collected data as a dataset with other agents that subscribe to that dataset.
Raw sensor data is only physically exchanged between agents if required. Some examples of when an agent would need large quantities of raw sensor data would be for visualization by a ground operator, for backup purposes, or for further processing such as extracting higher-level semantic information from the raw data. This could involve people identification or classification of building structures which would use specialized algorithms local to an agent.

\begin{figure}[tb]
  \centering
  \def\svgwidth{1.6\columnwidth}
\resizebox{0.9\linewidth}{!}{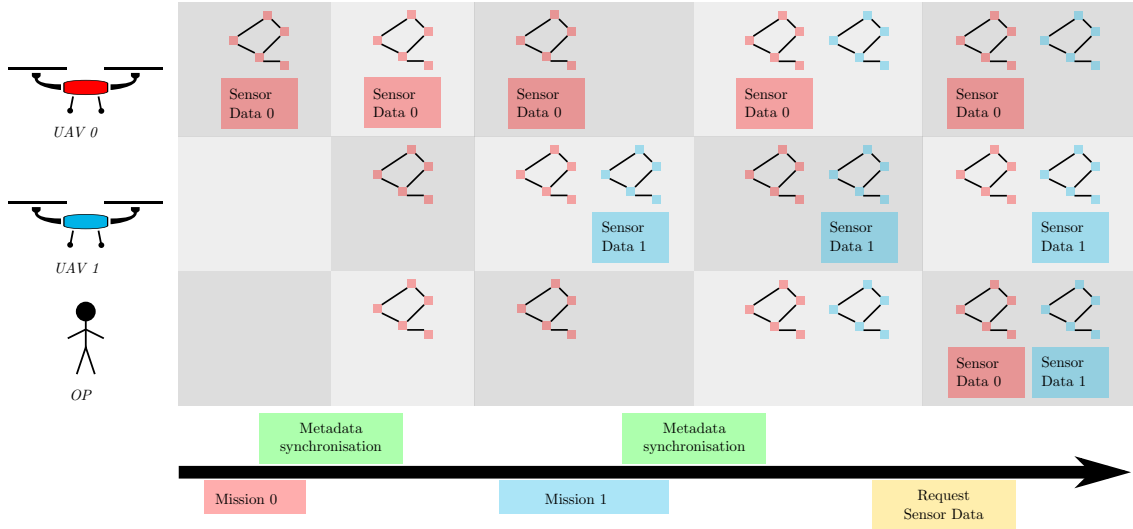}

  \caption{A timeline illustrating an example collaborative exploration scenario involving two UAVs (\textit{UAV 0}, \textit{UAV 1}) and a human operator (\textit{OP}). The UAVs perform two missions independently at different time points. The metadata describing collected sensor data is synchronised between all three agents after each mission is completed. The datasets collected during \textit{Mission 0} and \textit{Mission 1} are only available on a UAV which performed the respective mission.
 At the end the \textit{OP} would like to visualise both datasets and requests to download the sensor data from each UAV.
}
\label{fig:collaborative_exploration}
\end{figure}

\Cref{fig:collaborative_exploration} shows a typical data collection mission evolution from the perspective of the data/knowledge that is locally available to several interacting agents. The setup involves two UAVs and one human operator.
As can be seen, metadata about collected sensor data is synchronized between agents continuously while actual sensor data is exchanged only if an agent explicitly requests it. This is a \textit{by-need} strategy. These two processes are considered in detail in the following two sections.

\section{RDF Graph Synchronization}
\label{ssec:graph_synch_protocol}


In the HFKN Framework, high-level semantic information is represented as RDF Graphs, which are encoded as RDF Documents.
An RDF Document can contain one or more RDF Graphs. An RDF Dataset~\cite{CYGANIAK-W3C-2014}, not to be confused with our use of \textit{datasets} in~\Cref{sec:dataset}, is a collection of named RDF Graphs with no more than  one (unnamed) default graph.
Lower-level data, for example from sensors, is handled differently as described in the previous sections. This split allows for low-bandwidth synchronization of high-level knowledge and on-demand access to high-bandwidth low-level sensor data and intermediate data. This section focuses on algorithms in the HFKN Framework that deal with achieving a common view of the agents' shared knowledge in addition to its meta-knowledge about low-level sensor data. This is done through RDF Graph synchronization, which results in a common view of a team's shared RDF Graph collection and content.

Agents use RDF Graphs to store various types of information, such as agent capabilities, list of salient objects, metadata about datasets, general knowledge useful for decision making and planning, etc. Throughout a mission, agents make changes to those RDF Graphs. For instance, while collecting sensor data and processing it, an agent will need to update the metadata of the associated dataset. Achieving a common view of a shared RDF Graph between agents by simply exchanging it in its entirety is infeasible and impractical for several reasons.
First, it is inefficient when exchanging all of its content at all times. Moreover, agents may join and leave throughout the execution of a task or a mission. Communication between agents is also unreliable by nature since we deal with mobile robots and wireless communication links. This may result in partial transmissions or receiving incomplete information. Most importantly, a simple exchange of an entire set of RDF Graphs does not solve the consistency problem when multiple agents sharing the same RDF Graph make simultaneous updates.
Instead of communicating the whole RDF Graph content after every change, the HFKN Framework encodes changes as deltas representing incremental changes in an RDF Graph. These deltas are then used more efficiently in the synchronization processes. 

It is assumed that each agent has a dedicated RDF Document containing information about those RDF Graphs it will make publicly accessible. This information can of course be updated throughout an operational mission. As mentioned previously, a publish/subscribe mechanism is built into the system that allows agents to both publish (share) and subscribe to public documents. If agent A subscribes to an RDF Document from agent B, then that document becomes shareable by both agent A and agent B and can then be updated by both agents. Prior to this, the document can only be updated by Agent A (if there are no other agents that have already subscribed to this document). This generalizes to teams of agents collectively sharing and thus being able to update collections of RDF Graphs. 

The main problem then becomes how to guarantee that this collective sharing of RDF Graphs is guaranteed to stay synchronized (consistent) across all agents involved, while updates are made locally and concurrently by these agents. As mentioned, the added difficulty is due to the fact that there are communication breakdowns among agents and that agents both enter and leave the operational environment in question. Recall also, that some of these shared documents contain meta-information (dataset representations) about large quantities of sensor data collected during data gathering missions. The synchronization processes only keep this data consistent among agents indirectly through synchronization of the RDF Graphs containing meta-information. Consequently, data-transfer algorithms are required for physically transferring low-level sensor data among agents when that is required. 


Our approach is inspired by software systems used for code-versioning where new data is incremental and forms a graph in which new versions are nodes (called revisions). Incremental changes are differences between subsequent revisions (called deltas) and are the edges between revisions. This approach allows for recreating complete instances of the versioned information by starting at a specific revision, backtracking recursively through the parents to the initial revision, and finally applying all the deltas sequentially in a forward fashion. 
While in the traditional application, versioning is performed when needed as determined by the programmer, in the HFKN Framework, we are interested in continuous synchronization as soon as new information is available. 

Traditional code-versioning systems, such as Git, encode the source code as text files, where the deltas between two different code revisions are added, deleted, and merged text lines.
While the RDF Graphs could also be encoded as text, using such a representation becomes problematic.
Two semantically identical graphs can be encoded in an infinite number of ways depending on the actual text format used, for example, using different orders of RDF Triples or the number of spaces between the terms.
A system that uses text-based representation would essentially create new revisions based on these encoding differences even if the two versions of the RDF Graph are identical. A simple solution would be to adhere to a very strict encoding of an RDF Graph (order, spaces, etc.), but even such a solution could lead to merge conflicts that cannot be resolved automatically.

In the HFKN Framework, deltas representing differences between two RDF Graph revisions are computed at the syntactic structure level rather than at the encoding level. Deltas describe the evolution of an RDF graph in the form of added and/or deleted RDF Triples that are to be shared among all participating and interested agents in order to achieve shared, consistent, common knowledge. Using such an approach, the system can recognize the difference between two versions of an RDF Graph independently of its encoding. The basic concepts of revision management, such as merge and rebase used in the HFKN Framework, are directly inspired by those available in code-versioning systems. We have adapted both concepts to handle computing of deltas at a syntactic structure (RDF Triple) level.

Synchronization of RDF Graphs among agents in the HFKN Framework is achieved by exchanging messages with the goal of having a common and consistent view of the respective shared RDF Graphs. There are several assumptions and properties concerning participating agents, their roles,  communication capabilities, and exchanged messages. The following assumptions exist regarding the operation of agents:

\begin{itemize}
    \item Agents operate asynchronously and at arbitrary speeds.
    \item Agents may experience failures to operate or communicate.
    \item Agents store data locally in permanent storages.
    \item Agents operate in good faith and are truthful.
    \item Agents use synchronized clocks. 
    \item Agents are aware of other agents' existence based on periodically exchanged status messages.
\end{itemize}

Agents can take one or two roles in the RDF Graph synchronization mechanism:
\begin{itemize}
    \item Each agent has a role of a standard agent.
    \item One agent has an additional role of a merge master awarded in an election process if more than two agents exist. The merge master is part of the RDF Graph synchronization mechanism and is responsible for integrating all the changes from the other agents into the common view.
    There can be different merge masters for different documents.
\end{itemize}

Synchronizing RDF Graphs among a team of agents assuming the above-listed properties requires that each agent interested in a team's shared common knowledge, performs several tasks described in~\Cref{sec:rdf_doc_sync} and respective subsections.
The different algorithms allow for synchronizing different revisions of RDF Documents and handling the exchange of necessary information required for synchronization.
The responsibilities of a merge master and the process of electing the master among the agents are detailed in~\Cref{sec:election}.  

Agents cooperate by exchanging messages with the following properties:
\begin{itemize}
    \item Messages are sent asynchronously and without guarantees on arrival time bounds.
    \item Messages can be lost, arrive out of order, or be duplicated.
    \item Received messages are not corrupted.
    \item Agents exchange four types of messages: \verb|Status|, \verb|Revisions-request|, \verb|Revision|, \verb|Vote|.
\end{itemize}

These assumptions stem mainly from the fact that the communication between agents is achieved using wireless links, which are inherently unreliable.
Moreover, partially received or corrupted messages are ignored upon detection. A detailed description of the four types of exchanged messages is presented in~\Cref{sec:doc_sync_messages}.

In the remainder of this section we start by defining the necessary terminology in~\Cref{sec:doc_sync_terminology}, followed by the  description of the method for RDF Document synchronisation among agents in~\Cref{sec:rdf_doc_sync}. Detailed descriptions of two major algorithmic components of the proposed solution, namely \emph{merge} and \emph{rebase} algorithms, are presented in~\Cref{sec:merge,sec:rebase}, respectively.


\subsection{Terminology and Definitions}\label{sec:doc_sync_terminology}

In this section, we consider both terminology and definitions used in the RDF Graph Synchronization processes.
\begin{figure}[tb]
\centering
\includegraphics[width=\linewidth]{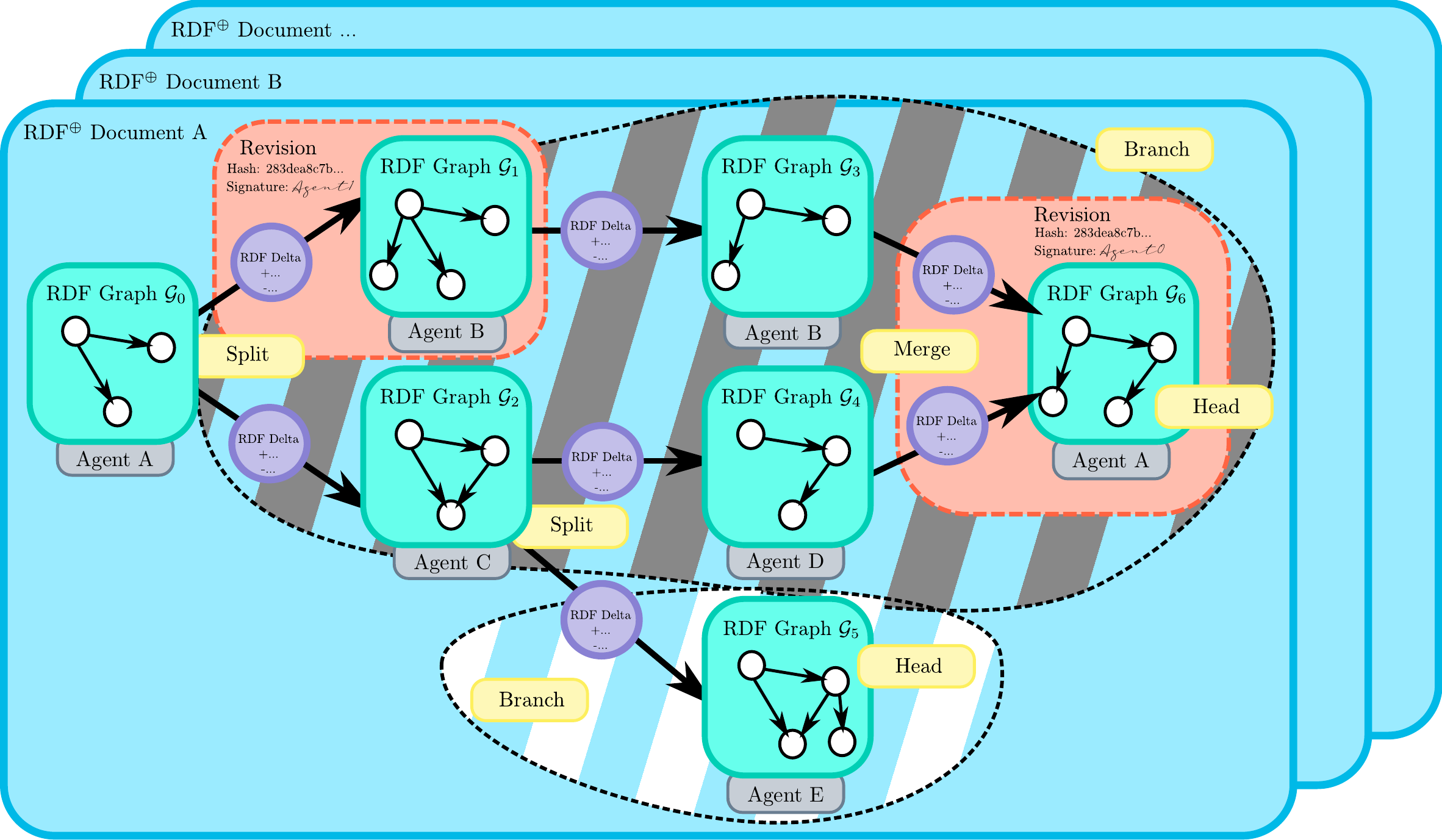}
\caption{Terminology and concepts in the HFKN RDF Graph Synchronisation.}
\label{fig_rdf_sync_terminology}
\end{figure}


\subsubsection{RDF\OP Document}
In \cite{CYGANIAK-W3C-2014}, an RDF Document is defined as an encoding of an RDF Graph. To facilitate RDF Graph synchronization, the RDF Document specification  is extended and used in the HFKN Framework to also specify the incremental changes of its associated RDF Graph. This history of changes is represented using a \emph{Graph of Revisions}, where vertices represent particular \emph{revisions} of an RDF Graph and edges represent incremental differences (i.e. \emph{deltas}) between revisions. We use the term RDF\OP Document for this extended specification. It includes the Graph of Revisions for the RDF Graphs in the document. 

\Cref{fig_rdf_sync_terminology} presents a schematic of an example RDF\OP Document. It depicts the main terms and concepts, as well as relations between them. At any given time, there exist several RDF\OP Documents (here: A and B) which need to stay synchronized between interested agents. A specific RDF\OP Document encodes an RDF Graph and its history in the form of incremental changes expressed using RDF Deltas, which are parts of Revisions (highlighted by red dashed lines in~\Cref{fig_rdf_sync_terminology}). 
In the example, there are seven RDF Graph versions created by five agents (Agent A...E) at different times during a mission execution. Different versions are connected with edges, which include RDF Deltas representing differences between them.
The graph containing the revisions can exhibit a \emph{split} which results in separate \emph{branches}. Two example branches are marked with striped areas in the figure. Branches can undergo a \emph{merge} operation when needed. The leaf nodes in the graph, that is the latest versions from the agents' perspectives, are called \emph{heads}. The operations of \emph{Split}, \emph{Branch} and \emph{Head}, associated with the construction of a Graph of Revisions, are considered in detail in~\Cref{ssec:graph_of_revisions}.

Using an RDF\OP Document it is possible to instantiate all past versions of the RDF Graphs associated with it, including the latest version (i.e. head), which should be in sync between all collaborating agents that share the document.
This structure allows for implementing an efficient mechanism for creating, updating, and synchronising RDF Graphs shared among a team of agents. Multiple agents can work in parallel on the same RDF Graph, and new agents can obtain the newest revision of an RDF Graph created by other agents.


An RDF\OP Document can be easily converted into a standard RDF Document and used in combination with other frameworks based on the use of Semantic Web technologies. Currently, our system allows for direct querying of RDF\OP Documents using SPARQL. 

A comprehensive specification of the encoding of an RDF\OP Document is provided in~\Cref{annex:rdf_document}.
The original definition of an RDF Document in~\cite{CYGANIAK-W3C-2014} uses text-based encoding of an RDF Graph. In the HFKN Framework, an RDF\OP Document is encoded using a binary representation (see \Cref{sec:Appendix1}, similar to~\cite{fernandez2013binary}). This representation is more suitable for representing sensor data such as images as well as storing it in a database, while being compatible with text-based Semantic Web frameworks.


\subsubsection{RDF Delta}\label{ssec:rdf_delta}

There have been several proposals for efficiently computing the differences between two RDF/S Knowledge Bases~\cite{10.1145/1993053.1993056} or RDF Graphs~\cite{Berners-Lee04delta:an,10.1007/978-3-540-76298-0_46}. These proposals are based on syntactic, semantic, or combinations of syntactic and semantic characteristics of the RDF Graphs in question. Our approach is based on one of the more straightforward definitions of deltas defined in terms of a plain set-theoretic semantics based on two operations of \textit{insertion} and \textit{deletion} of RDF Triples. Other approaches in the literature could be used that provide more efficient deltas without changing the basic operations of the HFKN framework, but we save this for future investigation.

An RDF Delta ($\Delta$) is an expression of the difference between two individual RDF Graphs (or two versions of the same RDF Graph).
Given two RDF Graphs $ \mathcal{G}_i $ and $ \mathcal{G}_j $, the difference can be summarized with the list of \emph{inserted} triples $ \mathcal{I}_{i,j} = \{ ... (s^I_k, p^I_k, o^I_k) ... \} $ and the list of \emph{removed} triples $ \mathcal{R}_{i,j} = \{ ... (s^R_k, p^R_k, o^R_k) ... \} $, such that:

\begin{align}
  \mathcal{G}_j     & = ( \mathcal{G}_i \setminus \mathcal{R}_{i,j} ) \cup \mathcal{I}_{i,j} \\
  \mathcal{R}_{i,j} & = \mathcal{G}_i \setminus \mathcal{G}_j \\
  \mathcal{I}_{i,j} & = \mathcal{G}_j \setminus \mathcal{G}_i
\end{align}

The \textit{removed} and \textit{inserted} lists are swapped when switching the order of graph indices $i$ and $j$ such that:

\begin{align}
  \mathcal{R}_{j,i} & = \mathcal{I}_{i,j}
  \label{eqn_inverse_delta_1} \\
  \mathcal{I}_{j,i} & = \mathcal{R}_{i,j}
  \label{eqn_inverse_delta_2}
\end{align}

We define a delta between $ \mathcal{G}_i $ and $ \mathcal{G}_j $ as:

\begin{equation}
  \Delta_{i,j} = (\mathcal{I}_{i,j}, \mathcal{R}_{i,j})
  \label{eqn_rdf_delta}
\end{equation}

RDF Deltas are encoded using a subset of SPARQL Update query~\cite{SEABORNE-W3C-2008} as presented in~\Cref{appendix:rdf_delta}.

\subsubsection{Graph of Revisions}\label{ssec:graph_of_revisions}


The history of changes of an RDF\OP Document is represented as a Graph of Revisions (GoR). Nodes in a GoR represent revisions, and edges represent deltas. The following content defines a \emph{revision}:
\begin{itemize}
  \item A list of RDF Deltas associated with parent revisions -  $\Delta_{i,j}$, where $i$ and $j$ are hashes (see below) of the parent and this revision, respectively. In the HFKN Framework, each revision has one or two parents (see \Cref{fig_rdf_sync_terminology}). Root revision is an exception and has no parents.
  \item The UUID of the author of the revision.
  \item The timestamp corresponding to the time when the revision was created\footnote{Expressed as Unix time.}.
    \item A hash used to identify the revision, and computed over its content using the SHA-512 algorithm\cite{SHA-2012} as follows\footnote{We use the SHA-512 of the delta for convenience, as deltas and revisions are stored separately in the database and associating a hash for the deltas allows more robust indexing.}:
        \begin{align}
            SHA_{512}(&uuid(author), timestamp, \nonumber \\&{\textstyle\bigcup} SHA_{512}(\Delta_{parent,revision}, uuid(parent)))
            \label{revision_hash}
        \end{align}
    
  \item A cryptographic signature computed over the hash using the RSA algorithm~\cite{BELLARE-ICTACT-1996}.
\end{itemize}

The hash allows for the unique identification of a specific revision as it is computed over all of its contents. The cryptographic signature has no direct use in the current version of the HFKN Framework since we assume that all agents are trustworthy. Nevertheless, it is included in preparation for extending the framework to handle the trust level between agents planned in  future work.

The root revision node in any GoR is defined as a \emph{null} revision. Specifically, the list of RDF Deltas is empty (i.e. it has no parents), and the values of the author's UUID, timestamp, and signature are null.
Published revisions are immutable, meaning that the deltas, parents, author, and timestamp cannot change after publication.
Before publication, revisions can be moved to a different branch, which means changing the hash and parent of a revision.

A \emph{head} is a revision that does not have any children.
\Cref{fig_rdf_sync_terminology} shows an example GoR with two heads: $\mathcal{G}_5$ and $\mathcal{G}_6$.
The set of heads is denoted as $\mathcal{H}$.
A \emph{split} is defined as a revision that has two or more children (e.g. $\mathcal{G}_0$ and $\mathcal{G}_2$ in \Cref{fig_rdf_sync_terminology}).
A \emph{branch} is defined as a set of revisions between a head and a split.
The branch includes the head but excludes the split.
An example GoR presented in \Cref{fig_rdf_sync_terminology} contains two branches $\{\mathcal{G}_1, \mathcal{G}_2, \mathcal{G}_3, \mathcal{G}_4, \mathcal{G}_6 \}$ and $\{\mathcal{G}_5 \}$.

If all the revisions are sequential, the GoR is linear. An example of such a case would be the GoR presented in \Cref{fig_rdf_sync_terminology} if we exclude $\mathcal{G}_2$, $\mathcal{G}_4$ and $\mathcal{G}_5$ revisions. Whenever RDF\OP Document changes occur concurrently (e.g., performed by two different agents), the resulting revisions have to be combined into a single consistent one. This can be done using two different approaches: \emph{merge} or \emph{rebase}. In \Cref{fig_rdf_sync_terminology} a result of a merge between two revisions $\mathcal{G}_3$ and $\mathcal{G}_4$ results in a new head revision $\mathcal{G}_6$.
More details about the merge and rebase procedures are presented in the following subsection. 

\subsubsection{Combining Branches: Merge / Rebase Overview}\label{sec:merge_rebase_intro}

Whenever an agent makes an update to an RDF\OP Document by adding or removing RDF Triples to the encoded RDF Graph, it computes a delta and creates a new version of the RDF\OP Document. It then broadcasts the delta to other agents that have subscribed to that RDF\OP Document. The delta can then be applied to the receiving agents' copy of the RDF\OP Document to maintain knowledge synchronization. This process is straightforward if only one agent performs document changes.

However, agents may create changes to one RDF\OP Document concurrently, either because they make a change simultaneously or because they are out of communication range. 
When this happens, the document will have several branches (i.e. heads) that need to be combined into a single one. This is depicted in \Cref{fig_rdf_sync_terminology} as a split and merge.

In this work, we propose two approaches for combining RDF Graph branches: \textit{merge}, which creates a new combined revision based on deltas of two other revisions, and \textit{rebase}, which moves the revisions from one branch to another. Merge is more general and is always applicable, while rebase can only be applied if the revisions are still \emph{local}, which means they have not been broadcasted yet. 

\begin{figure}
  \centering
  \subfloat[Two Branches]{
    \includegraphics[width=0.35\linewidth]{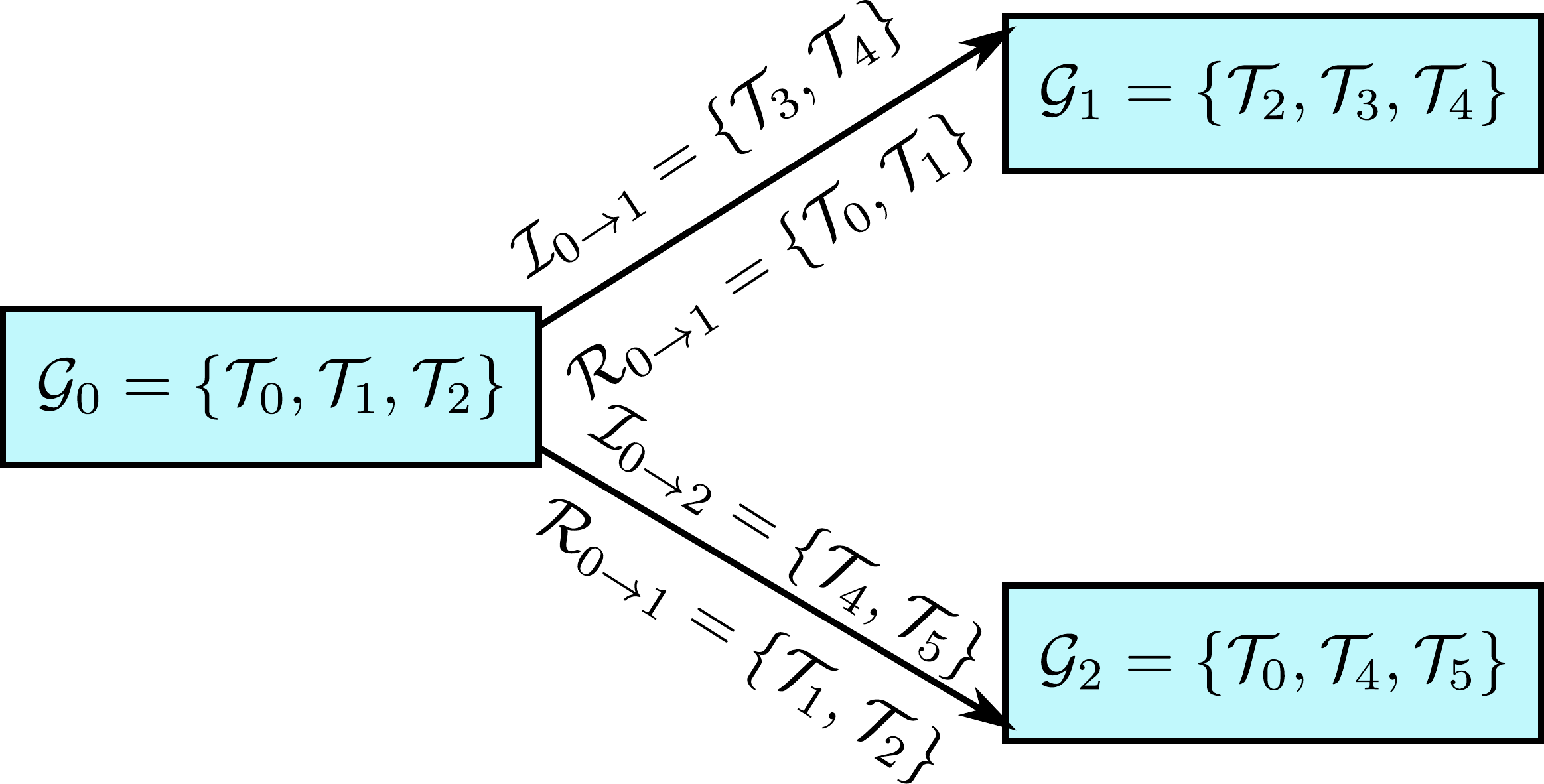}
    \label{fig:revision_example_before}
  }\\
  \subfloat[After Merge]{
    \includegraphics[width=0.49\linewidth]{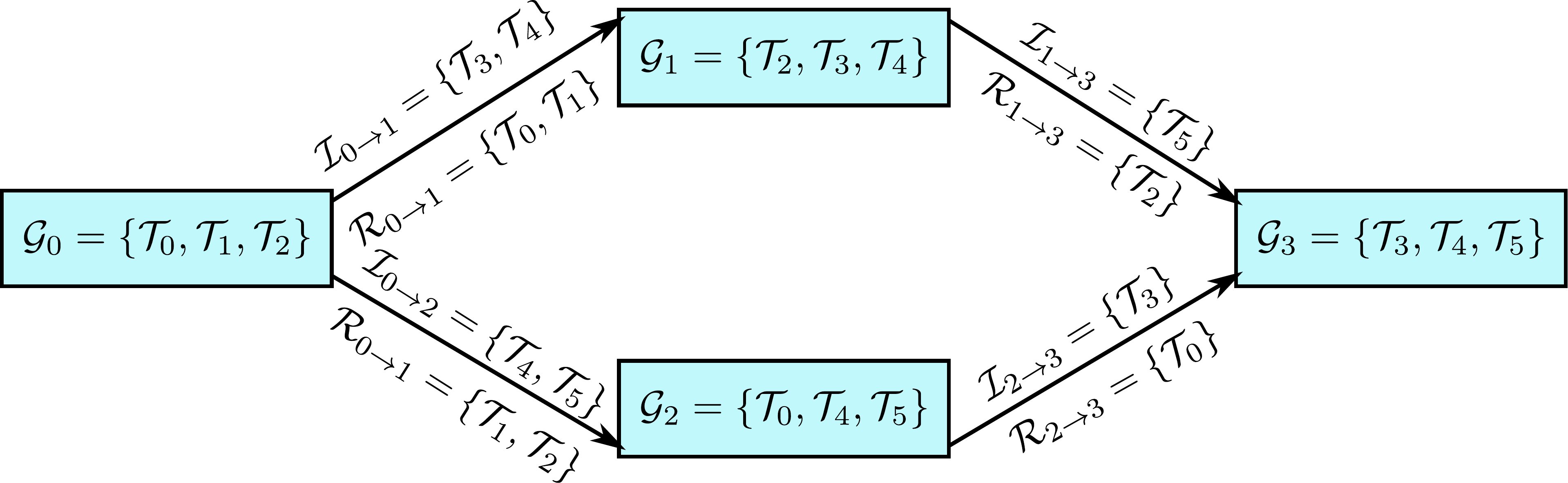}
    \label{fig:revision_example_merge}
  }
  \subfloat[After Rebase]{
    \includegraphics[width=0.49\linewidth]{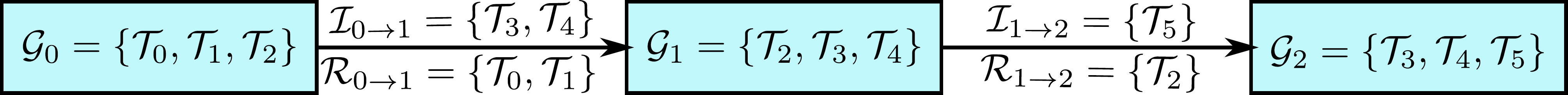}
    \label{fig:revision_example_rebase}
  }
  \caption{Example graphs of revisions before and after applying merge and rebase procedures. $\mathcal{G}$, $\mathcal{I}$ and $\mathcal{R}$ denote a revision, inserted and removed triples, respectively.}
  \label{fig:revision_example}
\end{figure}

Illustrative examples of combining revisions using the two proposed techniques are presented in~\Cref{fig:revision_example}, where three agents: A, B, and C,  create new revisions at different time points. An example scenario considering a merge procedure is depicted in \Cref{fig:revision_example_before,fig:revision_example_merge}. 
Agent A is responsible for performing the merge operation.
At first (i.e. time $t-1$), the document contains the triples $\mathcal{T}_0$, $\mathcal{T}_1$, $\mathcal{T}_2$, which correspond to revision $\mathcal{G}_0 = \{ \mathcal{T}_0, \mathcal{T}_1, \mathcal{T}_2 \}$.
Next, at time $t$, Agent B adds triples $\mathcal{T}_3$ and $\mathcal{T}_4$ and removes $\mathcal{T}_0$ and $\mathcal{T}_1$. This constitutes revision $\mathcal{G}_1$, $\mathcal{I}_{0\rightarrow 1} = \{\mathcal{T}_3, \mathcal{T}_4\}$ and $\mathcal{R}_{0\rightarrow 1} = \{\mathcal{T}_0, \mathcal{T}_1\}$, which corresponds to revision $\mathcal{G}_1 = \{ \mathcal{T}_2, \mathcal{T}_3, \mathcal{T}_4 \}$.

Concurrently, at time $t$, Agent C adds triples $\mathcal{T}_4$ and $\mathcal{T}_5$ and removes $\mathcal{T}_1$ and $\mathcal{T}_2$. This constitutes revision $\mathcal{G}_2$, $\mathcal{I}_{0\rightarrow 2} = \{\mathcal{T}_4, \mathcal{T}_5\}$ and $\mathcal{R}_{0\rightarrow 2} = \{\mathcal{T}_2, \mathcal{T}_3\}$, which correspond to revision $\mathcal{G}_2 = \{ \mathcal{T}_0, \mathcal{T}_4, \mathcal{T}_5 \}$.
The resulting graph of revisions is shown in \Cref{fig:revision_example_before}.

Finally, at time $t+1$, Agent A receives the deltas $(\mathcal{I}_{0\rightarrow 1}, \mathcal{R}_{0\rightarrow 1})$ and $(\mathcal{I}_{0\rightarrow 2}, \mathcal{R}_{0\rightarrow 2})$ from Agent B and Agent C, respectively.
It then creates a single revision $\mathcal{G}_3 = \{\mathcal{T}_3,\mathcal{T}_4,\mathcal{T}_5\}$, with two deltas by applying the merge operation:

\begin{itemize}
  \item $\mathcal{I}_{1\rightarrow 3} = \{ \mathcal{T}_5 \}$ $\mathcal{R}_{1\rightarrow 3} = \{ \mathcal{T}_2\}$
  \item $\mathcal{I}_{2\rightarrow 3} = \{ \mathcal{T}_3 \}$ $\mathcal{R}_{1\rightarrow 3} = \{ \mathcal{T}_0\}$
\end{itemize}

The structure of the resulting revision $\mathcal{G}_3$ is shown in \Cref{fig:revision_example_merge}.

The alternative \emph{rebase} process is illustrated in~\Cref{fig:revision_example_rebase}. At time $t$, Agent C creates revision $\mathcal{G}_2$ locally (i.e. does not publish it). Next, at time $t+1$, the agent receives a new revision $\mathcal{G}_1$ from Agent B. Agent C then can rebase $\mathcal{G}_2$ on top of $\mathcal{G}_1$ and create a revision such that  $\mathcal{I}_{1\rightarrow 2} = \{ \mathcal{T}_5 \}$ $\mathcal{R}_{1\rightarrow 2} = \{ \mathcal{T}_2\}$.
The structure of the revisions is shown in \Cref{fig:revision_example_rebase}. 






In summary, the HFKN Framework uses two techniques for combining  changes in RDF\OP Documents:

\begin{itemize}
  \item \textit{merge} combines two branches by creating a new revision with two RDF Deltas. The advantage of \textit{merge} is that it can be applied to any branches, and the new revision can be synchronized with other agents. The drawbacks are that it creates a new revision, which makes the revision graph more complex. Additionally, if only merge is used, and it is not performed in a timely fashion, the global synchronization between all agents may not be guaranteed in certain scenarios due to practical limitations (discussed in \Cref{ssec:sync_protocol}).

  \item \textit{rebase} moves a branch on top of another branch, resulting in a single branch with all the revisions. The advantage is that it reduces the complexity of the revision graph and minimizes the number of revisions. 
  In the HFKN Framework we limit the use of rebase only to local revisions to avoid the need for additional functionalities for guaranteeing that identical rebase operations are performed by all agents.
\end{itemize}

Further details of merge and rebase algorithms are presented in \Cref{sec:merge,sec:rebase}, respectively. The HFKN Framework uses a combination of merge and rebase operations in order to guarantee a global synchronization of RDF\OP Documents among all agents. Detailed discussion of the interactions between the two algorithms and their use is provided in the following subsection.

\subsubsection{Synchronization Protocol}
\label{ssec:sync_protocol}
Merge and rebase are two possible approaches for combining the changes in the RDF Graph made by different agents.
The Synchronization Protocol describes how these two approaches are used to achieve a common view of an RDF\OP Document.

A naive approach in which all agents individually apply only merge operations as soon as they have more than one head in their GoRs would result in a never-ending merge loop creating an infinite number of new merge revisions.
To solve this problem, distributed code-versioning systems rely on a central authority. Whenever a participant has a new version of their code that should be shared with other participants, the participant makes sure they are synchronized and consistent with the central authority.

This approach alone only works if the rate of changes of the RDF\OP Document is low, as each agent has to have enough time to ensure its GoR is consistent with the central authority\footnote{For reference, an active open source project such as Linux kernel has an average of less than ten commits per hour.}.
In the HFKN Framework, agents can make changes at a much higher rate, up to several times per second. Agents may not have time to ensure consistency of their GoR with the central authority before publishing their changes.
Thus, in the HFKN Framework, agents publish their changes without ensuring consistency, and merging the changes is the responsibility of a central authority. We call the central authority the \textit{Merge Master}. Although, the concept of central authority is directly inspired by distributed code-versioning systems, its use in our framework is novel. In the code-versioning systems, the central authority is static, while in the HFKN Framework, the role of the merge master is assigned dynamically in the process of election (\Cref{sec:election}). Each agent can be selected as the merge master at any time and the election is performed among all agents within communication range. This is done for redundancy and reliability reasons to deal with dynamic mission scenarios where agents join and leave missions, as well as to deal with limitations of wireless communication links, such as disconnections.

\begin{figure}[t]
\centering
\includegraphics[width=0.6\linewidth]{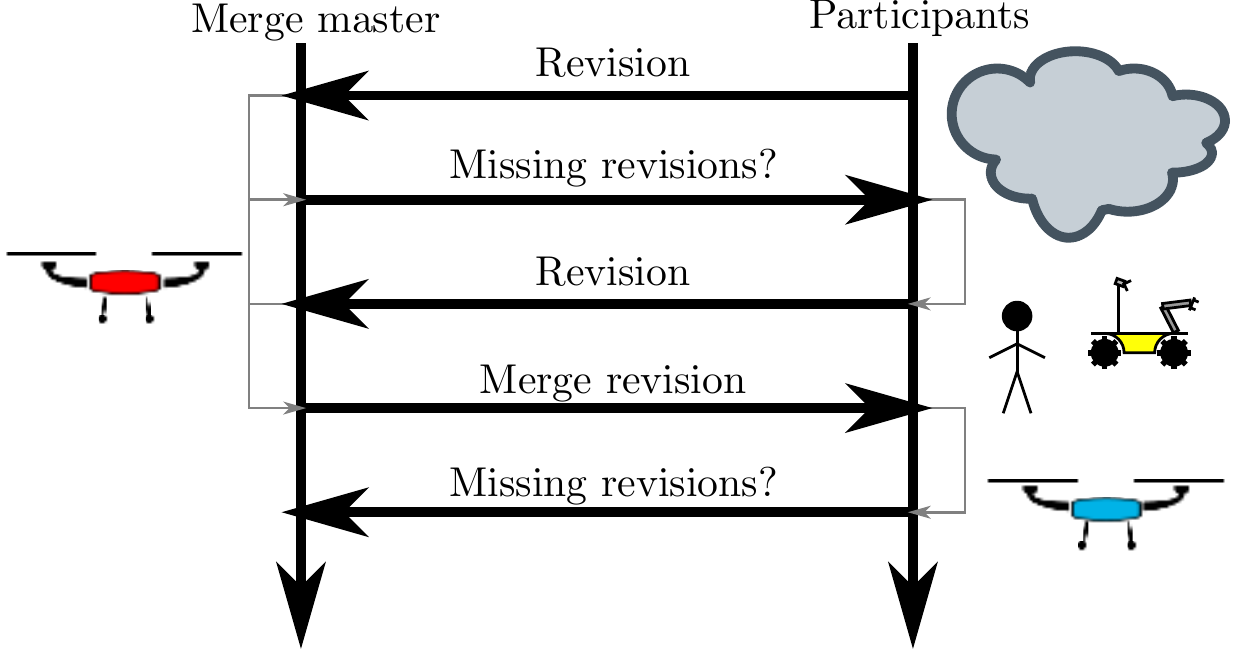}
\caption{Protocol for graph synchronization.}
\label{fig_graph_synchronisation}
\end{figure}

An overview of the synchronization protocol is shown in \Cref{fig_graph_synchronisation}.
When an agent makes a change, it broadcasts its new revision.
When the merge master receives the new revision, it checks if the merge operation can be performed. This is done by ensuring that its GoR has all parent revisions of the newly received one. If it does not, the merge master sends a request to other agents for all missing revisions.
Once the merge master has received all the revisions, it can then compute a merge revision by applying the merge algorithm. The resulting new merged revision is then broadcasted to other agents.
When an agent receives the merge revision, it checks if its GoR has all the parent revisions so that the merge revision can be directly incorporated. In case any revisions are missing from its GoR, the agent will send a request for the missing revisions.

If agents create new revisions too quickly in parallel, there is a risk that the merge master cannot cope with the load.
If that happens, the graph of revisions for each agent would include several branches which are never combined. Agents would only be able to make changes in their respective branches and a common view of the RDF Graph with all information from other agents would never be achieved. We call this the \textit{Never Synchronized Problem}.

\begin{figure}
  \subfloat[Only merge]{
    \includegraphics[width=\linewidth]{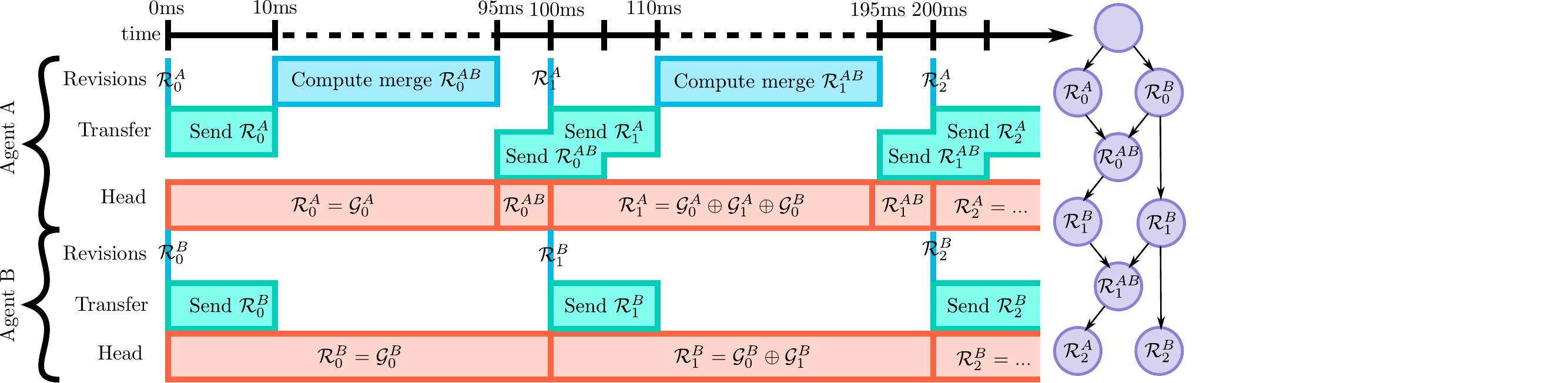}
      \label{fig:neversync_problem_only_merge}
  }\\
  \subfloat[Merge and rebase]{
    \includegraphics[width=\linewidth]{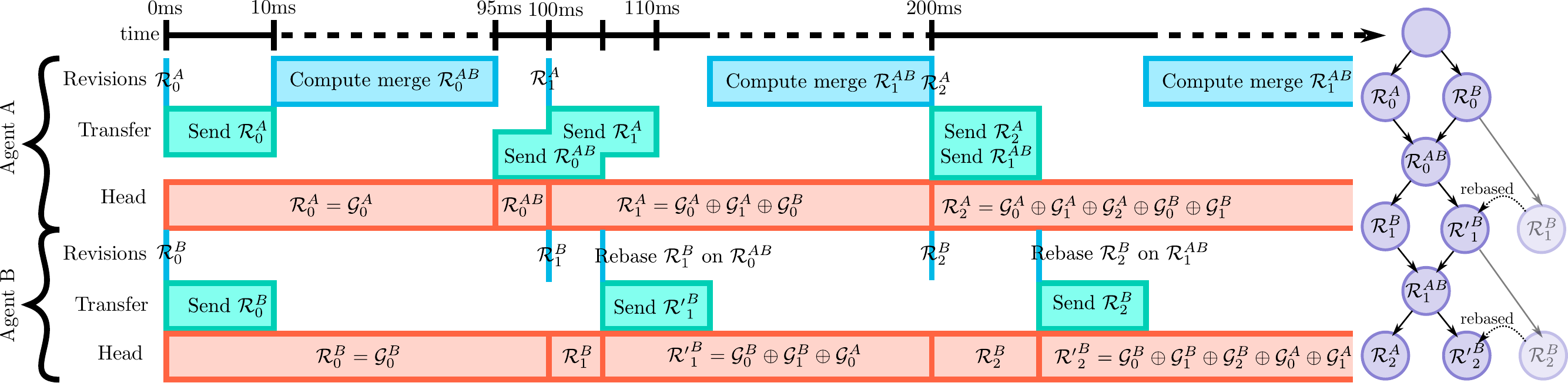}
      \label{fig:neversync_problem_merge_rebase}
  }
  \caption{Agent A is the merge master.
  The first line (Revisions) for each agent shows the revisions that are created. 
  $\mathcal{R}^A_i$ and $\mathcal{R}^B_i$ are created when the agents change the document, while $\mathcal{R}^{AB}_i$ represents a merge.
  The second line (Communication) represents the revisions that are sent by each agent and the delay in communication.
  The last line (Head) represents the current head for each agent and the content of that head.
  The final GoR is shown on the right side of the figure.
  As shown in \protect\subref{fig:neversync_problem_only_merge}, Agent B does not receive the changes created by Agent A before creating its new local revision.
  }
  \label{fig:neversync_problem}
\end{figure}

\paragraph{The Never Synchronized Problem} 
Let us consider the following hypothetical scenario which includes two agents: $A$ and $B$. Agent $A$ is the merge master. Both agents create a new revision every $100ms$. It takes $10ms$ to transmit revisions between agents and $85ms$ for $A$ to complete a merge.

The timeline of operations shown in \Cref{fig:neversync_problem_only_merge} is as follows:

\begin{itemize}
  \item At $t=0ms$, Agent A creates revision $\mathcal{R}^A_0$ and Agent B creates $\mathcal{R}^B_0$. The revisions are published.
  \item At $t=10ms$, Agent A receives $\mathcal{R}^B_0$ and initiates a merge.
  \item At $t=95ms$, Agent A completes the merge revision $\mathcal{R}^{AB}_0$ between $\mathcal{R}^A_0$ and $\mathcal{R}^B_0$.
  \item At $t=100ms$, Agent A creates revision $\mathcal{R}^A_1$ and Agent B creates $\mathcal{R}^B_1$.
  \item At $t=105ms$, Agent B receives $\mathcal{R}^{AB}_0$.
  \item At $t=110ms$, Agent A receives $\mathcal{R}^B_1$ and initiates a merge.
  \item ...
\end{itemize}


As can be seen in this example, Agent A gets access to the changes of Agent B, but Agent B does not receive the merge information in time before creating its new revision. This results in Agent B not being able to incorporate the RDF knowledge graph changes created by Agent A.


\paragraph{Rebase local changes and Merge public revisions} To avoid the Never Synchronized Problem, agents store local revisions for all the changes made until their GoRs are synchronized with the merge master.
The following procedure, which interleaves the use of merge and rebase algorithms, is used.
\begin{itemize}
  \item When an agent makes a change in an RDF Graph:
    \begin{itemize}
      \item The agent checks if its current revision inherits from the current revision of the merge master (it is known based on the status message). If that is the case, the agent creates a new public revision and publishes it.
      \item Otherwise, the agent creates a new local revision (i.e. the revision is not published).
    \end{itemize}
    The details of this behavior are described in \Cref{sec:handle_local_revisions}.
  \item When an agent receives a merge revision and if it has local revisions, the agent will apply the rebase algorithm using the local revisions and the latest available merge revision. This will result in a new revision that combines all changes that will be published.
        This is described in more detail in \Cref{sec:receive_revisions}.
\end{itemize}

The timeline for the new behavior of the two agents is shown in \Cref{fig:neversync_problem_merge_rebase}.
As we can see, both Agents A and B can make their local changes and access the other agent's changes. The timeline of operations now becomes as follows:

\begin{itemize}
  \item At $t=0ms$, Agent A creates revision $\mathcal{R}^A_0$ and Agent B creates $\mathcal{R}^B_0$.
        They believe they are synchronized and publish their revisions.
  \item At $t=10ms$, Agent A receives $\mathcal{R}^B_0$ and initiates a merge.
        At that point, Agent A and B believe they are not synchronized anymore.
  \item At $t=95ms$, Agent A complete the merge revision $\mathcal{R}^{AB}_0$ between $\mathcal{R}^A_0$ and $\mathcal{R}^B_0$.
        At that point, Agent A has access to $\mathcal{R}^A_0$ and $\mathcal{R}^B_0$.
  \item At $t=100ms$, Agent A creates revision $\mathcal{R}^A_1$ and Agent B creates a local revision $\mathcal{R}^B_1$.
  \item At $t=105ms$, Agent B receives $\mathcal{R}^{AB}_0$.
        Agent B rebases $\mathcal{R}^B_1$ on top of $\mathcal{R}^{AB}_0$.
        Agent B publishes $\mathcal{R}^B_1$.
        At that point Agent B has access to $\mathcal{R}^A_0$, $\mathcal{R}^B_0$ and $\mathcal{R}^B_1$.
  \item At $t=115ms$, Agent A receives $\mathcal{R}^B_1$ and initiates a merge.
  \item ...
\end{itemize}

By combining the merge and rebase methods for managing branches in GoRs, it is possible to maintain the RDF Graph synchronization between agents. Further details of the two methods and their relations are described in \Cref{sec:rdf_doc_sync}.

\subsubsection{Messages}\label{sec:doc_sync_messages}

All communication between agents allowing for maintaining a synchronized view of the RDF Graph is realized using the following messages:

\begin{itemize}
    \item The \verb|Status| message is used to discover agents subscribed to a document and that are within the communication range.
    It contains meta-information about the agent: its name, UUID, the public RSA key, and the state of the document.
    It includes the head revision hash and whether the agent believes and acts as a merge master.
    \item The \verb|Revision| message is used to exchange a revision between agents.
    It contains all the information needed to insert a new revision in the Graph of Revisions: author’s agent UUID, revision delta(s), timestamp, revision hash (\Cref{revision_hash}), and a cryptographic signature.
    
    \item The \verb|Revision-request| message is used to request a revision that is not available in an agent's Graph of Revisions. The message contains the requester's UUID and a list of the required revision hashes.
    \item The \verb|Vote| message is used during the election of the merge master. It contains the voter's UUID, candidate's UUID, the election round number, the vote timestamp and the election timestamp.
\end{itemize}

A \emph{status} message is sent periodically by all agents and has three purposes. First, it allows for detecting availability of other agents. Second, it allows the merge master to know other agent's head revisions and thus know whether it has all the necessary revisions before performing the merge operation. Third, it allows for knowing which agent has the role of the merge master and if an election is necessary (i.e. no or multiple masters are present). 

A \emph{revision} message is sent when a change to an RDF\OP Document is to be communicated to other agents or upon an explicit request. 

When an agent is aware of the existence of a revision through a status message or a parent of a known revision, it can use the \emph{revision-request} message to obtain the missing information.

A \emph{vote} message allows for initiating a merge master election process and casting votes. More information is presented in~\Cref{sec:election}.


    

\subsection{RDF\OP Document Synchronization}\label{sec:rdf_doc_sync}

As outlined in previous sections, agents acquire new knowledge throughout a mission execution, and its high-level description is represented as RDF Graphs encoded with their history in RDF\OP Documents. To achieve a common view of the continuously updated RDF\OP Documents among all participating agents, the HFKN Framework uses a number of algorithms and protocols that allow for efficient and robust RDF\OP Document synchronization.

\begin{figure}[tb]
    \centering
    \includegraphics[width=\linewidth]{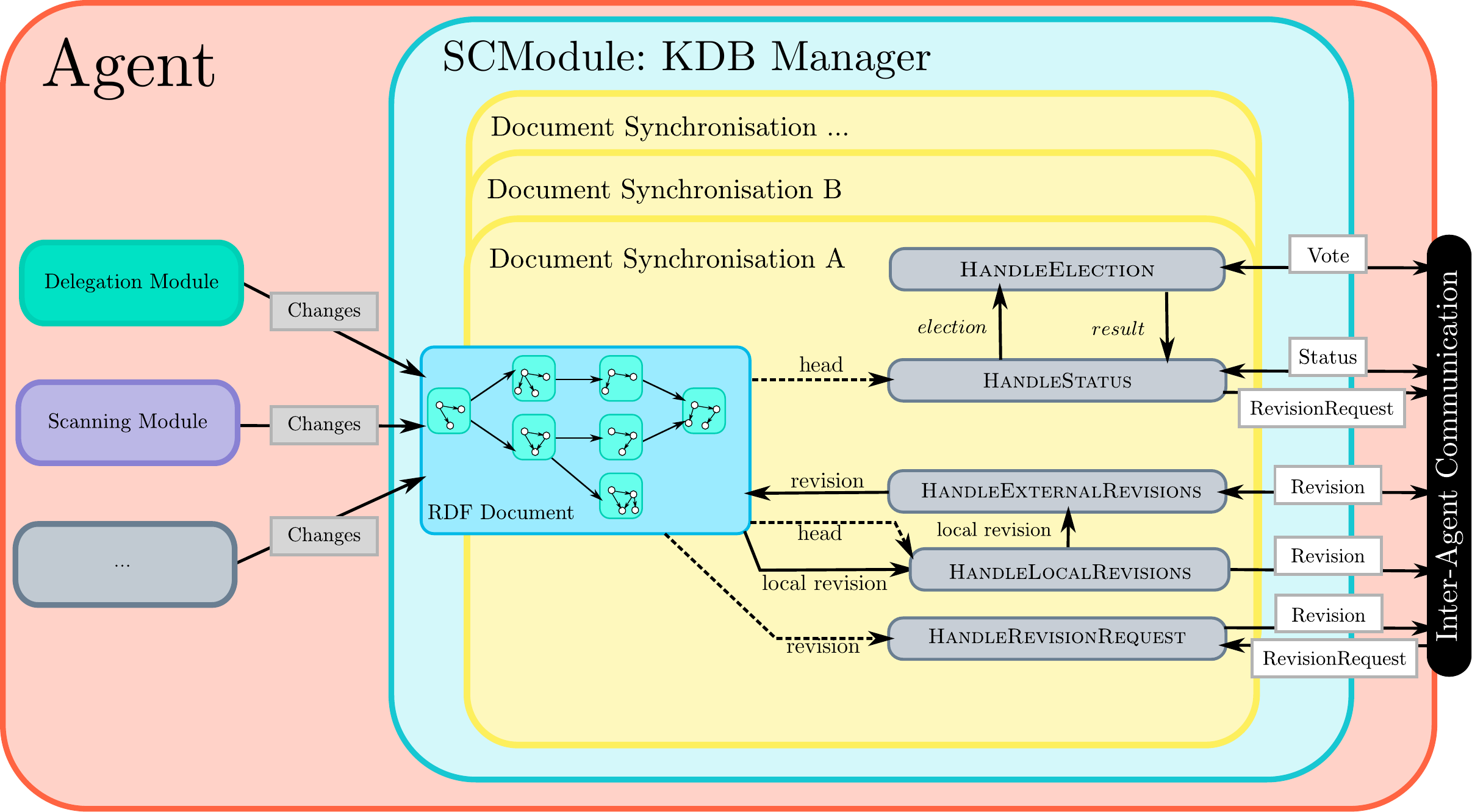}
    \caption{Schematic view of the HFKN RDF\OP Document synchronisation mechanism and processes involved.}
    \label{fig:rdf_document_synchronisation}
\end{figure}

A schematic view of the processes, functionalities, and the exchanged messages is presented in~\Cref{fig:rdf_document_synchronisation}. The view focuses on the perspective of one agent, which hosts several processes such as a Delegation or Scanning Modules and, most importantly, from the perspective of this work, the KDB Manager. The manager receives data from these modules and updates its RDF\OP Documents. For each document, the agent uses several processes (in gray) to handle RDF\OP Document synchronization as well as communication with other agents by exchanging messages (in white) to maintain the common view of the RDF\OP Document.

The following processes are used to achieve knowledge synchronization. Updates reflect the new information obtained by the agent itself to the RDF\OP Document by creating a new \emph{local} revision. The revision is either directly broadcasted to other agents or kept local. This decision is made by the \emph{HandleLocalRevisions} process. Upon receiving revisions from other agents, the \emph{HandleExternalRevisions} process is responsible for incorporating the new information into its version of the RDF\OP Document. The \emph{HandleRevisionRequest} process is responsible for providing revisions when other agents request them. The periodic sending of the status information as well as receiving the information from others is the responsibility of \emph{HandleStatus} process. Finally, initiating and participating in the merge master election is handled by the \emph{HandleElection} process.
The processes and the algorithms they employ are described in the following subsections.

\subsubsection{Handling Local Revisions}\label{sec:handle_local_revisions}

When an agent performs its tasks, the operation results often require updating the information contained in RDF\OP Documents. Examples of this include obtaining new sensor data such as images or LIDAR data, specifying the agent's capabilities, battery level, to name a few. When this need arises, a new revision is created. The decision whether it should be directly published to other agents is taken according to \Cref{alg:handlenewrevision}.


\begin{algorithm}[h]
  \DontPrintSemicolon
  \caption{{\sc HandleLocalRevisions}}
  \label{alg:handlenewrevision}
  \KwIn{$ \mathcal{G}_{new} $}
  \ForEach{$\mathcal{G}_i \in \mathcal{G}_{new} $}
  {
  \uIf{$\mathcal{G}_{merge master} \in Ancestors(\mathcal{G}_i)$ }{
     $Publish(\mathcal{G}_i)$\;\label{alg:handlenewrevision_publish}
  }
  \Else{
    $AddToLocalRevisions( \mathcal{G}_i )$\;\label{alg:handlenewrevision_private}
  }
  }
  
\end{algorithm}

The algorithm checks if the latest merge master's revision ($\mathcal{G}_{merge master}$), received from the merge master's status message, is one of the ancestors of the new revision ($\mathcal{G}_i$). The list of ancestors is computed by recursively listing all the parents of a given revision.
If that is the case, it means the current changes are ahead of the merge master, and the new revision can be published to other agents.
Otherwise, $\mathcal{G}_{i}$ is added to a \emph{list of local revisions} and will ultimately be published when the latest revision is received from the merge master, that is when they become synchronized.


\subsubsection{Handling External Revisions}\label{sec:receive_revisions}

When an external revision is received from another agent, it is incorporated into the appropriate RDF\OP Document through an update to its Graph of Revisions (GoR). The process is performed according to \Cref{alg:receiveloop}.


\begin{algorithm}[h]
  \DontPrintSemicolon
  \caption{{\sc HandleExternalRevisions}}
  \label{alg:receiveloop}
  \SetKw{Unless}{unless}
  \SetKw{And}{and}
  \SetKwBlock{Then}{then}{end}
  \While{true}{
    $\mathcal{G}_{new} \gets receiveNewRevision()$\;
    \Unless $HasRevision(parent(\mathcal{G}_{new}))$ \Then{
      $RequestRevision(parent(\mathcal{G}_{new}))$\;\label{alg:receiveloop_request_parent_revision}
    }
    $InsertRevision(\mathcal{G}_{new})$\;
    $\mathcal{G} \gets CurrentHeadRevision()$\;\label{alg:receive_loop_get_head}
    \If{$IsMergeRevision(\mathcal{G}_{new})$ \And  $CanRebase(\mathcal{G}, \mathcal{G}_{new})$}{\label{alg:receiveloop_check_rebase}
      $PublishRevisions(RebaseRevisions(\mathcal{G}, \mathcal{G}_{new}))$\label{alg:receiveloop_publish_revisions}
    }
    \If{$agentIsMergeMaster()$}{\label{alg:receive_loop_merge_master}
      $\mathcal{H} \gets RetrieveHeads()$\;
      \While{$|\mathcal{H}| \neq 1$}{
          $\mathcal{G}_i \gets \mathcal{H}[0]$\;
          $\mathcal{G}_j \gets \mathcal{H}[1]$\;
          $\mathcal{G}_{merge} \gets MergeRevision(\mathcal{G}_i, \mathcal{G}_j)$\;
          $InsertRevision(\mathcal{G}_{merge})$\;
          $PublishRevision(\mathcal{G}_{merge})$\;
          $\mathcal{H} \gets RetrieveHeads()$\;
      }
    }

  }  
\end{algorithm}

Upon receiving of a new revision, $\mathcal{G}_{new}$, the agent verifies that its GoR contains the parent revision(s) of $\mathcal{G}_{new}$. If it does not, it requests that revision.
If the revision is not already in the GoR, it is then inserted.

The algorithm continues in~\cref{alg:receive_loop_get_head} where the current head revision $\mathcal{G}$ is obtained followed by handling any local revisions in~\cref{alg:receiveloop_check_rebase}. If $\mathcal{G}_{new}$ is a merge revision (i.e. it has two RDF Deltas related to two parents) and a rebase operation can be performed (i.e. all revisions in $\mathcal{G}$ are local and $\mathcal{G}$ and $\mathcal{G}_{new}$ are connected in the agent's GoR) the following steps are performed. First, revisions $\mathcal{G}$ and $\mathcal{G}_{new}$ are combined by the rebase algorithm (\Cref{alg:rebaserevisions}, \Cref{sec:rebase}).
The resulting revision is then published to the other agents in~\cref{alg:receiveloop_publish_revisions}.


The algorithm continues in~\cref{alg:receive_loop_merge_master} with the tasks of the merge master. First, the agent waits until there is a need for merging of revisions, which means that there is more than one head in the graph of revisions. The \textit{MergeRevision} \Cref{alg:mergerevision} is applied to the first two heads in $\mathcal{H}$ and the result is then inserted in the graph of revisions and published using the \textit{InsertRevision} and \textit{PublishRevision} functions, respectively.
The algorithm continues until all revisions are merged, that is the number of heads is one.

\subsubsection{Handling Status}\label{sec:handle_status}

All agents participating in the RDF\OP Document synchronization process send their status information as well as receive the statuses from all other agents. The purpose of the status message as defined in~\Cref{sec:doc_sync_messages} is threefold. First, it allows for detecting the existence and availability of other agents.  Second, it enables the merge master to determine other agents' head revisions and thus conclude whether it has all the necessary revisions before performing the merge operation. Third, it is used to establish which agent has the role of the merge master and if an election is required (i.e. none or multiple masters present). The procedure for handling the status messages is presented in~\Cref{alg:statusloop}.

\begin{algorithm}[h]
  \DontPrintSemicolon
  \caption{{\sc HandleStatus}}
  \SetKw{Unless}{unless}
  \SetKwBlock{Then}{then}{end}
 \label{alg:statusloop}
  \While{true}{
  \ForEach{$\sigma \gets ReceiveStatus()$}
  {\label{alg:statusloop_receive_status}
    $\Gamma \gets Update(\Gamma, \sigma)$\;
    \Unless $HasRevision(RevisionOf(\sigma))$ \Then{\label{alg:statusloop_check_known_revision}
        $RequestRevision(RevisionOf(\sigma))$\;\label{alg:statusloop_request_revision}
    }
  }
  \If{$MinUUID(\Gamma) = AgentUUID$}{\label{alg:statusloop_lowest_uuid}
    $\mathcal{MM} \gets GetMergeMasters(\Gamma)$\;
    \If{$|\mathcal{MM}| \neq 1$}{\label{alg:statusloop_check_mms}
        $election \gets true$\;\label{alg:statusloop_start_election}
    }
  }
  $PublishStatus(\Gamma)$\;\label{alg:statusloop_publish_status}
  }

\end{algorithm}

When an agent receives a status message from another agent ($\sigma$ in~\cref{alg:statusloop_receive_status}) it first updates its own status $\Gamma$ and checks in~\cref{alg:statusloop_check_known_revision} if its GoR contains the revision designated by $\sigma$. If it does not, the agent requests the missing revision in \cref{alg:statusloop_request_revision}.

Subsequently, after incorporating all the received status messages, the agent checks whether there is a need for a new merge master election.
The check happens after updating the status to ensure the system reacts quickly to an incorrect number of merge masters.
First, the agent checks if its UUID is the lowest among all available agents.
This is done to minimize the number of initiated elections as only the agent with the lowest UUID can start the election process. 
Second, in~\cref{alg:statusloop_check_mms} the agent checks the number of agents which consider themselves merge masters ($\mathcal{MM}$) and if that number is different than one it initiates a new election in~\cref{alg:statusloop_start_election}.
This only happens when none or more than one agent consider themselves and act as merge masters.


Finally, in~\cref{alg:statusloop_publish_status} the updated version of the agent's status $\Gamma$ is used to publish the agent status message as defined in~\Cref{sec:doc_sync_messages}.


\subsubsection{Merge Master Election}\label{sec:election}

\begin{figure}[t]
\centering
\def\svgwidth{1\columnwidth}
\resizebox{0.7\linewidth}{!}{
\begingroup%
  \makeatletter%
  \providecommand\color[2][]{%
    \errmessage{(Inkscape) Color is used for the text in Inkscape, but the package 'color.sty' is not loaded}%
    \renewcommand\color[2][]{}%
  }%
  \providecommand\transparent[1]{%
    \errmessage{(Inkscape) Transparency is used (non-zero) for the text in Inkscape, but the package 'transparent.sty' is not loaded}%
    \renewcommand\transparent[1]{}%
  }%
  \providecommand\rotatebox[2]{#2}%
  \newcommand*\fsize{\dimexpr\f@size pt\relax}%
  \newcommand*\lineheight[1]{\fontsize{\fsize}{#1\fsize}\selectfont}%
  \ifx\svgwidth\undefined%
    \setlength{\unitlength}{312.64514112bp}%
    \ifx\svgscale\undefined%
      \relax%
    \else%
      \setlength{\unitlength}{\unitlength * \real{\svgscale}}%
    \fi%
  \else%
    \setlength{\unitlength}{\svgwidth}%
  \fi%
  \global\let\svgwidth\undefined%
  \global\let\svgscale\undefined%
  \makeatother%
  \begin{picture}(1,0.43615767)%
    \lineheight{1}%
    \setlength\tabcolsep{0pt}%
    \put(0,0){\includegraphics[width=\unitlength,page=1]{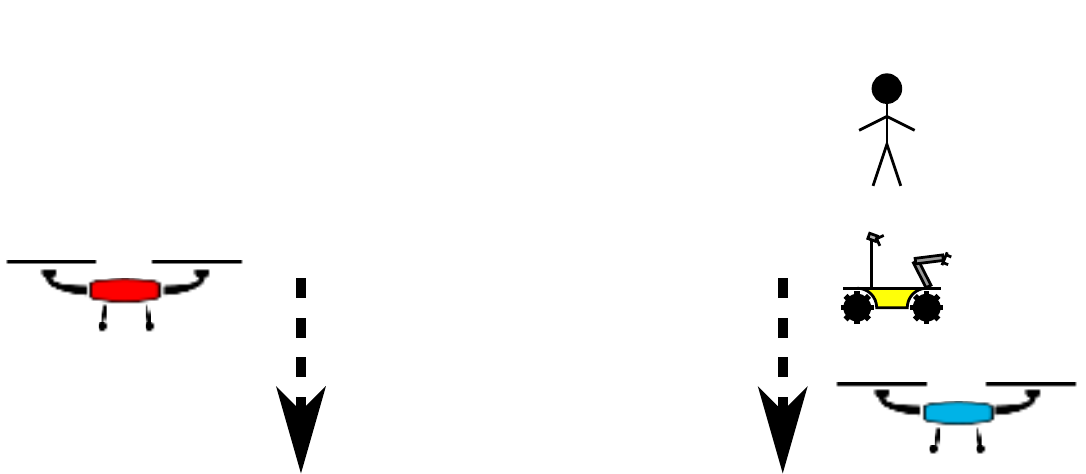}}%
    \put(0.22892569,0.41427167){\color[rgb]{0,0,0}\makebox(0,0)[lt]{\lineheight{1.25}\smash{\begin{tabular}[t]{l}Initiator\end{tabular}}}}%
    \put(0.65462041,0.41793176){\color[rgb]{0,0,0}\makebox(0,0)[lt]{\lineheight{1.25}\smash{\begin{tabular}[t]{l}Participants\end{tabular}}}}%
    \put(0,0){\includegraphics[width=\unitlength,page=2]{merge_master_election.pdf}}%
    \put(0.40490404,0.35885199){\color[rgb]{0,0,0}\makebox(0,0)[lt]{\lineheight{1.25}\smash{\begin{tabular}[t]{l}vote ($round_0$)\end{tabular}}}}%
    \put(0,0){\includegraphics[width=\unitlength,page=3]{merge_master_election.pdf}}%
    \put(0.40490348,0.26203976){\color[rgb]{0,0,0}\makebox(0,0)[lt]{\lineheight{1.25}\smash{\begin{tabular}[t]{l}votes  ($round_0$)\end{tabular}}}}%
    \put(0,0){\includegraphics[width=\unitlength,page=4]{merge_master_election.pdf}}%
    \put(0.39650739,0.12753089){\color[rgb]{0,0,0}\makebox(0,0)[lt]{\lineheight{1.25}\smash{\begin{tabular}[t]{l}votes  ($round_{i > 0}$)\end{tabular}}}}%
    \put(0,0){\includegraphics[width=\unitlength,page=5]{merge_master_election.pdf}}%
  \end{picture}%
\endgroup%
}

\caption{Overview of a \textit{merge master} election protocol. All agents elect a single master through casting votes. In case of ties, subsequent rounds of the election are performed.}
\label{fig_merge_master_election}
\end{figure}

When multiple agents are connected, they select a single merge master as discussed in~\Cref{sec:merge_rebase_intro}.
An overview of the procedure each agent follows to handle an election is presented in~\Cref{alg:electionloop} and graphically depicted in~\Cref{fig_merge_master_election}.

\begin{algorithm}[h]
  \DontPrintSemicolon
  \caption{{\sc HandleElection}}
  \label{alg:electionloop}
  \While{true}{
  
  $V \gets ReceiveVotes()$\;
  \If{$|V| > 0$}{
        $election \gets true$;\label{alg:electionloop_started}
  }
  \If{$election = true$}{
    $SendVote()$\;\label{alg:electionloop_first_vote}
    $V \gets V \cup ReceiveVotesWithTimeout()$\;\label{alg:statusloop_receive_votes}
    $\mathcal{MM} \gets GetMergeMasters(V)$\;
    \While{$|\mathcal{MM}| \neq 1$}{
        $SendVote()$\;\label{alg:electionloop_second_vote}
        $V \gets ReceiveVotesWithTimeout()$\;
        $\mathcal{MM} \gets GetMergeMasters(V)$;
    }
    $\Gamma \gets UpdateMergeMaster(\Gamma, \mathcal{MM})$\;
    $election \gets false$\;
  }
}
\end{algorithm}

When an agent starts an election process as specified in~\cref{alg:statusloop_start_election} of \Cref{alg:statusloop}, it publishes a vote message (\Cref{alg:electionloop} \cref{alg:electionloop_first_vote}).
Similarly, when an agent receives a vote, and is not in the election mode ($election = false$), then the agent knows that a new election has started (\cref{alg:electionloop_started}) and its vote is published in \cref{alg:electionloop_first_vote}. The structure of the vote message is defined in~\Cref{sec:doc_sync_messages} and the vote value is determined according to the following rules: 
\begin{itemize}
  \item pick the last agent for which the elector voted for to be a merge master if that agent is still connected, or
  \item pick the agent to which the elector has been connected to for the longest time.
\end{itemize}

Each agent receives the voting information (\cref{alg:statusloop_receive_votes}), which can then be used to determine the unique merge master in a decentralized manner.
To handle unreliability in communication, agents wait for a specific duration after the start of the election before they decide on the winner (\cref{alg:statusloop_receive_votes}).

In the case of ties, a second stage of the election is necessary (i.e. resolution), in which only the agents with the most votes are considered.
In the resolution stage, the agents vote (\cref{alg:electionloop_second_vote}) for a random candidate, and the procedure is repeated until a unique \textit{merge master} is selected. With each new round of the election, the election round number ($round_i$) is incremented in the \verb|Vote| message.

\subsubsection{Requesting Revisions}\label{sec:revision_request}

An agent can become aware of the existence of  revisions through a status message from another agent,
or by receiving a revision with parent revisions which does not exist in its GoR. In such a case, an agent can request the missing revision and complete its graph of revisions. This is performed in \Cref{alg:receiveloop} (\cref{alg:receiveloop_request_parent_revision}) or \Cref{alg:statusloop} (\cref{alg:statusloop_request_revision}). Notice that this also happens when a new agent joins the mission or becomes reconnected after a temporary communication loss.

A simple strategy for handling responses to requests which would require an agent to respond to all requests individually (assuming it has the requested revision) would be highly inefficient. Since the requested information in many cases can be already stored by multiple agents, this would lead to sending an unnecessary amount of duplicated messages.
Instead, to minimize the number of responses and exchanged messages, we propose to use the strategy presented in \Cref{alg:revision_request}.

\begin{algorithm}[h]
  \DontPrintSemicolon
  \caption{{\sc HandleRevisionRequest}}
  \label{alg:revision_request}
  \SetKw{And}{and}
  \SetKw{Or}{or}
  \SetKw{Not}{not}
  \While{true}{
    \ForEach{$\rho \gets ReceiveRevisonRequest()$}
    {\label{alg:revision_request_receive}
      \uIf{$Requester(\rho) = UUID(\mathcal{MM})$}{\label{alg:revision_request_from_mm}
        \If{$RevisionCreator(\rho) = UUID(self)$ \Or \Not $ConnectionBetween(RevisionCreator(\rho), UUID(self))$}{\label{alg:revision_request_from_mm_2}
          $Publish(Revision(\rho))$\;
        }
      }
      \ElseIf{$UUID(self) = UUID(\mathcal{MM})$}{\label{alg:revision_request_is_mm}
        $Publish(Revision(\rho))$\;
      }
    } 
  }
\end{algorithm}
 
After an agent receives a revision request (\cref{alg:revision_request_receive}), it responds by publishing the requested revision if the following conditions hold. First, the request came from the merge master (\cref{alg:revision_request_from_mm}). Second, the agent is the creator of the requested revision, or the agent is not connected with the revision creator (\cref{alg:revision_request_from_mm_2}).

In case the request came from an agent other than the merge master (\cref{alg:revision_request_is_mm}), the agent responds only if itself is the merge master.
By using \Cref{alg:revision_request}, the HFKN reduces the number of messages exchanged while guarantying that the requested revisions are sent.

\subsubsection{Illustrative Examples of Agent Behaviours}\label{sec:doc_sync_examples}

To illustrate the resiliency of our mechanism, we consider a few realistic scenarios that can occur during the synchronization process implemented in the HFKN Framework.

\paragraph{The merge master disappears without correctly publishing the last merged revision} 
When that happens, the other agents will keep updating their RDF\OP Documents locally by creating new local revisions (\Cref{alg:handlenewrevision} \cref{alg:handlenewrevision_private}).
Since the merge master is no longer available (e.g. due to communication failure or hardware malfunction) its status message is not broadcasted anymore.
Thus all participating agents will stop receiving the merge master's status message (\Cref{alg:statusloop} \cref{alg:statusloop_receive_status}) and the agent with the lowest UUID will then trigger an election (\Cref{alg:statusloop} \cref{alg:statusloop_start_election}). After the election process is completed, the new merge master will continue the synchronization process by merging the necessary branches.

\paragraph{There are multiple merge masters} The team of agents may become split into two (or more) groups due to communication range limitations. Each group in that case will have its own merge master elected. When communication between agents is re-established, all the agents will receive a status message from two (or more) merge masters (\Cref{alg:statusloop} \cref{alg:statusloop_receive_status}), and the agent with the lowest UUID will trigger an election (\Cref{alg:statusloop} \cref{alg:statusloop_start_election}).
After the election process is completed a single merge master is selected. And the synchronisation process continues normally.

\paragraph{A new agent joins a mission and has no previous knowledge of a particular RDF\OP Document} A common scenario during any mission execution may involve a team of agents collectively acquiring sensor data and updating shared RDF\OP Documents.
When a new agent joins the team and has no previous knowledge of the particular shared RDF\OP Document, its graph of revisions will be empty.
The new agent will receive the status message from the merge master (\Cref{alg:statusloop} \cref{alg:statusloop_receive_status}) with the hash value of the latest revision.
The new agent will then first request from the merge master the latest revision (\Cref{alg:statusloop} \cref{alg:statusloop_request_revision}), and then recursively all the past revisions (\Cref{alg:receiveloop} \cref{alg:receiveloop_request_parent_revision}).
The recursive request process will continue until the new agent has access to the entire GoR and can get the synchronized view of the shared RDF\OP Document.

\paragraph{There is no merge master} In this case, the synchronization process is suspended, which means that agents will not get the information from other agents. However, agents are still able to update their RDF\OP Documents locally, by creating new local revisions (\Cref{alg:handlenewrevision} \cref{alg:handlenewrevision_private}).
In parallel, a new election is started and after it is completed, the newly selected merge master will resume the process of merging revisions (\Cref{alg:receiveloop} \cref{alg:receive_loop_merge_master}).

\paragraph{Intermittent communication interruptions} Consider the following scenario involving four agents: A, B, C, and D. Agents A and B form \emph{Group 0}, and agents C and D form \emph{Group 1}. Agents within each group are assumed to have a reliable communication.
In this case, agents A and B will always receive each others status messages, and the same is true for agents C and D (\Cref{alg:statusloop} \cref{alg:statusloop_receive_status}).
Agents A and C, due to their UUID numbers are always elected as merge masters in their respective groups.

Consider a case in which the communication between \emph{Group 0} and \emph{Group 1} is being constantly interrupted, i.e. very unstable, groups are disconnected and re-connected.
When communication between groups is reestablished, the agents will have two merge masters in communication range, which will trigger an election (\Cref{alg:statusloop} \cref{alg:statusloop_start_election}), where either agent A or C is elected.
When communication between groups breaks down, one group will still have a merge master, while the other will require a new election (\Cref{alg:statusloop} \cref{alg:statusloop_start_election}).
If the communication links get constantly interrupted, there will be a situation where merge master election process will be triggered repeatedly.
This will slow down the synchronization process as merging of new changes is not performed during an election. However, the process of merging will resume immediately after the election process is completed (\Cref{alg:receiveloop} \cref{alg:receive_loop_merge_master}).

\subsection{Merge Algorithm}\label{sec:merge}

In case two concurrent changes have been made to the same RDF\OP Document resulting in two branches, it is necessary to combine the changes into a single revision. The merge algorithm is one of the two approaches used in the HFKN Framework that can be used for that purpose. 

\begin{figure}[tb]
\centering
\includegraphics[width=0.8\linewidth]{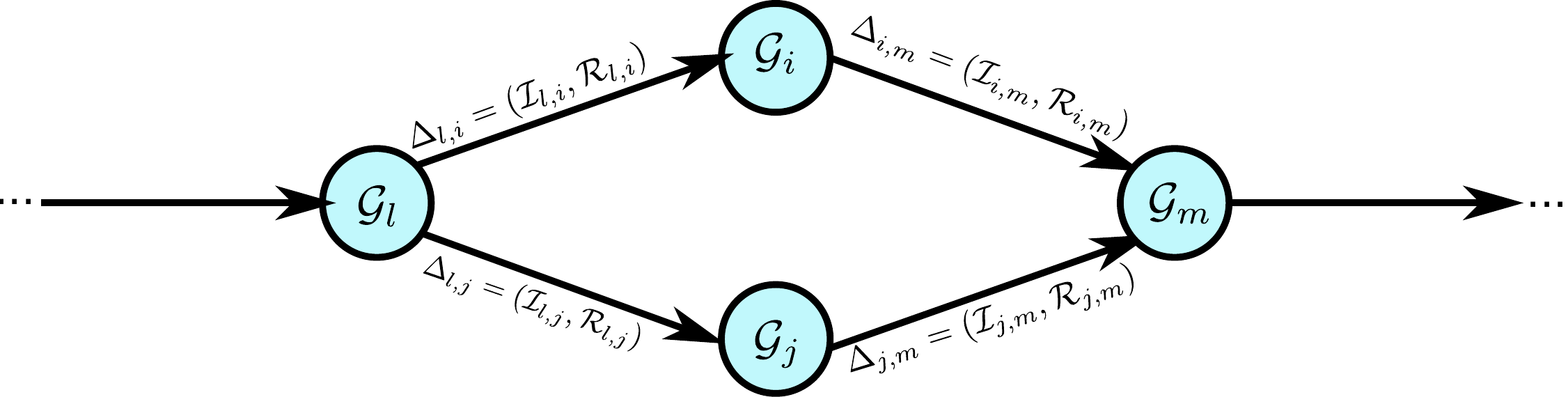}
\caption{Example simple merge problem with single revision in the concurrent paths.}
\label{fig_simpler_merge_problem}
\end{figure}

Consider a simple example of a GoR with two branches that contain a single change between revisions as shown in \Cref{fig_simpler_merge_problem}. Given two concurrent versions: $ \mathcal{G}_i $ with the delta $\Delta_{l,i} = (\mathcal{I}_{l,i}, \mathcal{R}_{l,i})$, and $ \mathcal{G}_j $ with delta $\Delta_{l,j} = (\mathcal{I}_{l,j}, \mathcal{R}_{l,j})$ split from the common ancestor $ \mathcal{G}_l $, the goal is to compute a new version $ \mathcal{G}_m $  with two deltas: $\Delta_{i,m} = (\mathcal{I}_{i,m}, \mathcal{R}_{i,m})$ and $\Delta_{j,m} = (\mathcal{I}_{j,m}, \mathcal{R}_{j,m})$.

In a general case $\mathcal{G}_m \subseteq \mathcal{G}_i \cup \mathcal{G}_j$. However, if a triple has been removed in the branch leading to $ \mathcal{G}_i $, that triple should not appear in $ \mathcal{G}_m $ even though it is available in $ \mathcal{G}_j $. Therefore the new revision is defined as:

\begin{equation}
  \mathcal{G}_m = ( \mathcal{G}_l \setminus (\mathcal{R}_{l,i} \cup \mathcal{R}_{l,j} ) ) \cup \mathcal{I}_{l,i} \cup \mathcal{I}_{l,j}
  \label{eqn_merge_graph}
\end{equation}

Finally, the wanted deltas: $\Delta_{i,m}$ and $\Delta_{j,m}$ can be computed using \Cref{eqn_merge_graph} taking into account the properties of \Cref{eqn_rdf_delta}:

\begin{align}
  \mathcal{I}_{i,m} & = \mathcal{I}_{l,j} \setminus \mathcal{I}_{l,i} \label{eq:I_i_m} \\
  \mathcal{R}_{i,m} & = \mathcal{R}_{l,j} \setminus \mathcal{R}_{l,i}  \label{eq:R_i_m}
\end{align}

\noindent where, indices $i$ and $j$ can be interchanged in \Cref{eq:I_i_m,eq:R_i_m}.

In the general case, there may be multiple revisions leading to $ \mathcal{G}_i $ and $ \mathcal{G}_j $, and the idea is to reduce those multiple revisions into a single one and apply  \Cref{eqn_merge_graph,eq:I_i_m,eq:R_i_m}.
To reduce multiple revisions, we use the following $combine$ function:

\begin{align}
  combine(\mathcal{I}_{i,i+1}, \mathcal{R}_{i,i+1}, \mathcal{I}_{i+1,i+2}, \mathcal{R}_{i+1,i+2}) & \rightarrow \mathcal{I}_{i,i+2}, \mathcal{R}_{i,i+2} \label{eq:combine} \\
      \mathcal{I}_{i,i+2} & = ( \mathcal{I}_{i,i+1} \setminus \mathcal{R}_{i+1,i+2} ) \cup \mathcal{I}_{i+1,i+2} \label{eq:combine_2}\\
      \mathcal{R}_{i,i+2} & = \mathcal{R}_{i,i+1} \cup \mathcal{R}_{i+1,i+2} \label{eq:combine_3}
\end{align}

The $combine$ function is derived from:

\begin{align}
  \mathcal{G}_{i+2} & = ( \mathcal{G}_{i+1} \setminus \mathcal{R}_{i+1,i+2} ) \cup \mathcal{I}_{i+1,i+2} \\
                    & = ( (( \mathcal{G}_i \setminus \mathcal{R}_{i,i+1} ) \cup \mathcal{I}_{i,i+1}) \setminus \mathcal{R}_{i+1,i+2} ) \cup \mathcal{I}_{i+1,i+2} \\
                    & = ( \mathcal{G}_i \setminus (\mathcal{R}_{i,i+1} \cup \mathcal{R}_{i+1,i+2}) ) \cup (\mathcal{I}_{i,i+1} \setminus \mathcal{R}_{i+1,i+2} \cup \mathcal{I}_{i+1,i+2} ) \\
                    & = ( \mathcal{G}_i \setminus \mathcal{R}_{i,i+2} ) \cup \mathcal{I}_{i,i+2}
\end{align}

The \textit{MergeRevision} algorithm is shown in \Cref{alg:mergerevision}.
The algorithm uses the \textit{CombineMany} operation (\Cref{alg:combinemany}) which applies recursively the combine function (\Cref{eq:combine,eq:combine_2,eq:combine_3}) to reduce a path of revisions into a single RDF Delta.

\begin{algorithm}
	\DontPrintSemicolon
  \caption{{\sc MergeRevision}}
  \label{alg:mergerevision}
  \KwIn{$ \mathcal{G}_i $, $ \mathcal{G}_j $}
  \KwOut{$ \mathcal{G}_m $, $\Delta_{i,m}$, $\Delta_{j,m}$}
  
  $\mathcal{G}_l \gets CommonAncestor(\mathcal{G}_i, \mathcal{G}_j $)\;
  $\mathcal{I}_{l,i},\mathcal{R}_{l,i} \gets CombineMany(RevisionsBetween(\mathcal{G}_l, \mathcal{G}_i))$\;
  $\mathcal{I}_{l,j},\mathcal{R}_{l,j} \gets CombineMany(RevisionsBetween(\mathcal{G}_l, \mathcal{G}_j))$\;
  $ \mathcal{G}_m \gets ( \mathcal{G}_l \setminus (\mathcal{R}_{l,i} \cup \mathcal{R}_{l,j} ) ) \cup \mathcal{I}_{l,i} \cup \mathcal{I}_{l,j} $\;
  \Return{$ \mathcal{G}_m $, $(\mathcal{I}_{l,j}, \mathcal{R}_{l,j})$, $(\mathcal{I}_{l,i}, \mathcal{R}_{l,i})$}
\end{algorithm}

\begin{algorithm}
	\DontPrintSemicolon
  \caption{{\sc CombineMany}}
  \label{alg:combinemany}
  \KwIn{$revisions = { \Delta_{i,i+1}, \Delta_{i+1,i+2}, \Delta... , \Delta_{j-1,j}}$}
  \KwOut{$\Delta_{i,j}$}
  \lIf{$\vert revisions \vert = 1$}{\Return{$\Delta_{i,i+1}$\Comment*[f]{end of recursion}}}
  $\Delta_{i,i+2} \gets combine(\Delta_{i,i+1}, \Delta_{i+1,i+2})$\;
  \Return{$CombineMany(\Delta_{i,i+2}, \Delta_{i+2,i+3}, \Delta..., \Delta_{j-1,j})$}
\end{algorithm}

\subsubsection{Complexity}
\label{ssec:complexity_merge}

The time complexity of \Cref{alg:mergerevision} and \Cref{alg:combinemany} is considered below. Note that when integrated with the full HFKN Framework, there is additional communication and other types of overhead involved. This is considered in~\Cref{sec:validate_synch}. 

\paragraph{Complexity of a Single Merge}

Assume that each revision has a maximum of $n^{max}_t$ triple removals or additions. The distance in the graph of revisions between two RDF Graph revisions, e.g. $\mathcal{G}_{i}$ and $\mathcal{G}_{j}$ is denoted as $d(\mathcal{G}_{i}, \mathcal{G}_{j})$.
The maximum value of $d$ is the total number of revisions, i.e. $|\mathcal{G}|$.
The $combine(\mathcal{G}_{i}, \mathcal{G}_{i+1})$ function in~\Cref{eq:combine} needs to be executed in a loop over the $n^{max}_t$ changes in $\mathcal{G}_{i+1}$ resulting in the worst-case time complexity of:

\begin{equation}
  O( (n^{max}_t)^2 )
\end{equation}

It is possible to avoid iterating over all the triples by using a hash table representation to index the triples.
A hash table has a worst-case time complexity of $O(n)$ and an average-case time complexity of $O(1)$~\cite{cormen01introduction}.

The complexity of the $combine$ function can be reduced to the following average-case time complexity (over all the possible inputs to the $combine$ function):

\begin{equation}
  O(n^{max}_t)
\end{equation}

$CombineMany(\mathcal{G}_l, \mathcal{G}_i)$ (\Cref{alg:combinemany}) applies the $combine$ function $d(\mathcal{G}_l, \mathcal{G}_i)$ times and has therefore a worst-case time complexity of:

\begin{equation}
  O( (n^{max}_t)^2 d(\mathcal{G}_l, \mathcal{G}_i) ) \sim O( (n^{max}_t)^2 |\mathcal{G}| )
\end{equation}

with an average-case time complexity (i.e. using hash table representation) of:

\begin{equation}
  O( (n^{max}_t)^2 d(\mathcal{G}_l, \mathcal{G}_i) ) \sim O( n^{max}_t |\mathcal{G}| )
\end{equation}

$MergeRevision(\mathcal{G}_i, \mathcal{G}_j)$(\Cref{alg:mergerevision}) applies the $CombineMany$ algorithm two times. Therefore, it has the same time complexity.

\paragraph{Merge K Revisions with the Same Ancestor}

Assume $K$ agents make a concurrent change to the same revision $\mathcal{G}_0$ (see \Cref{ssec:evaluting_merge_algorithm}).
At the $k$-th merge, the number of revisions is equal to $|\mathcal{G}| = k + 1$. The total worst-case time complexity of the merge of the $K$ revisions is given by:

\begin{equation}
  O( (n^{max}_t)^2 \sum\limits_{k=1}^{K} k+1 ) \sim O( (n^{max}_t)^2 K^2 )
\end{equation}

with an average-case time complexity (i.e. using hash table representation) of:

\begin{equation}
  O( n^{max}_t K^2 )
\end{equation}

\paragraph{General case}

In the general case, when merging two RDF Graph revisions, for example $\mathcal{G}_i$ with $\mathcal{G}_j$ in \Cref{fig_simpler_merge_problem}, the merge algorithm has to first find the shortest path between the two revisions and then apply the merge.
The shortest path is found using a \textit{Bidirectional-Dijkstra} algorithm~\cite{10.1145/322358.322360}.
In a sparse graph, the complexity of the Dijkstra algorithm is bounded by:

\begin{equation}
  O((|E|+|\mathcal{G}|) log(|\mathcal{G}|))
\end{equation}

Where $|\mathcal{G}|$ is the number of revisions and $|E|$ is the number of edges between each revision.
In a GoR, each revision has a maximum of $2$ parents, therefore $|E| \leq 2 |\mathcal{G}|$.
The complexity of finding the shortest path between two RDF Graph revisions, for example $\mathcal{G}_i$ and $\mathcal{G}_j$ becomes:

\begin{equation}
  O(3 |\mathcal{G}| log(|\mathcal{G}|)) = O(|\mathcal{G}| log(|\mathcal{G}|))
  \label{eqn:diskjtra_complexity}
\end{equation}

The complexity of computing the delta is given by:

\begin{equation}
    O((n^{max}_t)^2 d(\mathcal{G}_i, \mathcal{G}_j)) = O((n^{max}_t)^2 |\mathcal{G}|)
\end{equation}

The worst-case time complexity of a single merge is therefore given by:

\begin{equation}
  O(|\mathcal{G}| log(|\mathcal{G}|) + (n^{max}_t)^2 |\mathcal{G}|)
\end{equation}

As the graph grows, it is likely that $log(|\mathcal{G}|) >> (n^{max}_t)^2$ and therefore the worst-case complexity is given by~\Cref{eqn:diskjtra_complexity}.

\subsection{Rebase Algorithm}\label{sec:rebase}

\begin{figure}[tb]
  \centering
  \includegraphics[width=0.7\linewidth]{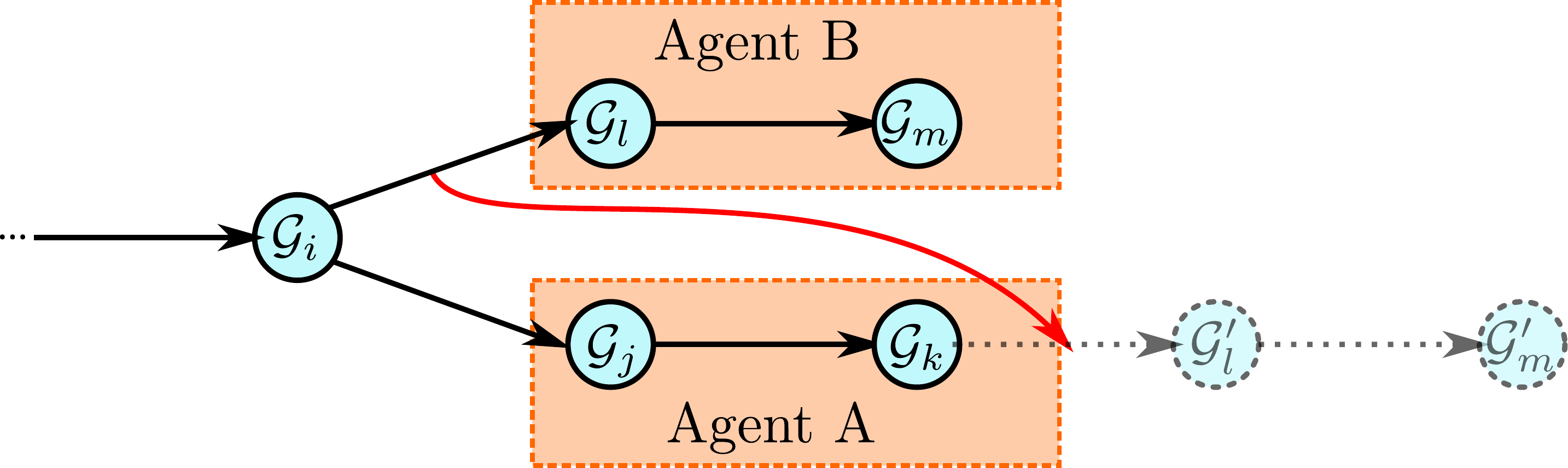}
  \caption{An illustrative example of applying the rebase operation to Graph of Revisions with two branches created by two agents.}
  \label{fig:rebase}
\end{figure}

An alternative to merging revisions is the \textit{rebase} operation. Rebase changes a parent of a revision, essentially moving a branch in the Graph of Revisions. In the HFKN Framework, the rebase is only applied to linear branches. 
A linear branch is defined as a branch that does not contain any split or merge revisions. This is not a limiting factor as changes made locally by an agent to an RDF\OP Document are always linear in nature. The rebase process applied in a simple example GoR with two branches is presented in~\Cref{fig:rebase}. In this scenario, two agents (A and B) created two branches in an RDF\OP Document. The branch created by Agent B is done locally (i.e. not broadcasted to other agents). The rebase operation applied by Agent B moves the source branch $\mathcal{G}_l$ (created by Agent B) to the destination branch $\mathcal{G}_k$ (created by Agent A).

The rebase process follows \Cref{alg:rebaserevisions}.
In \cref{alg:rebaserevisions_ancestor}, the algorithm looks for the common ancestor, using the Dijkstra search.
In case of the example shown in \Cref{fig:rebase}, the common ancestor is $\mathcal{G}_i$.
$\mathcal{G}_{next}$ holds the next parent revision when moving, it is initialised in \cref{alg:rebaserevisions_current_revision} to $\mathcal{G}_k$, so that the revision $\mathcal{G}_l$ is moved on $\mathcal{G}_k$ in \cref{alg:rebaserevisions_create_new}.
$\mathcal{G}_{next}$ is updated to the latest created revision in \cref{alg:rebaserevisions_update_g_next} (e.g. to $\mathcal{G}'_l$ for the first iteration).
Finally, the local revisions that have been rebased are removed in \cref{alg:rebaserevisions_delete_all}.

\begin{algorithm}
  \DontPrintSemicolon
  \caption{{\sc RebaseRevisions}}
  \label{alg:rebaserevisions}
  \KwIn{$ \mathcal{G}_m $, $ \mathcal{G}_k $}
  \KwOut{$ \mathcal{G}'_{l \rightarrow m} $}
  
  $\mathcal{G}_i \gets CommonAncestor(\mathcal{G}_m, \mathcal{G}_k $)\;\label{alg:rebaserevisions_ancestor}
  $\mathcal{G}_{i \rightarrow m} \gets RevisionsBetween(\mathcal{G}_i, \mathcal{G}_m)$\;
  $\mathcal{G}'_{i \rightarrow m} = \emptyset $\;
  $\mathcal{G}_{next} \gets \mathcal{G}_k$\;\label{alg:rebaserevisions_current_revision}
  \ForEach{$\mathcal{G} \in \mathcal{G}_{i \rightarrow m}$}{
    $\mathcal{G}' \gets InsertRevisionCopyAfter(\mathcal{G}, \mathcal{G}_{next}$)\;\label{alg:rebaserevisions_create_new}
    $\mathcal{G}_{next} \gets \mathcal{G}'$\;\label{alg:rebaserevisions_update_g_next}
    $\mathcal{G}'_{i \rightarrow m} = \mathcal{G}'_{i \rightarrow m} \bigcup \{ \mathcal{G}' \}$\;
  }
  $RemoveRevisions(\mathcal{G}_{i \rightarrow m})$\;\label{alg:rebaserevisions_delete_all}
  \Return{$\mathcal{G}'_{i \rightarrow m}$}
\end{algorithm}

After applying the rebase algorithm presented above, all local revisions are combined into the final GoR, which means the history of changes is preserved. This will directly influence the time required to synchronize changes among all agents as the number of messages used in the overall synchronization mechanism is proportional to the number of newly rebased revisions.
If preserving the history of all revisions is less important than how fast the shared information is synchronized among all agents, an alternative approach to the default rebase algorithm implemented in the HFKN Framework can be used.
The alternative adds an operation, called \textit{squashing} before the default rebase algorithm is applied.
The \textit{squashing} procedure combines all of the revisions in the source branch (i.e. $\mathcal{G}_l$ to $\mathcal{G}_m$ created by Agent B) into a single revision, thus reducing the number of revisions in the final GoR. \textit{Squashing} is performed using the \textit{CombineMany} operation (\Cref{alg:combinemany}). Performance of both variants of the rebase algorithm is evaluated in \Cref{ssec:evaluting_rebase_algorithm}.

\subsubsection{Complexity}
\label{ssec:complexity_rebase}

Following the notation from \Cref{fig:rebase}, when rebasing revision $\mathcal{G}_m$ on $\mathcal{G}_k$ with a common ancestor $\mathcal{G}_i$, the complexity of performing the rebase operation is given by:

\begin{equation}
  O(n_t^{max} d(\mathcal{G}_i, \mathcal{G}_m))
\end{equation}

However, it is necessary to first compute the path between $\mathcal{G}_m$ and $\mathcal{G}_k$, which means the complexity of rebase is also bounded by the complexity of the shortest path algorithm given by \Cref{eqn:diskjtra_complexity}.

\subsection{Synchronization Timing and Weak Consistency}
\label{ssec:synch_consistency}

Given two RDF\OP documents, we say that those documents are 
\textit{weakly consistent} relative to each other if the RDF Triples in the documents are the same. The HFKN framework does not require logical consistency. Weak consistency is guaranteed in the following sense. Each agent shares a subset of its RDF Graphs with other agents. For each subset of RDF Graphs shared by an agent, the union of those subsets is guaranteed to be weakly consistent after termination of synchronization processes associated with the synchronization algorithms subject to certain timing and computational assumptions. These assumptions are clarified in the following scenario.

Consider the following scenario in which $m$ RDF\OP Documents are shared among a team of agents.  The team has made $n$ asynchronous and/or concurrent changes to each of the shared RDF Graphs and no other changes occur.
Under these assumptions, the HFKN RDF\OP Document synchronization mechanism will ensure that the changes are synchronized among all team members within a certain time $T_{Total}$ dependent on computational capabilities of each agent and the performance of communication links. The time is defined as:
\begin{equation}
T_{Total} = m\times T_S + m\times T_M(n)+m\times T_C(n)+m\times T_U(n)
\end{equation}

where $T_S$ is the maximum time to select a merge master for each RDF\OP Document. $T_M(n)$ is the maximum time needed to merge all changes introduced by each agent to each RDF\OP Document to produce a consistent version of common graphs. Time $T_C(n)$ is the maximum time required to transfer necessary revisions to all agents. Time $T_U(n)$ is the maximum time required for each agent to apply the updates received from the merge master to its copy of the RDF\OP Document.

In a general case, for all scenarios where the asynchronous and concurrent changes to the shared RDF\OP Documents finish at time $t$, the HFKN Synchronization mechanism will ensure that all changes are synchronized among all participating agents at time $t+T_{Total}$.


\section{Dataset Transfer Protocol}\label{sec:data_exchange}

RDF Graph synchronization is used to exchange semantic information and metadata automatically between agents. Bandwidth intensive data, such as point clouds or images are not exchanged automatically. Instead, this kind of lower-level data is encapsulated in datasets and associated with metadata represented as RDF Graphs, as presented in \Cref{sec:dataset}.
The metadata is synchronized between agents using the synchronization protocol described in~\Cref{ssec:graph_synch_protocol}. When an agent needs to access the actual bandwidth intensive information, a dataset transfer protocol is used. The focus of this section is on how this is achieved.

An agent is aware of the existence and the location of a dataset from the automatically synchronised RDF Graphs. The process of downloading the data for local use is achieved through the use of the delegation framework. When the need for data transfer arises, a Task Specification Tree (TST) is generated consisting of two types of nodes:


\begin{itemize}
    \item The \verb|Send| node is instantiated by the agent that has the requested data and is able to provide it to others,
    \item The \verb|Receive| node is instantiated by the agents that require the data.
\end{itemize}
\begin{figure}
\centering
\includegraphics[width=0.7\linewidth]{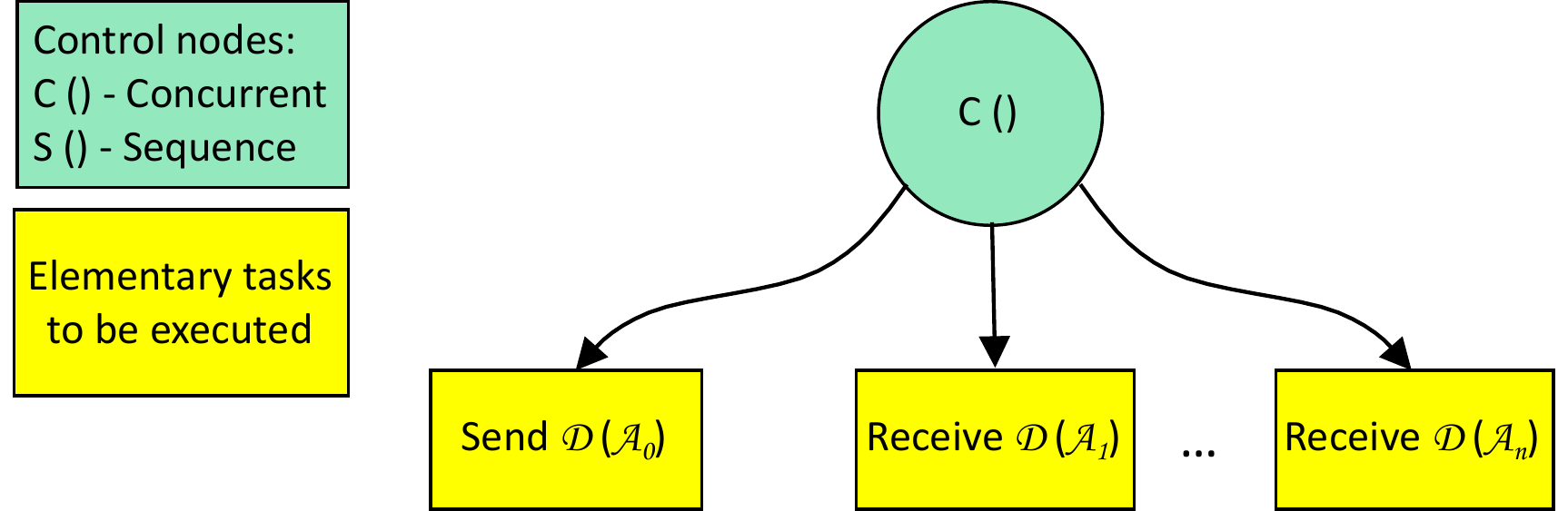}
\caption{A generic Task Specification Tree (TST) used to transfer the dataset $\mathcal{D}$ from agent $\mathcal{A}_0$ to agents $\mathcal{A}_1,..,\mathcal{A}_n$.
The Send and Receive nodes are executed concurrently by agents.}
\label{fig:tst_data_exchange}
\end{figure}

During the delegation process (see \Cref{sec:2}), which is usually initiated by the receiving agent, a sending agent is selected. As the data can be available from several agents, the criteria for choosing a sender can include constraints such as the battery level, the available bandwidth, or if the agent is expected to stay in communication range. 
The delegation process results in the generation of the TST shown in Figure~\ref{fig:tst_data_exchange}. In some cases, multiple agents may want to receive the same dataset. For this reason the TST can contain multiple \verb|Receive| nodes as can be seen in the figure. The data transmitted between agents is split into chunks and transferred sequentially.

\begin{figure}[tb]
\centering
\includegraphics[width=0.7\linewidth]{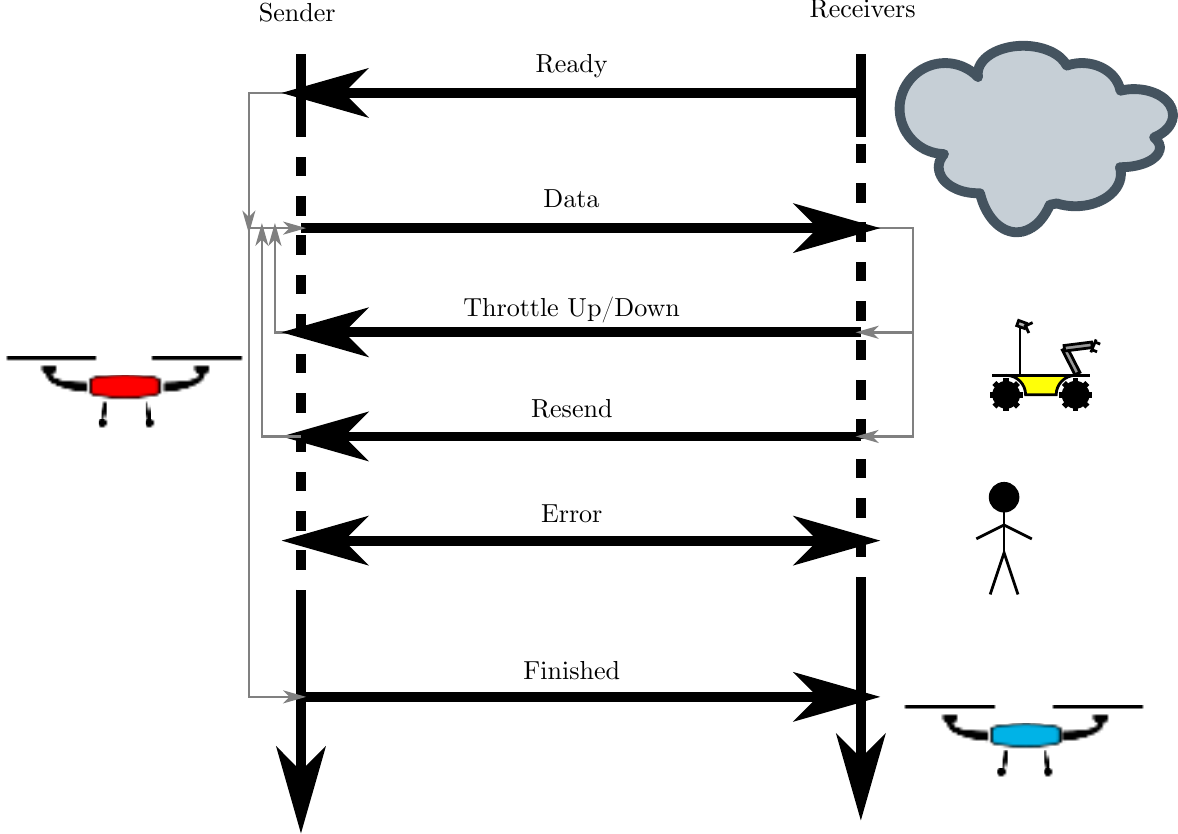}
\caption{Protocol for transferring datasets.}
\label{fig_data_synchronisation}
\end{figure}

An  overview of the data exchange protocol is presented in \Cref{fig_data_synchronisation}. There is only one sender that executes \Cref{alg:senddataset} and one or more receivers that execute \Cref{alg:receivedataset}.

The protocol uses the following messages:

\begin{itemize}
  \item The \verb|Ready| message is used to indicate readiness to receive data. 
  \item The \verb|Data| message contains a \textit{sequence} number, which allows receivers to check whether they obtained all the data, and an array of bytes containing a \textit{data} chunk of the dataset.
  \item The \verb|ResendRequest| message allows receivers to request missing data chunks.
  \item The \verb|Error| message allows participants to indicate an unrecoverable failure and abort the transfer.
  \item The \verb|ThrottleUp| and \verb|ThrottleDown| messages allow the receiver to indicate if the sender can send messages faster or slower.
  \item The \verb|Finished| message is sent by the sender to indicate that all the data has been sent. This message contains the last \textit{sequence} number, so that receivers can check they received all the data.
\end{itemize}

\begin{algorithm}[h]
  \DontPrintSemicolon
  \caption{{\sc SendDataset}}
  \label{alg:senddataset}
  \SetKw{Or}{or}
  \SetKwFor{Until}{until}{}{end}
  $WaitForReady()$\;\label{alg:senddataset_waitready}
  $finished \gets false$\;
  $\tau \gets 0$\;
  \Until{$finished$  }{
    $V \gets RetrieveData()$\;\label{alg:senddataset_retriever}
    \uIf{$V = \emptyset$}{\label{alg:senddataset_finished}
      $finished \gets true$\;
    }
    \Else{
      $SendData(V)$\;\label{alg:senddataset_send}
    }
    \While{$M \gets ReceiveResendRequest()$}{\label{alg:senddataset_receivemissing}
      $V \gets RetrieveData(M)$\;
      $SendData(V)$\;
    }
    \While{$ReceiveThrottleUp()$}{\label{alg:senddataset_throttle_up}
      $\tau \gets max(0, \tau - 1)$
    }
    \While{$ReceiveThrottleDown()$}{\label{alg:senddataset_throttle_down}
      $\tau \gets \tau + 1$
    }
    \If{$ReceiveError()$}{\label{alg:senddataset_failure}
      $finished \gets true$\;
    }
    $Sleep(\tau)$\;\label{alg:senddataset_sleep}
  }
  $SendFinished()$\;\label{alg:senddataset_send_finished}
\end{algorithm}

The agent responsible for providing data operates according to \Cref{alg:senddataset}.
It knows the list of receivers from the TST (\Cref{fig:tst_data_exchange}), and waits for them to be ready (\cref{alg:senddataset_waitready}).
The sender retrieves the data from its database (\cref{alg:senddataset_retriever}) and sends it to the receivers (\cref{alg:senddataset_send}) until all the data has been sent (\cref{alg:senddataset_finished}), then it sends the finished message (\cref{alg:senddataset_send_finished}).
The sender checks if it has received a request for missing data (\cref{alg:senddataset_receivemissing}) and adjusts the transfer speed up (\cref{alg:senddataset_throttle_up}) or down (\cref{alg:senddataset_throttle_down}) if needed.
If the sender receives an error message, it aborts the process (\cref{alg:senddataset_failure}).
At the end of the loop, the sender waits (\cref{alg:senddataset_sleep}) for the duration $\tau$, to make sure that the receivers can handle the reception of data messages. 

\begin{algorithm}[h]
  \DontPrintSemicolon
  \caption{{\sc ReceiveDataset}}
  \label{alg:receivedataset}
  \SetKw{Or}{or}
  \SetKwFor{Until}{until}{}{end}
  $SendReady()$\;\label{alg:receivedataset_sendready}
  $failure \gets false$\;
  \Until{$failure$ \Or $ReceiveFinished()$ }{\label{alg:receivedataset_until}
    $V \gets ReceiveData()$\;\label{alg:receivedataset_receivedata}
    \uIf{$Missing(V)$}{\label{alg:receivedataset_checkmissing}
      $RequestMissing(V)$\;
    }
    \uIf{$Valid(V)$}{\label{alg:receivedataset_checkvalidity}
      $InsertData(V)$\;\label{alg:receivedataset_insert}
    }
    \Else{
      $SendError()$ ;\label{alg:receivedataset_sendfailure}
      $failure \gets true$\;
    } 
    \uIf{$ReceptionQueueSize() > 5$}{\label{alg:receivedataset_throttle_down}
      $SendThrottleDown()$\;
    }
    \ElseIf{$ReceptionQueueSize() = 0$}{\label{alg:receivedataset_throttle_up}
      $SendThrottleUp()$\;
    }
    \If{$ReceiveError()$}{\label{alg:receivedataset_receive_failure}
      $failure \gets true$\;\label{alg:receivedataset_endloop}
    }
  }
  \uIf{$failure$}{
    $AbortInsertion()$\;\label{alg:receivedataset_abort_insertion}
  }
  \Else{
    $Finalise()$\;\label{alg:receivedataset_finalise}
  }
\end{algorithm}


An agent receiving the data operates according to  \Cref{alg:receivedataset}.
It starts by sending the ready message (\cref{alg:receivedataset_sendready}).
It then executes the main loop (\cref{alg:receivedataset_until}-\ref{alg:receivedataset_endloop}) that handles the data transfer including error handling, requesting missing data etc. The loop terminates either when the agent receives all data (i.e. receiving of the finished message) or when an error occurs (\cref{alg:receivedataset_until}).
The process starts with receiving of the data message (\cref{alg:receivedataset_receivedata}). 
Based on the sequence number of the received message, a check is done if there is any missing data (\cref{alg:receivedataset_checkmissing}), and if that is the case the agent requests it.
Before the agent inserts the received data into its database, the validity of the message is checked (\cref{alg:receivedataset_checkvalidity}).
In case the data is invalid the error message is send and the execution of the main loop is interrupted (\cref{alg:receivedataset_sendfailure}).
The receiving agent can request the sending agent to change the rate at which data messages are sent.
This is based on the amount of data messages placed in the reception queue. If the queue has more than 5 messages waiting to be processed, the receiver sends a throttle down message (\cref{alg:receivedataset_throttle_down}).
Otherwise, if the queue is empty it sends a throttle up message (\cref{alg:receivedataset_throttle_up}).
At the end of the loop, the agent checks if it has received an error message from other agents (\cref{alg:receivedataset_receive_failure}).
Once the main loop execution is finished, the agent can finalise the insertion of data (\cref{alg:receivedataset_finalise}). In case the transfer was not successful, the agent removes the received data from its database (\cref{alg:receivedataset_abort_insertion}).

\section{Empirical Evaluation of RDF Graph Synchronization}\label{sec:validate_synch}

The RDF Graph Synchronization mechanism,
described in \Cref{ssec:graph_synch_protocol},
provides the backbone for keeping distributed knowledge collected by a team of agents synchronized, up-to-date, and accessible for use by the team. In this section, we describe simulation experiments designed to evaluate the performance of this backbone. \Cref{sec:case_study} will provide a field robotics experiment with the full HFKN Framework, using a number of deployed UAVs and human operators. 

Several aspects of the synchronization mechanism are evaluated. The first set of experiments focuses on the performance of the merge algorithm used by the merge master to synchronize RDF Graphs. The algorithm's efficiency is evaluated in a simulated scenario where a large number of new revisions are created, with varying size of RDF Triple changes (additions or deletions). The experiments solely focus on measuring the timing of the algorithm excluding any extraneous factors related to communication links. The results are presented in \Cref{ssec:evaluting_merge_algorithm}. Similar evaluation was performed for the rebase algorithm with the results presented in \Cref{ssec:evaluting_rebase_algorithm}.

Additionally, a more complex simulation was performed to evaluate the HFKN Framework from a more practical perspective. A scenario, in which all functionalities of the RDF Graph synchronization mechanism are used, was created. It includes multiple agents creating, changing and synchronizing multiple RDF Graphs. During the experiment network disconnections were introduced to showcase the resilience of the HFKN Framework.
The evaluation results presented in \Cref{ssec:evaluating_sync_protocol} include an in-depth analysis of the HFKN Framework behaviour in this complex scenario that include message exchange statistics, timeline for synchronization events and the history of all revisions over the course of the experiment.

Finally, a set of experiments was conducted to evaluate the scalability of the RDF Graph synchronization mechanism. In particular, the focus was to find the maximum rate at which new revisions can be created and still synchronized in practice among all agents. This rate depends on multiple factors (e.g. number of agents, number of RDF\OP Documents etc.) which were taken into account during the experiment design. The results of the evaluation in scenarios including missions with up to 20 agents are presented in \Cref{ssec:evaluating_max_sync_rate}.


\subsection{Evaluation of the Merge Algorithm}
\label{ssec:evaluting_merge_algorithm}
The performance of the merge algorithm presented in \Cref{sec:merge} was evaluated in an experiment focused on the timing requirements of the merge operation itself. This means that the overhead of communication required for the RDF Graph synchronization is not included.
The experimental setup can be described by the following scenario. 
Consider 1000 agents making concurrent changes (in form of adding or removing RDF Triples) to a single RDF Graph. The graph is to be synchronized among all agents.
Each change results in a new revision of the RDF\OP Document, such that its GoR will contain multiple branches.
The merge algorithm is used to combine newly created revisions into one merge revision which includes all the changes introduced by all agents. 

\begin{figure}[t]
\centering
\includegraphics[width=0.35\linewidth]{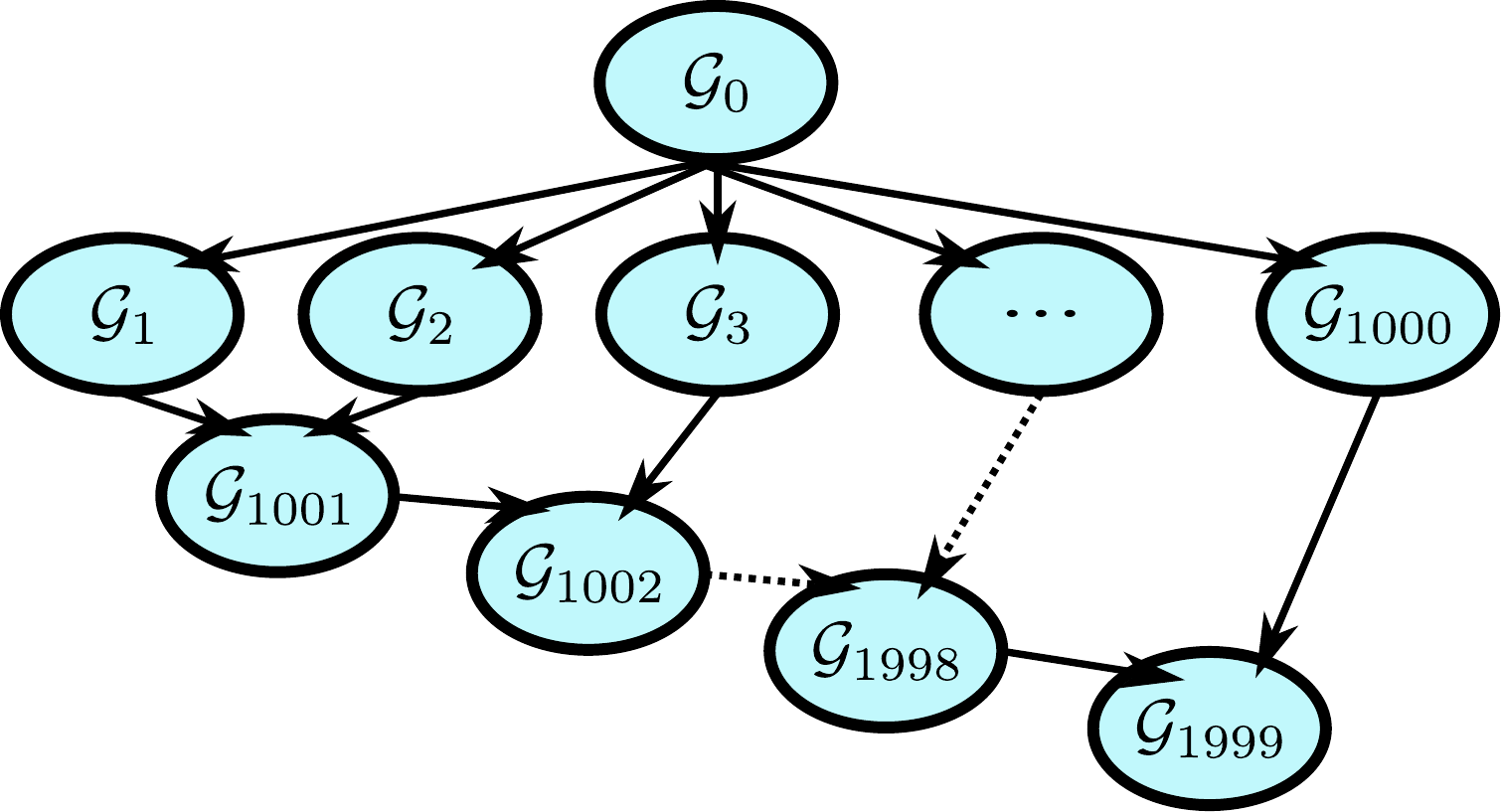}
\caption{The final graph of revisions in the experimental scenario used for the performance evaluation of the merge algorithm. }
\label{fig:merge_many_concurrent_edit}
\end{figure}

\Cref{fig:merge_many_concurrent_edit} depicts the GoR after the experiment is finished.
All generated revisions ($\mathcal{G}_1 \ldots \mathcal{G}_{1000}$) share the same parent revision $\mathcal{G}_0$. Since the merge operation is applied to two branches at a time, it has to be applied 999 times in this scenario in order to create the final merge revision which includes all the changes (i.e. $\mathcal{G}_{1999}$).
In order to quantify the influence of the size of the changes applied to the RDF Graph on the performance of merge algorithm, two cases of the experiment were performed, each containing changes consisting of 10 or 100 RDF Triples.
The experiment was performed on a computer with an \texttt{Intel(R) Xeon(R) CPU E5-1620 v2 @ 3.70GHz} with six cores and 16GB of RAM.\footnote{All algorithmic code is implemented in C++.}

The results of the experiment are presented in~\Cref{fig:merge_many_concurrent_edit_results}. They confirm that the merge operation time complexity is linear in the number of revisions that need to be combined when merging (see the red curve depicting the time required for a single merge as a function of the number of merge operations).
This observation follows the complexity analysis presented in \Cref{ssec:complexity_merge}.
Additionally, the update size for each revision (i.e. 10 vs 100 tuples) increases the time required by a constant factor. For instance, the average time in the case of the last merge operation (i.e. 999) increases from $0.5s$ to $5s$ for 10 and 100 tuples per revision, respectively. 
Inconsistencies in timing when considering single merge operations (spikes present in the red plot) are most likely caused by the overhead of accessing the database and variations of the computer system performance.
The total complexity of merging is quadratic in the number of revisions to merge as can be seen in \Cref{fig:merge_many_concurrent_edit_results} (blue curves).

\input{MergeManyConcurrentEditResults.tex}

\subsection{Evaluation of the Rebase Algorithm}
\label{ssec:evaluting_rebase_algorithm}

The purpose of this experiment is to evaluate the performance of the rebase operation described in \Cref{sec:rebase}. As in \Cref{fig:rebase}, we consider two agents $A$ and $B$ creating new revisions in parallel that correspond to two branches in the graph of revisions of the RDF\OP Document.
The rebase source branch (top of \Cref{fig:rebase}) created by agent $B$ has a varying number of revisions with varying number of changes in each revision.
The content of the rebase destination branch created by agent $A$ does not influence the performance of the rebase algorithm, therefore the destination branch contains only a single revision in this experiment.
Two variations were evaluated:
\begin{itemize}
  \item move all the individual revisions from the source  to the destination branch (c.f. \Cref{fig:rebase}, revisions $G_l$ and $G_m$ are kept separate). This is currently the default behaviour of the rebase operation as used in the HFKN Framework, see \Cref{alg:rebaserevisions}.
  \item combine all revisions from the source branch into a single one before applying the rebase algorithm, using the squashing operation.
\end{itemize}

The number of revisions in the source branch was varying from $1$ to $40$ and the number of changes (additions and removals) in each revision from $10$ to $50$. The experiment was performed on a single computer omitting the communication links between agents. Other aspects, such as access to the database, were considered. We measured the timing for each variant of the rebase and each variation of the number of revisions and changes.

\begin{figure}[tb]
  \begin{center}
    \subfloat{
      \tikzsetnextfilename{rebase_raw_legend}
      \begin{tikzpicture}[scale=0.6]
        \begin{axis}[
            width=0.2\textwidth, height=0.2\textwidth,
            hide axis,
            xmin=0, xmax=10, ymin=0, ymax=0.2,
            legend columns=6,
            hide axis,
            xmin=0, xmax=10, ymin=0, ymax=0.2,
            no markers,
            cycle list/Set1,
          ]
          \addlegendimage{empty legend}
          \addlegendentry{\hspace{-1.5cm}\makebox[0pt][l]{\textbf{Number of Changes per Revision}}}
          \pgfplotsinvokeforeach{1,2,3,4}{\addlegendimage{empty legend}\addlegendentry{}}
          \addlegendimage{empty legend}\addlegendentry{\makebox[0.5cm][l]{}}
          \pgfplotsinvokeforeach{1,2,3,4,5}{\addplot coordinates {(0,-1)};}
          \addlegendentry{10}
          \addlegendentry{20}
          \addlegendentry{30}
          \addlegendentry{40}
          \addlegendentry{50}
        \end{axis}
      \end{tikzpicture}
    } \\
    \addtocounter{subfigure}{-1}
    \subfloat[Without squashing]{
      \tikzsetnextfilename{rebase_raw_without_squashing} 
      \begin{tikzpicture}[scale=0.65]
        \begin{axis}
        [
            width=0.7\linewidth, 
            grid=major, 
            grid style={dashed,gray!30}, 
            xlabel=Number of revisions, 
            ylabel=Average rebase time,
            axis y line=left,
            axis x line=bottom,
            y unit=\si{\ms},
            xmin=1,
            xmax=40,
            ymin=20,
            no markers,
            cycle list/Set1,
            legend pos=north west,
          ]
           \addplot+ [discard if not={changes}{10}] table[x=revisions,y=average_time,col sep=space] {results/benchmark_rebase_raw_3.txt}; 
           \addplot+ [discard if not={changes}{20}] table[x=revisions,y=average_time,col sep=space] {results/benchmark_rebase_raw_3.txt}; 
           \addplot+ [discard if not={changes}{30}] table[x=revisions,y=average_time,col sep=space] {results/benchmark_rebase_raw_3.txt}; 
           \addplot+ [discard if not={changes}{40}] table[x=revisions,y=average_time,col sep=space] {results/benchmark_rebase_raw_3.txt}; 
           \addplot+ [discard if not={changes}{50}] table[x=revisions,y=average_time,col sep=space] {results/benchmark_rebase_raw_3.txt}; 
        \end{axis}
      \end{tikzpicture}
      \label{fig:rebase_no_squashing}
      }
    \subfloat[With squashing]{
      \tikzsetnextfilename{rebase_raw_with_squashing} 
      \begin{tikzpicture}[scale=0.65]
        \begin{axis}
        [
            width=0.7\linewidth, 
            grid=major, 
            grid style={dashed,gray!30}, 
            xlabel=Number of revisions, 
            ylabel=Average rebase time,
            axis y line=left,
            axis x line=bottom,
            y unit=\si{\ms},
            xmin=1,
            xmax=40,
            ymin=20,
            no markers,
            cycle list/Set1,
            legend pos=north west,
          ]
           \addplot+ [discard if not={changes}{10}] table[x=revisions,y=average_time,col sep=space] {results/benchmark_rebase_raw_5.txt}; 
           \addplot+ [discard if not={changes}{20}] table[x=revisions,y=average_time,col sep=space] {results/benchmark_rebase_raw_5.txt}; 
           \addplot+ [discard if not={changes}{30}] table[x=revisions,y=average_time,col sep=space] {results/benchmark_rebase_raw_5.txt}; 
           \addplot+ [discard if not={changes}{40}] table[x=revisions,y=average_time,col sep=space] {results/benchmark_rebase_raw_5.txt}; 
           \addplot+ [discard if not={changes}{50}] table[x=revisions,y=average_time,col sep=space] {results/benchmark_rebase_raw_5.txt}; 
        \end{axis}
      \end{tikzpicture}
      \label{fig:rebase_with_squashing}
      }
  \end{center}
  \caption{Results of the performance evaluation of the rebase algorithm. The time is averaged over 20 rebase operations.}
  \label{fig:rebase_raw}
\end{figure}
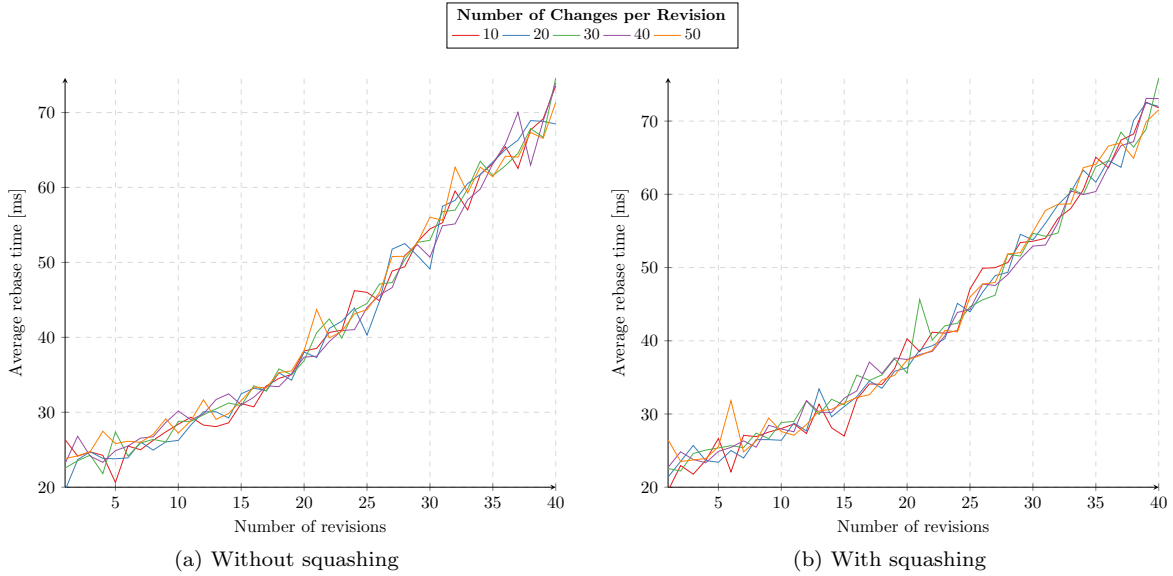

\Cref{fig:rebase_raw} shows the average time it takes for the rebase algorithm for the different configurations. 
The rebase operation has a constant time cost of about $23ms$ when only one revision is rebased. Each additional revision takes $1.3ms$ on average, which results in $72ms$ for the case of rebasing 40 revisions as shown in \Cref{fig:rebase_raw}.
Additionally, the rebase with squashing (\Cref{fig:rebase_with_squashing}) or without (\Cref{fig:rebase_no_squashing}) has very little influence on the overall performance of the rebase operation.
Squashing reduces the number of revisions in the graph, which means that the number of messages exchanged between agents is reduced. This in turn minimizes the number of messages which have to be transmitted in the overall process of the RDF Graph synchronization mechanism. Therefore performing squashing should be preferred if the quickest synchronization between all agents is the main priority.
It is important to note that in the evaluation the dominating factor is the overhead caused by accessing the database to fetch the revisions. For this reason, the number of changes and whether squashing is performed has minimal influence on the overall performance of the rebase operation.

\subsection{Evaluation of the RDF Graph Synchronization}
\label{ssec:evaluating_sync_protocol}
The HFKN Framework uses a number of algorithms and protocols which allow for efficient and robust RDF Graph synchronization described in detail in \Cref{sec:rdf_doc_sync}.
Those functionalities are part of the KDB Manager which is executed by each participating agent. Throughout a mission execution the new information, which is acquired by each agent, results in changes to the RDF Graphs shared among all agents. In order to achieve a common consistent view of the available information among agents the RDF Graphs are synchronized. 
Unlike the evaluations described thus far,
in this subsection the evaluation of the whole RDF Graph synchronization mechanism is presented.

This experiment was designed to demonstrate the ability of the HFKN Framework to handle concurrent changes in an RDF Graph shared among agents and robustness against communication interruptions.
A typical data-gathering mission is simulated where twelve agents explore an environment and new datasets are created and exchanged. Note that the focus is put only on how this affects modifications to the shared RDF Graph. Thus for this experiment, sensor readings were not simulated, and no sensor data was transferred between agents. Only the RDF Graph synchronization protocol as presented in  \Cref{ssec:graph_synch_protocol} was used.

 \begin{figure}[tb]
\centering
\includegraphics[width=0.55\linewidth]{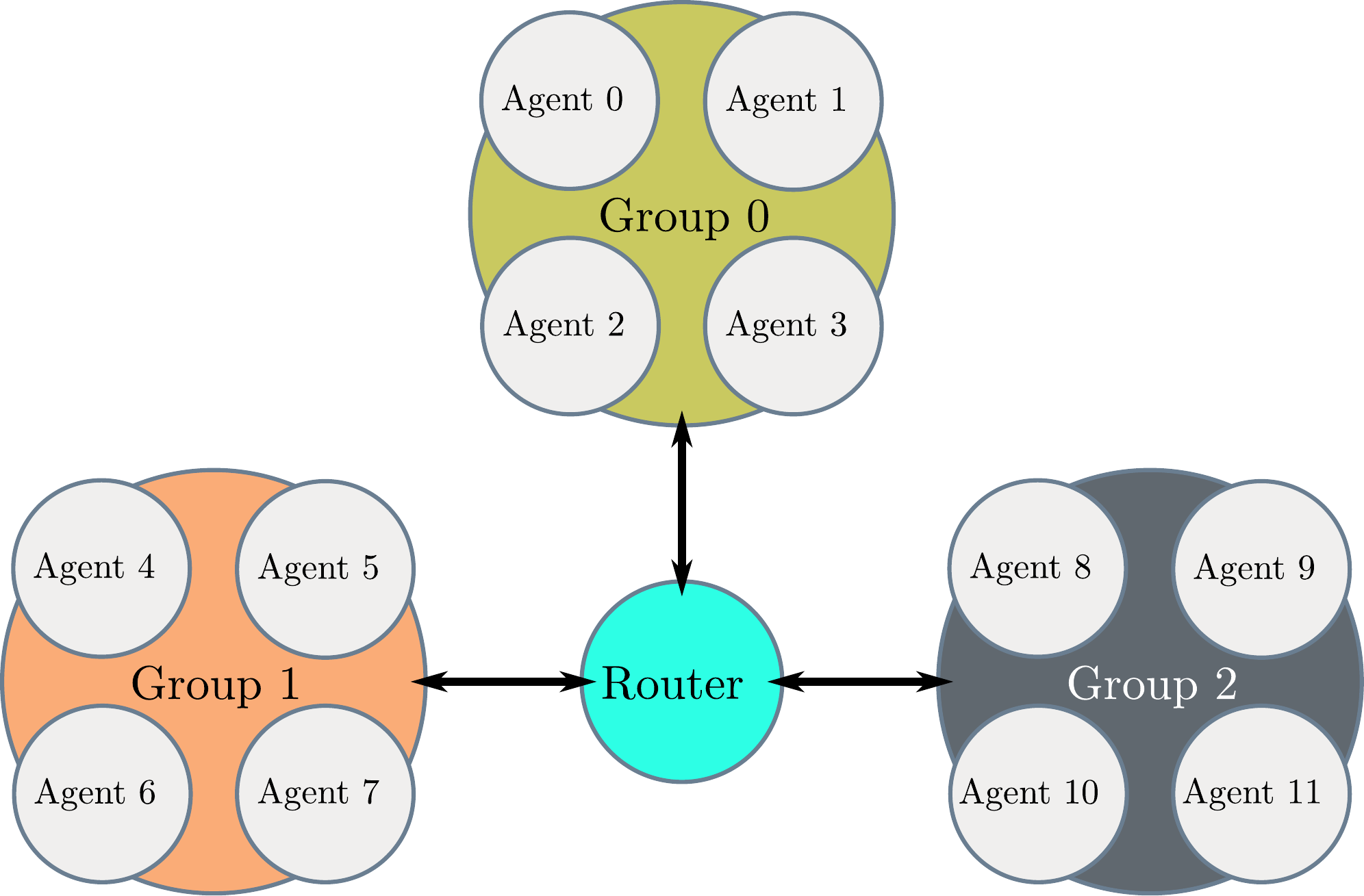}
\caption{The experimental setup in a mission involving 12 agents distributed over three computers.}
\label{fig_graph_synchronisation_experiment_setup}
\end{figure}

Twelve agents are split into three groups of four as shown in \Cref{fig_graph_synchronisation_experiment_setup}.
Each group is simulated on one computer.
One of them uses an \texttt{Intel(R) Xeon(R) CPU E5-1620 v2 @ 3.70GHz} with six cores, two other computers have an \texttt{Intel(R) i7-7567U} CPU.
All three computers use 16GB of RAM. The machines were connected through a router to easily demonstrate what happens when agents get disconnected from each other.

\begin{figure}[p]
\vspace*{-2cm}
\makebox[\linewidth]{
    \includegraphics[width=0.95\linewidth]{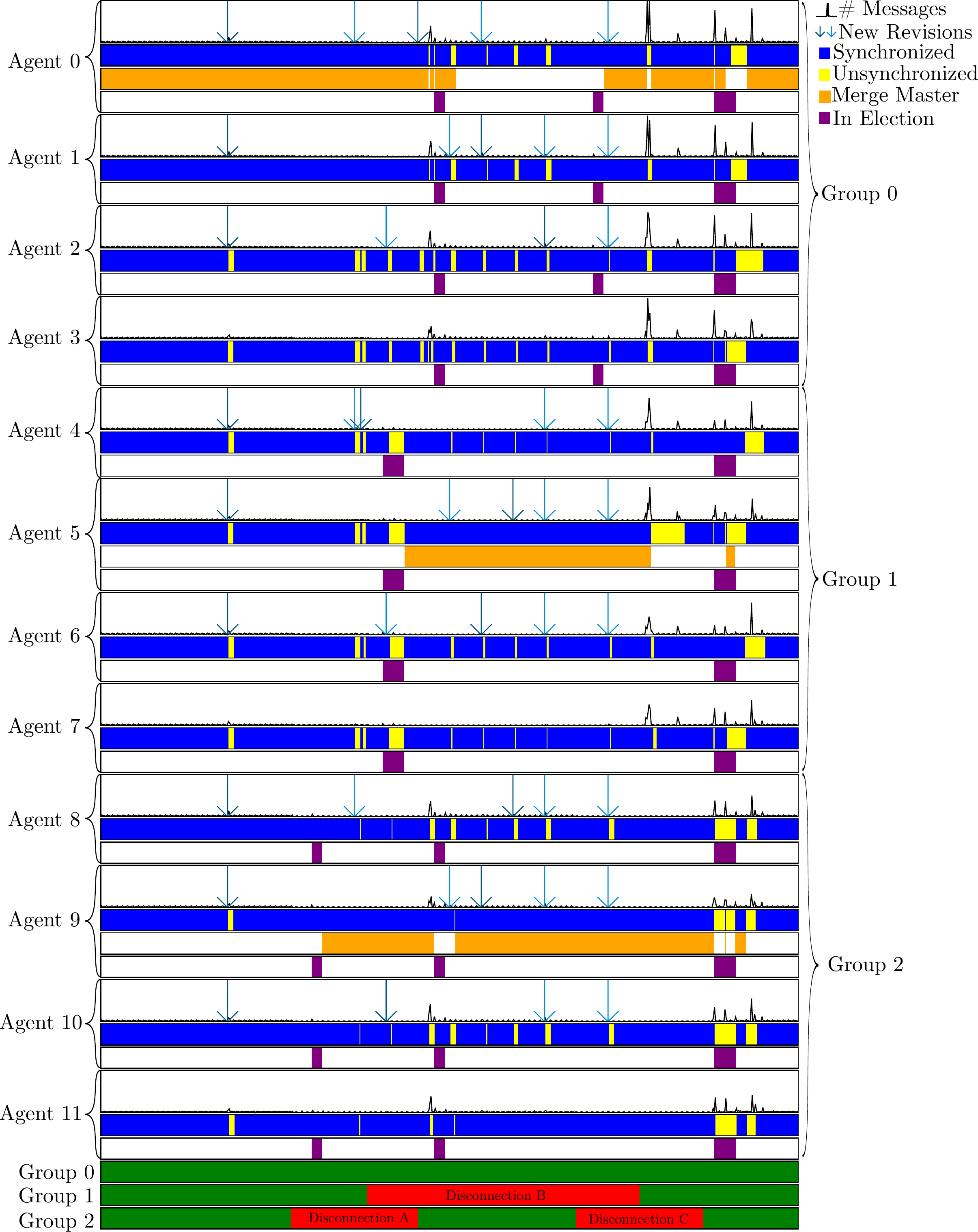}
}
\caption{Experimental results: The top row of each agent shows arrows to indicate when new triples are inserted into the agent's RDF Graph and the plots to indicate the number of messages exchanged at any given time. Blue indicates when the agent is synchronized with the merge master and yellow when it is unsynchronised. Purple indicates an election period. Orange indicates when an agent has the role of a merge master. Green indicates when a team computer is connected to the router and red when it is disconnected.}

\label{fig_graph_synchronisation_experiment_revs}
\end{figure}

The timeline summarizing the experiment is shown in  \Cref{fig_graph_synchronisation_experiment_revs}.
The agents within a group are assumed to always be able to communicate. To demonstrate the ability to handle network interruptions, we simulate disconnection and re-connection between groups.
Computers used for \textit{Group 1} and \textit{Group 2} get disconnected for some period of time from the router as shown in red in \Cref{fig_graph_synchronisation_experiment_revs}. This results in creating up to three distinct groups that elect their own merge masters (c.f. the orange color in the figure) until the connection is reestablished and a single merge master for all three groups is elected.
During the course of the experiment several agents make simultaneous changes to the RDF Graph which need to be synchronized. The new revisions, which are the result of the changes, are marked by blue arrows.


The results show that the agents are quick to synchronize the RDF Graphs with each other (c.f. the unsynchronised times in \Cref{fig_graph_synchronisation_experiment_revs}) even in the context of multiple concurrent edits or connection and disconnection behavior.

Each agent received an average of 1959 messages, out of which 1584 (approx. 81\%) were status messages (see \Cref{sec:doc_sync_messages}).
Agents received the maximum of 105 messages/s and the average of 2.25 messages/s.

\begin{figure*}[tb]
\centering
\includegraphics[width=0.99\linewidth]{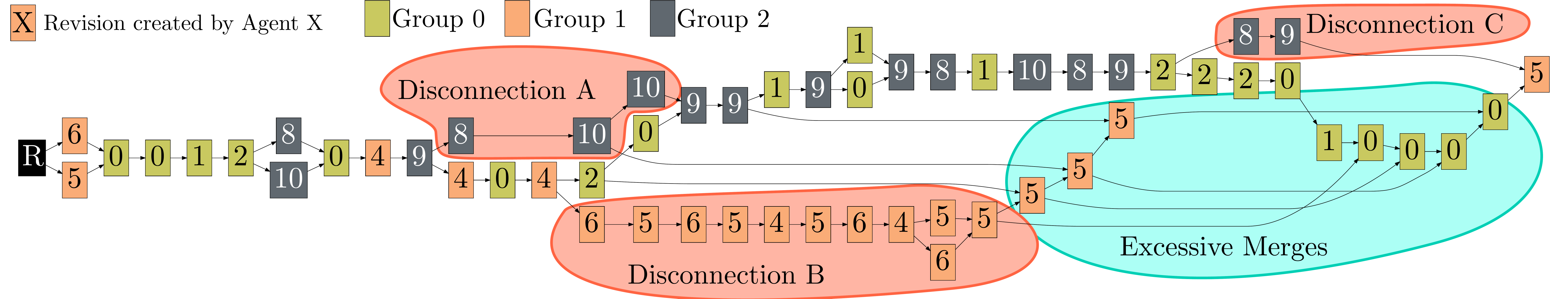}
\caption{The final Graph of Revisions for the experiment. Numbers indicate which agent created the revision. The colors of boxes show the three groups of agents. Three disconnection periods are marked with read areas. The turquoise area indicates the period of excessive merges.}
\label{fig_exp_revisions}
\end{figure*}

The final Graph of Revisions of the experiment presented in \Cref{fig_exp_revisions} shows the relation between the different revisions and their creators (numbers in boxes). 
The red areas show the revisions that were created during the three disconnections: A, B and C (c.f. bottom of~\Cref{fig_graph_synchronisation_experiment_revs}).
The turquoise area shows an excessive number of merges that were computed during the reconnection.
This happens in particular at the end of \textit{Disconnection B}, when \textit{Group 0} and \textit{Group 1} reconnect.
The merge master, Agent 5, follows an eager strategy for merging, and starts the process as soon as it receives revisions from \textit{Group 0}. After an election occurs and Agent 0 is selected, the same problem repeats.
In normal operation, the merge eagerness allows for faster propagation of knowledge between agents, but in future work, different strategies for when to merge will be investigated.

\subsection{Evaluating the Maximum Revision Creation Rate}
\label{ssec:evaluating_max_sync_rate}
In this experiment we are interested in the scalability of the RDF Graph synchronization mechanism when many agents make concurrent changes to several documents. 
The evaluation presented in this section is focused on finding the limits on how fast the changes created concurrently by multiple agents in multiple RDF Graphs can in practice be merged and synchronized without unnecessary delays.
In other words, we are interested in the \emph{practical} rate of changes below which the merge master can perform its task in a timely manner. Above this limit the queue of changes grows without the possibility to be merged on time, before more changes are created. 
There exists a wide range of factors which influence this maximum rate: 
the number of RDF\OP Documents, the number of revisions per document, the number of participating agents, communication links' properties, agent's hardware etc.
For this experiment, we focus on the factors that directly affect the computational cost of merging, such as the number of documents, the amount of changes to the documents and the number of participating agents.



Agents perform changes to the RDF Graphs concurrently and locally as they carry out their mission tasks in real world environments. In a general case, the agents' KDB Managers create new revisions for each change in a document which is then sent to the merge master.
In practice, it is the performance of executing the tasks of the merge master which limits the rate at which the RDF Graphs can be merged and synchronized in a timely manner.
This issue is related to the Never Synchronized Problem described in \Cref{ssec:sync_protocol} from which one can conclude that the rate of publishing of local revisions needs to be smaller than the time it takes to perform the merge by the master. Also note that the fact that rebase is used does not solve this practical bottleneck. It only handles the behaviour of an agent regarding local revisions while waiting for the result of a merge. 

Effectively, if agents create changes to shared documents above the maximum rate at which a merge master can perform its tasks in a timely manner, a delay will be introduced in the synchronization process. The delay duration will vary depending on the amount of excessive changes that still need to be combined. In an extreme case, when all agents always create changes over the maximum rate, the delay will keep growing and the full synchronization will never be achieved. However, in a more practical setting agents join and leave missions as they have limits on how long they can operate (e.g. fuel, battery power), and the maximum load imposed on the merge master will vary over time. After a period of too frequent updates, when the rate becomes lower and manageable by the master, the changes will eventually be merged, thus the information will be synchronized among all agents.
One way to improve maximum synchronization rate would be to include the computational performance of the agents when electing a new merge master, but this will be pursued in future work.

With the above in mind, the evaluation was performed using the following procedure loop:
\begin{itemize}
 \item Each agent creates a new revision with at most $ C $ insertions or deletions of triples for each document.
 \item Each agent broadcasts its revisions to others and waits for the master to finish the merge.
 \item The master performs the merge procedure.
\end{itemize}

To isolate and evaluate the performance of the tasks of the merge master, in this experiment the agents are \emph{not allowed} to make local changes until the merge master finishes the merge. When the merge master finishes the synchronization process and sends the new revision, 
the cycle may then start again. 
In the experiment all documents are merged by a single master. This is a worst-case scenario as, in general, separate documents can have different merge masters thus the merge performance can be improved.

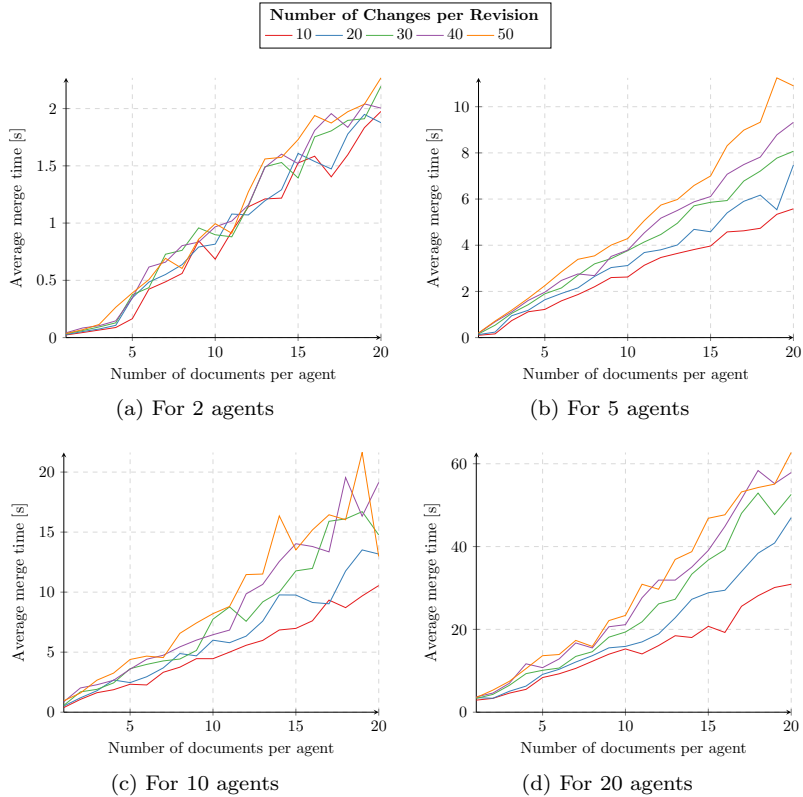
\begin{figure}[tb]
  \begin{center}
    \subfloat{
      \tikzsetnextfilename{brtl_legend}
      \begin{tikzpicture}[scale=0.6]
        \begin{axis}[
            width=0.2\textwidth, height=0.2\textwidth,
            hide axis,
            xmin=0, xmax=10, ymin=0, ymax=0.2,
            legend columns=6,
            hide axis,
            xmin=0, xmax=10, ymin=0, ymax=0.2,
            no markers,
            cycle list/Set1,
          ]
          \addlegendimage{empty legend}
          \addlegendentry{\hspace{-1.5cm}\makebox[0pt][l]{\textbf{Number of Changes per Revision}}}
          \pgfplotsinvokeforeach{1,2,3,4}{\addlegendimage{empty legend}\addlegendentry{}}
          \addlegendimage{empty legend}\addlegendentry{\makebox[0.5cm][l]{}}
          \pgfplotsinvokeforeach{1,2,3,4,5}{\addplot coordinates {(0,-1)};}
          \addlegendentry{10}
          \addlegendentry{20}
          \addlegendentry{30}
          \addlegendentry{40}
          \addlegendentry{50}
        \end{axis}
      \end{tikzpicture}
    } \\
    \addtocounter{subfigure}{-1}
    \subfloat[For 2 agents]{
      \tikzsetnextfilename{brtl2} 
      \begin{tikzpicture}[scale=0.6]
        \begin{axis}[
            width=0.7\linewidth, 
            grid=major, 
            grid style={dashed,gray!30}, 
            xlabel=Number of documents per agent, 
            ylabel=Average merge time,
            axis y line=left,
            axis x line=bottom,
            y unit=\si{\s},
            xmin=1,
            xmax=20,
            ymin=0,
            no markers,
            cycle list/Set1,
            legend pos=north west,
            y filter/.code={\pgfmathparse{#1/(1000)}\pgfmathresult}
          ]
          \addplot table[x=documents,y=average_time,col sep=comma] {results/benchmark_many_documents_many_agents/documents_to_merge_time/2_10.csv}; 
          \addplot table[x=documents,y=average_time,col sep=comma] {results/benchmark_many_documents_many_agents/documents_to_merge_time/2_20.csv}; 
          \addplot table[x=documents,y=average_time,col sep=comma] {results/benchmark_many_documents_many_agents/documents_to_merge_time/2_30.csv}; 
          \addplot table[x=documents,y=average_time,col sep=comma] {results/benchmark_many_documents_many_agents/documents_to_merge_time/2_40.csv}; 
          \addplot table[x=documents,y=average_time,col sep=comma] {results/benchmark_many_documents_many_agents/documents_to_merge_time/2_50.csv}; 
        \end{axis}
      \end{tikzpicture}
      }
    \subfloat[For 5 agents]{
      \tikzsetnextfilename{brtl5} 
      \begin{tikzpicture}[scale=0.6]
        \begin{axis}[
            width=0.7\linewidth, 
            grid=major, 
            grid style={dashed,gray!30}, 
            xlabel=Number of documents per agent, 
            ylabel=Average merge time,
            axis y line=left,
            axis x line=bottom,
            y unit=\si{\s},
            xmin=1,
            xmax=20,
            ymin=0,
            no markers,
            cycle list/Set1,
            legend pos=north west,
            y filter/.code={\pgfmathparse{#1/(1000)}\pgfmathresult}
          ]
          \addplot table[x=documents,y=average_time,col sep=comma] {results/benchmark_many_documents_many_agents/documents_to_merge_time/5_10.csv}; 
          \addplot table[x=documents,y=average_time,col sep=comma] {results/benchmark_many_documents_many_agents/documents_to_merge_time/5_20.csv}; 
          \addplot table[x=documents,y=average_time,col sep=comma] {results/benchmark_many_documents_many_agents/documents_to_merge_time/5_30.csv}; 
          \addplot table[x=documents,y=average_time,col sep=comma] {results/benchmark_many_documents_many_agents/documents_to_merge_time/5_40.csv}; 
          \addplot table[x=documents,y=average_time,col sep=comma] {results/benchmark_many_documents_many_agents/documents_to_merge_time/5_50.csv}; 
        \end{axis}
      \end{tikzpicture}
      }
      
    \subfloat[For 10 agents]{
      \tikzsetnextfilename{brtl9} 
      \begin{tikzpicture}[scale=0.6]
        \begin{axis}[
            width=0.7\linewidth, 
            grid=major, 
            grid style={dashed,gray!30}, 
            xlabel=Number of documents per agent, 
            ylabel=Average merge time,
            axis y line=left,
            axis x line=bottom,
            y unit=\si{\s},
            xmin=1,
            xmax=20,
            ymin=0,
            no markers,
            cycle list/Set1,
            legend pos=north west,
            y filter/.code={\pgfmathparse{#1/(1000)}\pgfmathresult}
          ]
          \addplot table[x=documents,y=average_time,col sep=comma] {results/benchmark_many_documents_many_agents/documents_to_merge_time/10_10.csv}; 
          \addplot table[x=documents,y=average_time,col sep=comma] {results/benchmark_many_documents_many_agents/documents_to_merge_time/10_20.csv}; 
          \addplot table[x=documents,y=average_time,col sep=comma] {results/benchmark_many_documents_many_agents/documents_to_merge_time/10_30.csv}; 
          \addplot table[x=documents,y=average_time,col sep=comma] {results/benchmark_many_documents_many_agents/documents_to_merge_time/10_40.csv}; 
          \addplot table[x=documents,y=average_time,col sep=comma] {results/benchmark_many_documents_many_agents/documents_to_merge_time/10_50.csv}; 
        \end{axis}
      \end{tikzpicture}
      }
    \subfloat[For 20 agents]{
      \tikzsetnextfilename{brtl20} 
      \begin{tikzpicture}[scale=0.6]
        \begin{axis}[
            width=0.7\linewidth, 
            grid=major, 
            grid style={dashed,gray!30}, 
            xlabel=Number of documents per agent, 
            ylabel=Average merge time,
            axis y line=left,
            axis x line=bottom,
            y unit=\si{\s},
            xmin=1,
            xmax=20,
            ymin=0,
            no markers,
            cycle list/Set1,
            legend pos=north west,
            y filter/.code={\pgfmathparse{#1/(1000)}\pgfmathresult}
          ]
          \addplot table[x=documents,y=average_time,col sep=comma] {results/benchmark_many_documents_many_agents/documents_to_merge_time/20_10.csv}; 
          \addplot table[x=documents,y=average_time,col sep=comma] {results/benchmark_many_documents_many_agents/documents_to_merge_time/20_20.csv}; 
          \addplot table[x=documents,y=average_time,col sep=comma] {results/benchmark_many_documents_many_agents/documents_to_merge_time/20_30.csv}; 
          \addplot table[x=documents,y=average_time,col sep=comma] {results/benchmark_many_documents_many_agents/documents_to_merge_time/20_40.csv}; 
          \addplot table[x=documents,y=average_time,col sep=comma] {results/benchmark_many_documents_many_agents/documents_to_merge_time/20_50.csv}; 
        \end{axis}
      \end{tikzpicture}
    }
  \end{center}
  \caption{Results of the experimental evaluation with focus on the scalability aspects of the HKFN Framework. Plots depict the average merge time as a function of a number of documents per agent for 2, 5, 10 and 20 agents involved, with a varying number of changes in each document revision.}
  \label{fig:average_merge_time}
\end{figure}

\begin{figure}[tb]
  \begin{center}
    \subfloat{
      \tikzsetnextfilename{brtl_a_legend}
      \begin{tikzpicture}[scale=0.6]
        \begin{axis}[
            width=0.2\textwidth, height=0.2\textwidth,
            hide axis,
            xmin=0, xmax=10, ymin=0, ymax=0.2,
            legend columns=6,
            hide axis,
            xmin=0, xmax=10, ymin=0, ymax=0.2,
            no markers,
            cycle list/Set1,
          ]
          \addlegendimage{empty legend}
          \addlegendentry{\hspace{-1.5cm}\makebox[0pt][l]{\textbf{Number of Changes per Revision}}}
          \pgfplotsinvokeforeach{1,2,3,4}{\addlegendimage{empty legend}\addlegendentry{}}
          \addlegendimage{empty legend}\addlegendentry{\makebox[0.5cm][l]{}}
          \pgfplotsinvokeforeach{1,2,3,4,5}{\addplot coordinates {(0,-1)};}
          \addlegendentry{10}
          \addlegendentry{20}
          \addlegendentry{30}
          \addlegendentry{40}
          \addlegendentry{50}
        \end{axis}
      \end{tikzpicture}
    }

    \subfloat[For 2 documents]{
      \tikzsetnextfilename{brtl_a2} 
      \begin{tikzpicture}[scale=0.6]
        \begin{axis}[
            width=0.7\linewidth, 
            grid=major, 
            grid style={dashed,gray!30}, 
            xlabel=Number of agents, 
            ylabel=Average merge time,
            axis y line=left,
            axis x line=bottom,
            y unit=\si{\s},
            xmin=1,
            xmax=20,
            ymin=0,
            no markers,
            cycle list/Set1,
            legend pos=north west,
            y filter/.code={\pgfmathparse{#1/(1000)}\pgfmathresult}
          ]
          \addplot table[x=agents,y=average_time,col sep=comma] {results/benchmark_many_documents_many_agents/agents_to_merge_time/2_10.csv}; 
          \addplot table[x=agents,y=average_time,col sep=comma] {results/benchmark_many_documents_many_agents/agents_to_merge_time/2_20.csv}; 
          \addplot table[x=agents,y=average_time,col sep=comma] {results/benchmark_many_documents_many_agents/agents_to_merge_time/2_30.csv}; 
          \addplot table[x=agents,y=average_time,col sep=comma] {results/benchmark_many_documents_many_agents/agents_to_merge_time/2_40.csv}; 
          \addplot table[x=agents,y=average_time,col sep=comma] {results/benchmark_many_documents_many_agents/agents_to_merge_time/2_50.csv}; 
        \end{axis}
      \end{tikzpicture}
      }
    \subfloat[For 5 documents]{
      \tikzsetnextfilename{brtl_a5} 
      \begin{tikzpicture}[scale=0.6]
        \begin{axis}[
            width=0.7\linewidth, 
            grid=major, 
            grid style={dashed,gray!30}, 
            xlabel=Number of agents, 
            ylabel=Average merge time,
            axis y line=left,
            axis x line=bottom,
            y unit=\si{\s},
            xmin=1,
            xmax=20,
            ymin=0,
            no markers,
            cycle list/Set1,
            legend pos=north west,
            y filter/.code={\pgfmathparse{#1/(1000)}\pgfmathresult}
          ]
          \addplot table[x=agents,y=average_time,col sep=comma] {results/benchmark_many_documents_many_agents/agents_to_merge_time/5_10.csv}; 
          \addplot table[x=agents,y=average_time,col sep=comma] {results/benchmark_many_documents_many_agents/agents_to_merge_time/5_20.csv}; 
          \addplot table[x=agents,y=average_time,col sep=comma] {results/benchmark_many_documents_many_agents/agents_to_merge_time/5_30.csv}; 
          \addplot table[x=agents,y=average_time,col sep=comma] {results/benchmark_many_documents_many_agents/agents_to_merge_time/5_40.csv}; 
          \addplot table[x=agents,y=average_time,col sep=comma] {results/benchmark_many_documents_many_agents/agents_to_merge_time/5_50.csv}; 
        \end{axis}
      \end{tikzpicture}
      }
      
    \subfloat[For 10 documents]{
      \tikzsetnextfilename{brtl_a9} 
      \begin{tikzpicture}[scale=0.6]
        \begin{axis}[
            width=0.7\linewidth, 
            grid=major, 
            grid style={dashed,gray!30}, 
            xlabel=Number of agents, 
            ylabel=Average merge time,
            axis y line=left,
            axis x line=bottom,
            y unit=\si{\s},
            xmin=1,
            xmax=20,
            ymin=0,
            no markers,
            cycle list/Set1,
            legend pos=north west,
            y filter/.code={\pgfmathparse{#1/(1000)}\pgfmathresult}
          ]
          \addplot table[x=agents,y=average_time,col sep=comma] {results/benchmark_many_documents_many_agents/agents_to_merge_time/10_10.csv}; 
          \addplot table[x=agents,y=average_time,col sep=comma] {results/benchmark_many_documents_many_agents/agents_to_merge_time/10_20.csv}; 
          \addplot table[x=agents,y=average_time,col sep=comma] {results/benchmark_many_documents_many_agents/agents_to_merge_time/10_30.csv}; 
          \addplot table[x=agents,y=average_time,col sep=comma] {results/benchmark_many_documents_many_agents/agents_to_merge_time/10_40.csv}; 
          \addplot table[x=agents,y=average_time,col sep=comma] {results/benchmark_many_documents_many_agents/agents_to_merge_time/10_50.csv}; 
        \end{axis}
      \end{tikzpicture}
      }
    \subfloat[For 20 documents]{
      \tikzsetnextfilename{brtl_a20} 
      \begin{tikzpicture}[scale=0.6]
        \begin{axis}[
            width=0.7\linewidth, 
            grid=major, 
            grid style={dashed,gray!30}, 
            xlabel=Number of agents, 
            ylabel=Average merge time,
            axis y line=left,
            axis x line=bottom,
            y unit=\si{\s},
            xmin=1,
            xmax=20,
            ymin=0,
            no markers,
            cycle list/Set1,
            legend pos=north west,
            y filter/.code={\pgfmathparse{#1/(1000)}\pgfmathresult}
          ]
          \addplot table[x=agents,y=average_time,col sep=comma] {results/benchmark_many_documents_many_agents/agents_to_merge_time/20_10.csv}; 
          \addplot table[x=agents,y=average_time,col sep=comma] {results/benchmark_many_documents_many_agents/agents_to_merge_time/20_20.csv}; 
          \addplot table[x=agents,y=average_time,col sep=comma] {results/benchmark_many_documents_many_agents/agents_to_merge_time/20_30.csv}; 
          \addplot table[x=agents,y=average_time,col sep=comma] {results/benchmark_many_documents_many_agents/agents_to_merge_time/20_40.csv}; 
          \addplot table[x=agents,y=average_time,col sep=comma] {results/benchmark_many_documents_many_agents/agents_to_merge_time/20_50.csv}; 
        \end{axis}
      \end{tikzpicture}
      }
  \end{center}
  \caption{Results of the experimental evaluation with focus on the scalability aspects of the HKFN Framework. Plots depict the average merge time as a function of a number of agents involved for 2, 5, 10 and 20 documents, with a varying number of changes in each document revision.}
  \label{fig:average_merge_time_agents}
\end{figure}
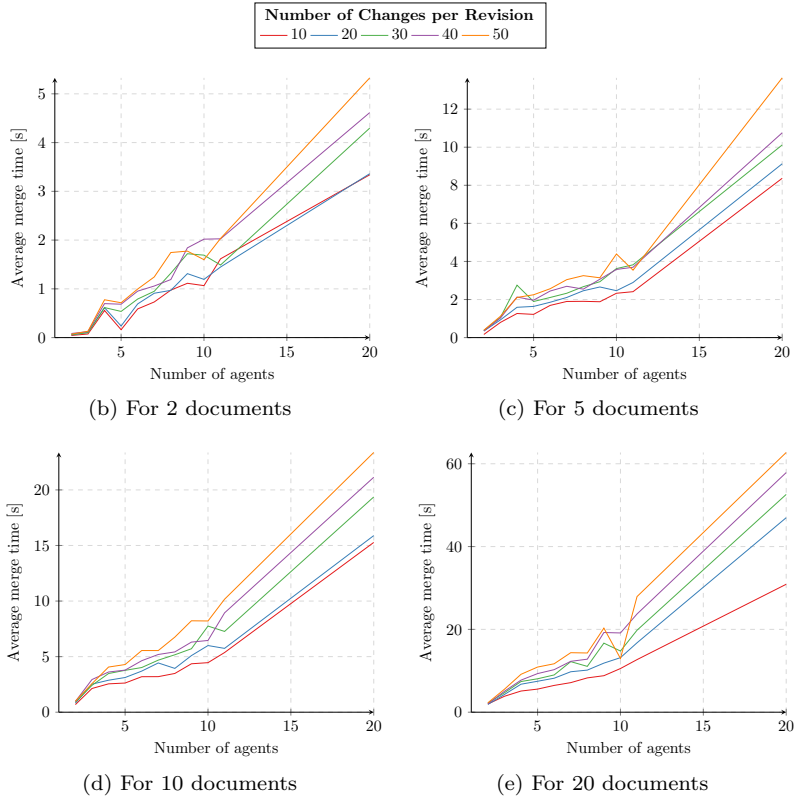

Using the procedure described above the influence of the following parameters was evaluated: 
the number of agents $N=\{2,5,10,20\}$, each making $C=\{10,20,30,40,50\}$ changes resulting in $N$ revisions to each of $M=\{1,\ldots,20\}$ RDF\OP Documents. 
The procedure loop described above is performed 30 times, on a computer equipped with a six core \texttt{Intel(R) Xeon(R) CPU E5-1620 v2 @ 3.70GHz}. 
The influence of the communication links' properties is excluded in this experiment in the name of reproducability between single merges in the loop. Average time for 30 iterations is recorded and reported in \Cref{fig:average_merge_time,fig:average_merge_time_agents}. 

The average time to perform a merge for 2, 5, 10, and 20 agents is presented in~\Cref{fig:average_merge_time}. For each case the plots show the time as a function of the number of documents and changes per revision. 
For example, for $N=2, M=20, C=40$, during the merge period, local changes can not occur faster than every two seconds, as the merge master would not be able to keep up. 
\Cref{fig:average_merge_time_agents} presents the same results but allows to evaluate how the system scales with the number of agents. 

We believe that the performance shown is sufficient in typical use cases as agents usually generate smaller changes at much lower rates. The merge is sporadic in a typical mission because there is a longer period of time between creating new revisions.
For example, in the case study presented in \Cref{ssec:evaluating_sync_protocol} involving 12 agents, the merge master only performed 13 merges.
Even in the biggest case, where $N=20,C=50,M=20$ the merge procedure took only $60$ seconds.



\subsubsection{Discussion of Results}

To provide additional examples and put the results of the experiment presented above in an extended and practical context, we discuss three possible use cases.

\paragraph{Use case 1: dataset discovery}

If we consider our dataset system (\Cref{sec:dataset}), a typical scan mission takes between 5 to 20 minutes, and agents would create two revisions of the RDF Graph describing the metadata of the dataset. Assuming the mission involves $N$ agents, $2 N$ revisions would be required. 
First revision would be created at the start of the scanning mission and the second one at the end, once the dataset is complete.
In this scenario, agents work with a single RDF Graph.

The results in \Cref{fig:average_merge_time} show that for 20 agents, for a single document, if all agents make a change simultaneously, the merge takes $2.9s$, meaning that agents can make up to $0.3$ changes per second on average.
In this scenario, each agent makes at most 2 new revisions per 5 minutes, which is well below what the system can handle.
Extrapolating these results using the fact that the merge and rebase algorithms scale quadratically in the worst-case, the HFKN framework should be able to handle missions of this type with up to 143 participating agents.


\paragraph{Use case 2: victim detection}

In this scenario, we consider a fleet of UAVs exploring an operational environment and searching for victims of a natural or human-made disaster.
Each UAV adds the location and selected information about the identified victims in a single document.
Each victim is represented by two triples, one for the victim's location and one for the state (injured, unconscious, hungry, etc.).

Let's consider the mission is performed in one of the most dense cities in the world is Manila\footnote{Source: \url{https://en.wikipedia.org/wiki/List_of_cities_proper_by_population_density}} with $41515$ person per square kilometer (i.e. $0.04$ person per square meter). 
One can assume that the UAVs are flying at an altitude of $13.7m$ and are equipped with cameras using a wide-angle lens with $60\deg$ field of view. Each image acquired by a single UAV covers an area of approximately $15.8\times15.8\approx 250^2$. Assuming each UAV flies at $15.8m/s$ speed, it can scan $250m^2/s$. Ideally, this would result in detecting $250\times 0.04 = 10 people/s$ in the city of Manila. Consequently, each UAV would generate $2\times10 = 20\: RDF\; Triples/s$. Instead of creating a new revision per one detected victim, we buffer the information until we detect 5 victims which results in creating $10/5=2$ revisions per second.

According to the results presented in \Cref{fig:average_merge_time}, this type of mission would be feasible when using up to 20 agents working on the same RDF Graph.
During a rescue operation, it is critical to conclude the exploration as quickly as possible.
Given example of rescue mission in Manila performed using a small fleet of 20 agents, would result in rather long exploration time of $2h20m$ (Manila has a surface of $42.88 km$, and each agent covers $250m^2/s$).
Since 20 agents is the limit, in this scenario, for collaboration on a single document, to use more agents a divide and conquer approach is used, and several documents covering smaller areas would be created.



\paragraph{Use case 3: exploration with a fleet of UAVs}

In this scenario, we consider $N$ UAVs scanning an area with color and thermal cameras, both delivering images at a rate of $10Hz$. Additionally, locations of the platforms are recorded at $30Hz$.
The UAVs collaborate on three documents: one for the color camera, one for the thermal camera, and one for the location data ($M=3$).

Each image is represented by three RDF Triples: one to specify which agent acquired the image, one to specify the acquisition timestamp, and one to point to the location of image data. Since images are delivered at a rate of $10Hz$, and there are three triples used, agents will create $30$ changes per second for each of the two camera sensors. Therefore the total number of changes to the documents about the camera data is 60 per second.

The location information is represented by three RDF Triples: one to specify whose agent location it relates to, the timestamp, and the actual position value.
In that case, agents will create 90 changes per second to represent the location information.
The total number of changes the agents will create per second is $60+90$, that is 150 changes per second in the three documents ($C=150$).

The agents collaborate on three documents, and they report images as well as locations twice per second which results in revisions with 30 and 45 changes, respectively.

According to \Cref{fig:average_merge_time}, this would work up to 10 agents. With 20 agents, it would take more than two seconds to merge the changes created during one second. As before this limitation can be overcome using a divide and conquer approach.

\subsection{Summary}

In this section, we have presented an empirical evaluation of the RDF Graph Synchronization mechanism.
We have confirmed that merge and rebase algorithms have a quadratic complexity (see \Cref{ssec:evaluting_merge_algorithm} and \Cref{ssec:evaluting_rebase_algorithm}).
We have also demonstrated the behavior of the system when agents get disconnected from each other (in \Cref{ssec:evaluating_sync_protocol}).
The experiment shows that when disconnections happen, agents can still be synchronized with a sub-group of agents and that when agents reconnect, they can handle the synchronization.
Finally, we have demonstrated in  \Cref{ssec:evaluating_max_sync_rate} that the HFKN Framework scales well when the number of agents and documents increases.

The results confirm that the RDF Graph Synchronization mechanism is suitable for sharing low bandwidth semantic knowledge among agents.
The results of \Cref{ssec:evaluating_max_sync_rate} also indicate that the framework could be used with higher bandwidth data, such as storing the results of victim detection.
Future work will include evaluations of different applications.

All the experiments presented in this section are synthetic and show the theoretical performance.
In the following section, we present a field experiment of the HFKN Framework deployed onboard actual robots.

\section{Field Robotics Case Study and Experiment}\label{sec:case_study}

The proposed HFKN Framework has been implemented in prototype and validated through a series of experiments both in simulation (\Cref{sec:validate_synch}) and in real mission scenarios using multiple UAV platforms. The latter is the topic of this section. We first provide a short description of the UAV platforms used in the field study and then proceed to a description of  
a multi-agent data collection mission that uses these UAV platforms. We then conclude with a description of the rich set of query mechanisms that are part of the HFKN Framework and have also been used in the field study.

\subsection{UAV Platforms used in the Field Test}

\begin{figure}[tb]
\centering
\includegraphics[width=1.0\linewidth]{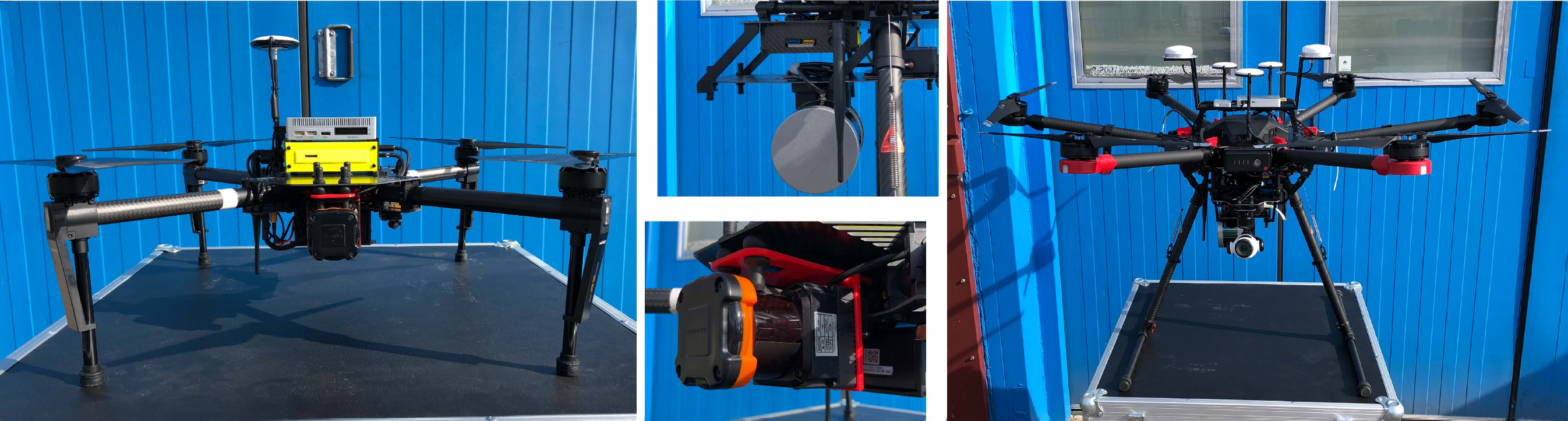}
\caption{Experimental platforms: DJI Matrice 100 equipped with a Hokuyo UTM-30LX LIDAR sensor (left), DJI Matrice 600 Pro equipped with a Velodyne Puck LIDAR sensor (right).}
\label{fig_platforms}
\end{figure}
Two types of UAV platforms were used for experimentation in the case study. The first, shown on the left of Figure~\ref{fig_platforms}, is a modified DJI Matrice 100. 
It has a maximum takeoff weight of 3.6kg and 1.2kg of payload capacity. The platform measures 100cm between propeller tips. 
It can fly with speeds up to 22m/s and has a maximum flight endurance of 22 minutes. 
The platform is equipped with a Hokuyo UTM-30LX LIDAR, which is a single scan device with a guaranteed range of 30m (60m maximum).  

The second type of platform, shown on the right of Figure~\ref{fig_platforms}, is a modified DJI Matrice 600 Pro.
It has a 15.1kg maximum takeoff weight, 6kg of payload capacity, maximum flight speed of 18m/s, and 35 minutes of flight time using 5.5kg of payload.
It measures 167cm between propeller tips.
The GPS system on-board uses a Real-Time Kinematic (RTK) positioning technique to deliver centimeter accuracy measurements.
The platform is equipped with a Velodyne PUCK LIDAR sensor, which has an effective range of 100m and uses 16 scan channels.
A LIDAR mounting mechanism developed and deployed on the DJI Matrice 600 Pro allows for choosing the sensor orientation depending on the applications or missions at hand.

Both platforms are equipped with the same type of onboard computer system.
It is an \texttt{Intel NUC Kaby Lake i7-7567U CPU} platform in a custom enclosure equipped with 16GB of RAM and 500GB SSD of storage.
The systems interface with the platforms and run the software modules associated with the \textit{Delegation Module} and the \textit{SCModule}. 
The communication with the ground station for both platforms is realised using 5GHz WiFi connections.

\subsection{Field-Robotics Case Study: Collaborative 3D Modeling}\label{ssec:collab_exploration}

In this field experiment, we want to demonstrate how multiple agents can explore different parts of an operational environment, share the resulting collected information, and reuse existing information from other missions. The mission leverages functionalities from the HFKN Framework and the Delegation Framework and each robotic agent has a  Delegation Module and a SCModule. For human agents, it is assumed that they can interface to the collaborative system through Command and Control centers, or if in the field, through laptops, smartphones, or tablet systems with access to both the Delegation Module and SCModule.

We have divided the experiment into four scan missions\footnote{A \emph{Scan mission} is a generic type of mission in which agents collect data with their sensors, such as camera or LIDAR, over an area.} shown in \Cref{fig:scan_missions} with different types of team interaction and use of collected distributed information:


\begin{figure}
  \centering
  \def\svgwidth{0.6\columnwidth}
\resizebox{0.5\linewidth}{!}{
\begingroup%
  \makeatletter%
  \providecommand\color[2][]{%
    \errmessage{(Inkscape) Color is used for the text in Inkscape, but the package 'color.sty' is not loaded}%
    \renewcommand\color[2][]{}%
  }%
  \providecommand\transparent[1]{%
    \errmessage{(Inkscape) Transparency is used (non-zero) for the text in Inkscape, but the package 'transparent.sty' is not loaded}%
    \renewcommand\transparent[1]{}%
  }%
  \providecommand\rotatebox[2]{#2}%
  \newcommand*\fsize{\dimexpr\f@size pt\relax}%
  \newcommand*\lineheight[1]{\fontsize{\fsize}{#1\fsize}\selectfont}%
  \ifx\svgwidth\undefined%
    \setlength{\unitlength}{339.74999135bp}%
    \ifx\svgscale\undefined%
      \relax%
    \else%
      \setlength{\unitlength}{\unitlength * \real{\svgscale}}%
    \fi%
  \else%
    \setlength{\unitlength}{\svgwidth}%
  \fi%
  \global\let\svgwidth\undefined%
  \global\let\svgscale\undefined%
  \makeatother%
  \begin{picture}(1,0.67991173)%
    \lineheight{1}%
    \setlength\tabcolsep{0pt}%
    \put(0,0){\includegraphics[width=\unitlength,page=1]{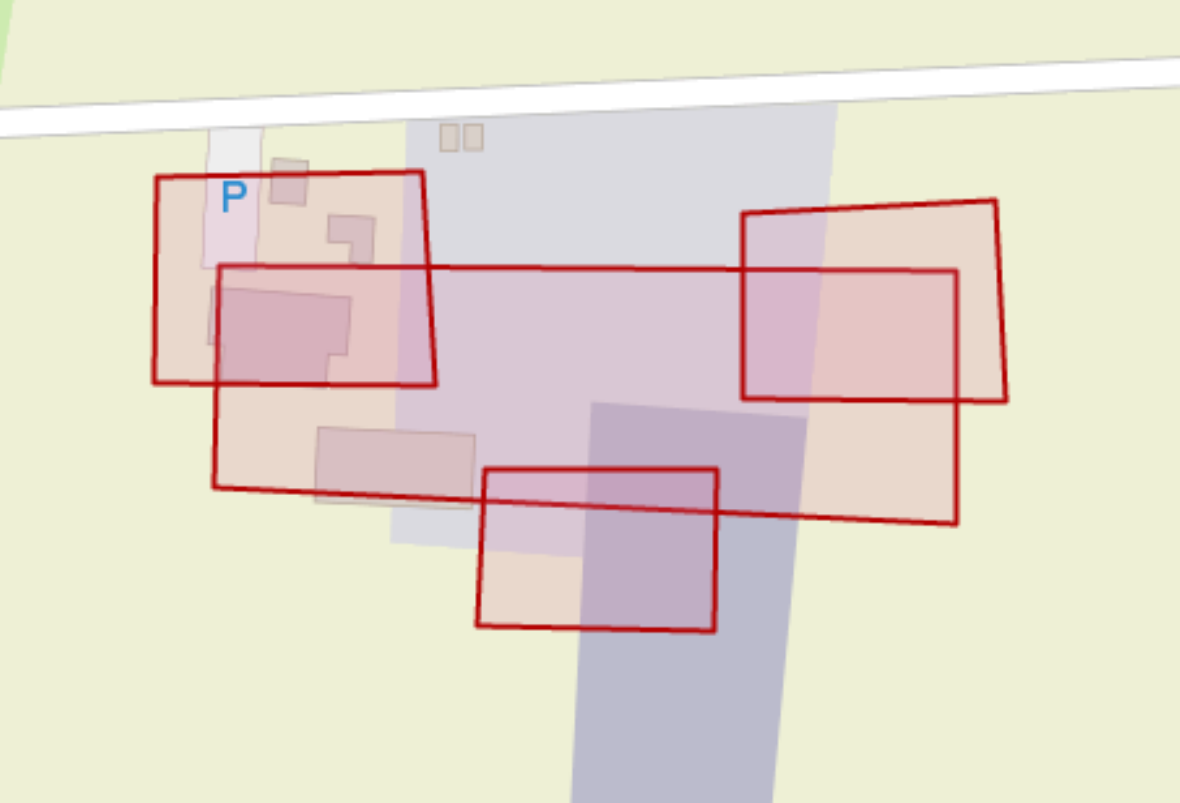}}%
    \put(0.26821176,0.48502053){\color[rgb]{0,0,0}\makebox(0,0)[lt]{\lineheight{1.25}\smash{\begin{tabular}[t]{l}$\mathbf{C}$\end{tabular}}}}%
    \put(0.76017021,0.4645223){\color[rgb]{0,0,0}\makebox(0,0)[lt]{\lineheight{1.25}\smash{\begin{tabular}[t]{l}$\mathbf{A}$\end{tabular}}}}%
    \put(0.50315341,0.19331442){\color[rgb]{0,0,0}\makebox(0,0)[lt]{\lineheight{1.25}\smash{\begin{tabular}[t]{l}$\mathbf{B}$\end{tabular}}}}%
    \put(0.48896227,0.35572377){\color[rgb]{0,0,0}\makebox(0,0)[lt]{\lineheight{1.25}\smash{\begin{tabular}[t]{l}$\mathbf{D}$\end{tabular}}}}%
  \end{picture}%
\endgroup%
}

  \caption{Division into four scan missions $\mathbf{A}$, $\mathbf{B}$, $\mathbf{C}$ and $\mathbf{D}$}
  \label{fig:scan_missions}
\end{figure}

\begin{itemize}
 \item For the first three Scan missions ($\mathbf{A}$, $\mathbf{B}$ and $\mathbf{C}$), three different UAVs ($UAV_0$, $UAV_1$, $UAV_2$) are tasked by a ground operator $OP$  to explore non-overlapping areas of the operational environment of interest. This may be done sequentially or concurrently since the regions are non-overlapping.  Missions are set up and executed autonomously through the use of the Delegation Framework. 
 \item At the end of mission $\mathbf{C}$, the ground operator $OP$ requests the transfer and download of the resulting scanned data from $UAV_2$. This is done using the Dataset Transfer Protocol.
 \item After viewing a 3D model generated for region $\mathbf{C}$, the ground operator determines that more information is required for a larger region that overlaps with the three previously specified regions, $\mathbf{A}$, $\mathbf{B}$ and $\mathbf{C}$.
 \item The ground operator $OP$ sets up the fourth mission using the Delegation Framework to gather a full scan of region $\mathbf{D}$ by $UAV_0$. Since raw data for regions $\mathbf{A}$, $\mathbf{B}$ and $\mathbf{C}$ already exists, the fourth mission by $UAV_0$ is automatically tasked to gather data from the region $\mathcal{D'}_D=\mathcal{D}_D\setminus (\mathcal{D}_A \cup \mathcal{D}_B \cup\mathcal{D}_C) $. 
 
\end{itemize}

For each Scan mission attempted, the following steps are applied by an agent:

\begin{itemize}
  \item First, check if any relevant scanned data is already available in the agent's KDB. This is done by executing a SPARQL query on the metadata for the dataset, checking for any intersections between the raw data areas previously covered by the agent and the target dataset.
  \item If no such data (or only partial data) is available in the requesting agent's KDB, an attempt is made to set up a task to copy any relevant data for the geographic region from any remote agents on the team. The copy is done through a combination of queries and use of the Dataset Transfer Protocol described previously.
  \item After doing this, a check is made to see if data for the geographic region of interest is complete with full coverage. If not, the agent can query its KDB to compute those remaining regions in the geographic area not yet covered.  The agent can now specify a Scan mission to gather the remaining data.
  \item Before executing the Scan mission, the agent creates a dataset for the geographic area of the mission. During the execution, the agent collects data and associates it with the dataset.
  \item At the end of the mission, the agent would have the necessary raw data in its KDB for the geographic region in question. Additionally, due to the periodic synchronization of knowledge in each of the team agent's KDB, they would all have a new dataset with information about where the raw data resides in other agents.
\end{itemize}

Before mission $\mathbf{A}$, $\mathbf{B}$ and $\mathbf{C}$, no data for these regions is stored, either by the individual agents or the team as a whole, so there is no exchange of raw data.
Each UAV collects the requisite data associated with its respective mission during the execution of each mission and stores it in its local KDB. We will refer to the resulting datasets as $\mathcal{D}_A$, $\mathcal{D}_B$ and $\mathcal{D}_C$. No raw data is exchanged between $UAV_0$, $UAV_1$, $UAV_2$ at the end of their missions.  Metadata for $\mathcal{D}_A$, $\mathcal{D}_B$, and $\mathcal{D}_C$ has been synchronized in the KDBs of all team members.  

At the end of mission $\mathbf{C}$, by request of operator $OP$, raw data for the dataset $\mathcal{D}_C$ is copied to the operator $OP$'s ground station using the data exchange protocol.
The association between available datasets before the start of mission $\mathbf{D}$ is shown in \Cref{table_dataset_association_m123}

For the mission $\mathbf{D}$, the $OP$ operator queries its datasets' metadata and finds that datasets $\mathcal{D}_A$, $\mathcal{D}_B$ and $\mathcal{D}_C$ overlap with the operator's area of interest.
Since $OP$ already has a copy of $\mathcal{D}_C$, it only needs to download raw data from $\mathcal{D}_A$ and $\mathcal{D}_B$ and set up a mission for the remaining region $\mathcal{D}_D\setminus (\mathcal{D}_A \cup \mathcal{D}_B \cup\mathcal{D}_C) $.

\begin{figure}[tb]
\centering
\includegraphics[width=1.0\linewidth]{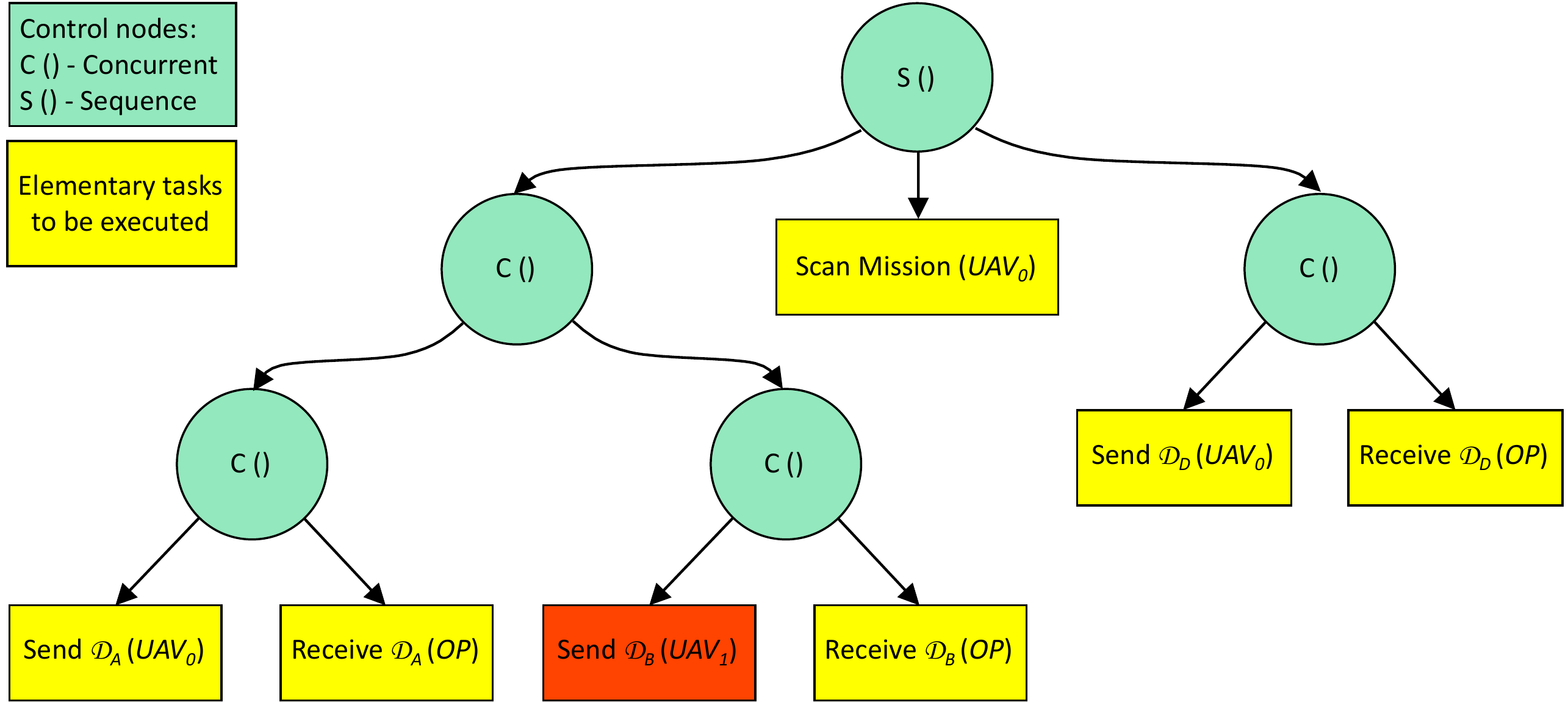}
\caption{TST for mission $\textbf{D}$. Yellow indicates leaf nodes, red indicates the transfer node that failed.}
\label{fig_tst_mission_4}
\end{figure}

The TST generated for mission $\mathbf{D}$ is shown in \Cref{fig_tst_mission_4}. First, the ground operator's station should download concurrently datasets $\mathcal{D}_A$ and $\mathcal{D}_B$. Then, it delegates the final UAV Scan mission to collect the remaining scan data to $UAV_0$. This TST is the actual TST generated by the delegation framework and intended to be executed by the ground station agent and $UAV_0$ autonomously (The \textsf{S} and \textsf{C} in the control nodes of the TST refer to Sequential nodes and Concurrent nodes, respectively).

To demonstrate our system's robustness, we simulated a communication failure between agent $UAV_1$ and $ OP $, so that $\mathcal{D}_B$ was not transferred to $ OP $ during mission execution.
$OP$ is made aware of this. It would either need to attempt a new data exchange mission or start a new exploration mission to make sure that $\mathcal{D}_D$ covers the requested area.


\begin{table}
  \centering
  \subfloat[Dataset content after mission's $\mathbf{A}$, $\mathbf{B}$ and $\mathbf{C}$ are completed.]{
    \begin{tabular}{|c|c|c|c|c|}
      \hline
       & $\mathcal{D}_A$ & $\mathcal{D}_B$ & $\mathcal{D}_C$ & $\mathcal{D}_D$ \\
      \hline
      $UAV_0$ & X & & &  \\
      \hline
      $UAV_1$ & & X & &  \\
      \hline
      $UAV_2$ & & & X &  \\
      \hline
      $OP$ & & & X &  \\
      \hline
    \end{tabular}
    \label{table_dataset_association_m123}
  }
  \hspace*{1cm}
  \subfloat[Dataset content after mission $\mathbf{D}$ is completed.]{
    \begin{tabular}{|c|c|c|c|c|}
      \hline
       & $\mathcal{D}_A$ & $\mathcal{D}_B$ & $\mathcal{D}_C$ & $\mathcal{D}_D$ \\
      \hline
      $UAV_0$ & X & & & X \\
      \hline
      $UAV_1$ & & X & &  \\
      \hline
      $UAV_2$ & & & X &  \\
      \hline
      $OP$ & X & & X & X \\
      \hline
    \end{tabular}
    \label{table_dataset_association_m4}
  }
  \caption{The association between available datasets after execution of different missions.}
\end{table}

\Cref{table_dataset_association_m4} shows the raw dataset associations to team agents after mission $\mathbf{D}$ is completed. $UAV_0$ has performed two scan missions: for region $\mathbf{A}$ and for the latest request from the operator. That is why it has a copy of the raw data associated with datasets $\mathcal{D}_A$ and $\mathcal{D}_D$. At the end of mission $\mathbf{D}$, all metadata associated with all datasets generated is synchronized, and general queries about the operational region can be made by the agents of the team, where each has a partially shared situation awareness by virtue of each agent's sharing respective metadata about datasets.

\begin{figure}
\centering
\subfloat[Mission A]{
\includegraphics[width=0.48\linewidth]{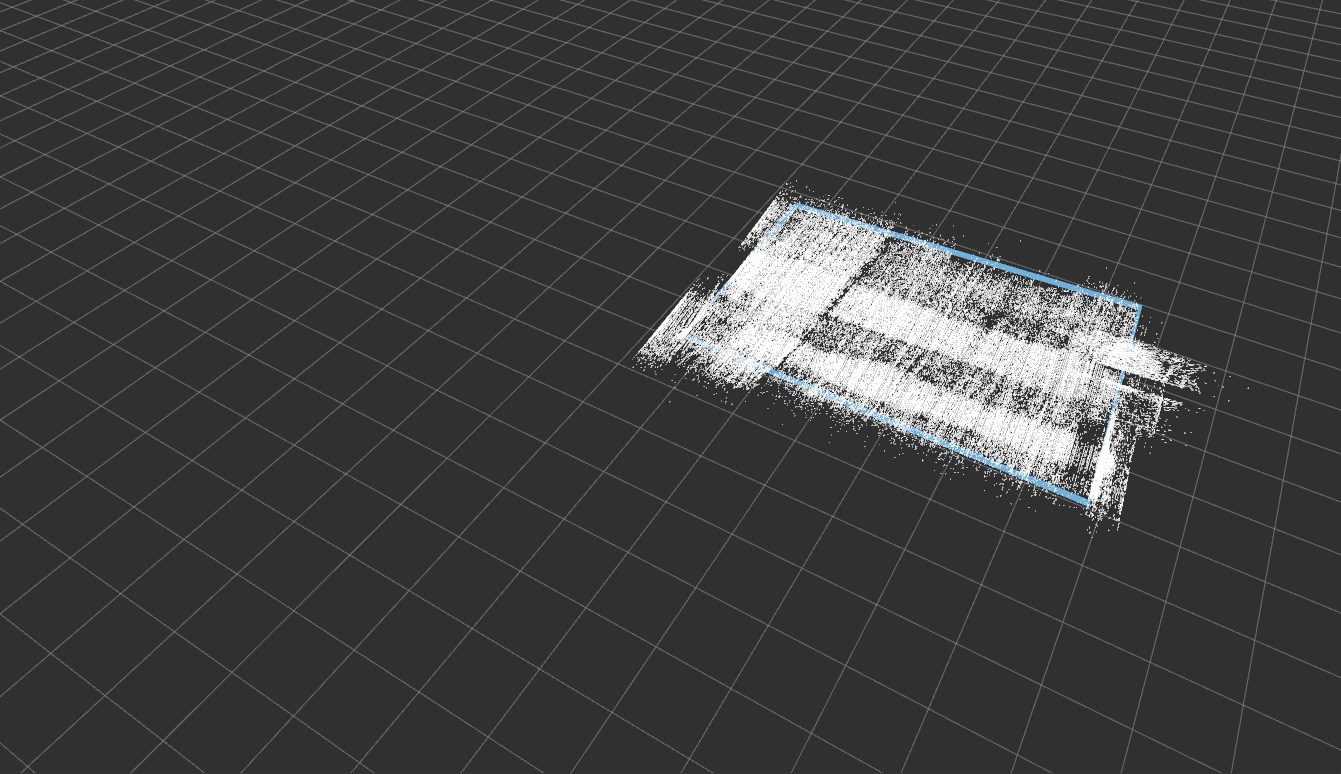}
\label{fig_pc_mission_a}
}
\subfloat[Mission B]{
\includegraphics[width=0.48\linewidth]{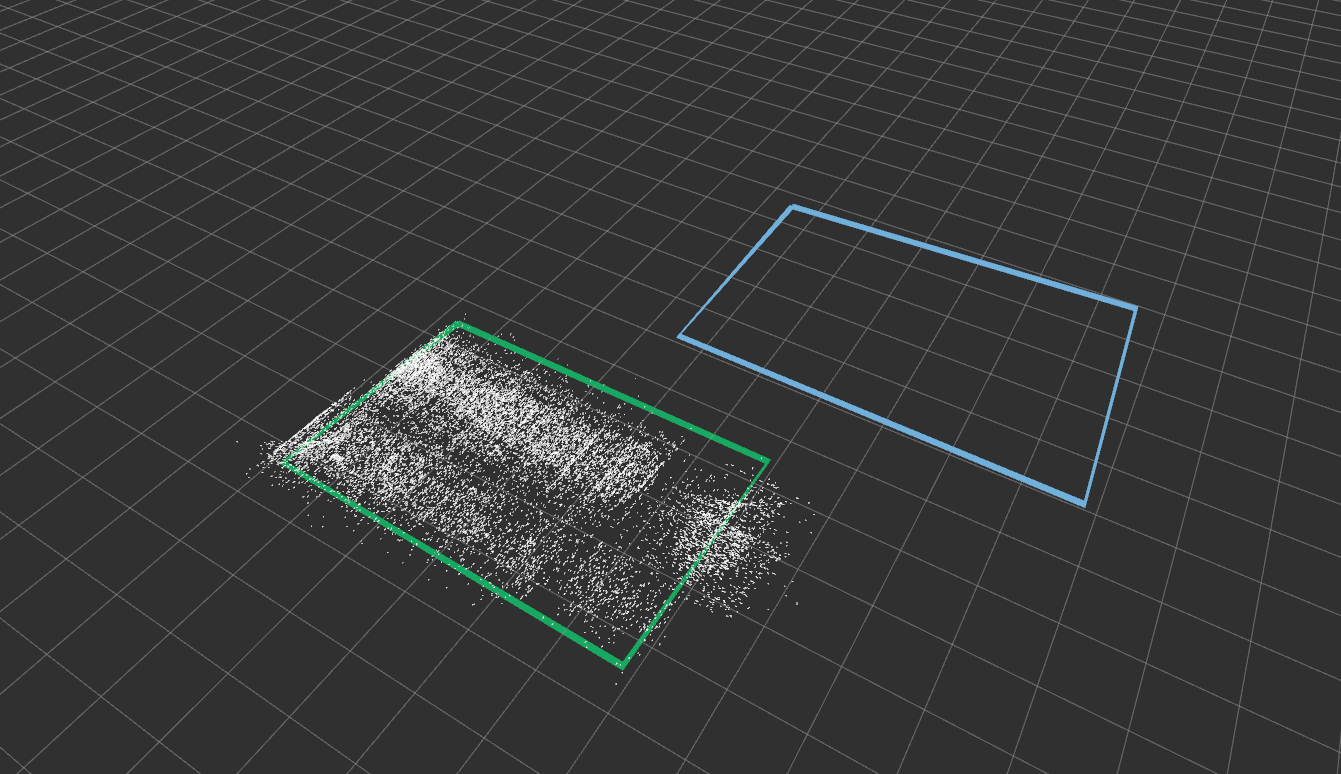}
\label{fig_pc_mission_b}
}

\subfloat[Mission C]{
\includegraphics[width=0.48\linewidth]{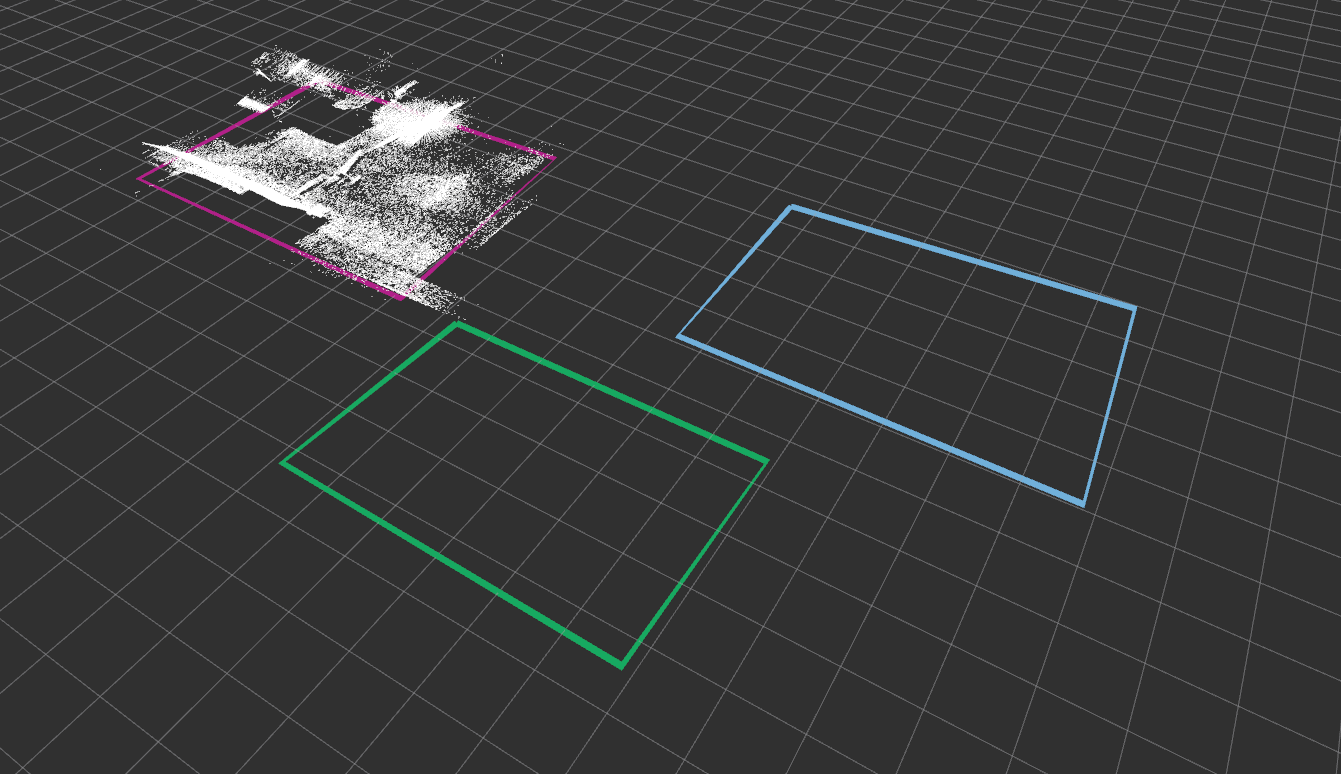}
\label{fig_pc_mission_c}
}
\subfloat[Mission D]{
\includegraphics[width=0.48\linewidth]{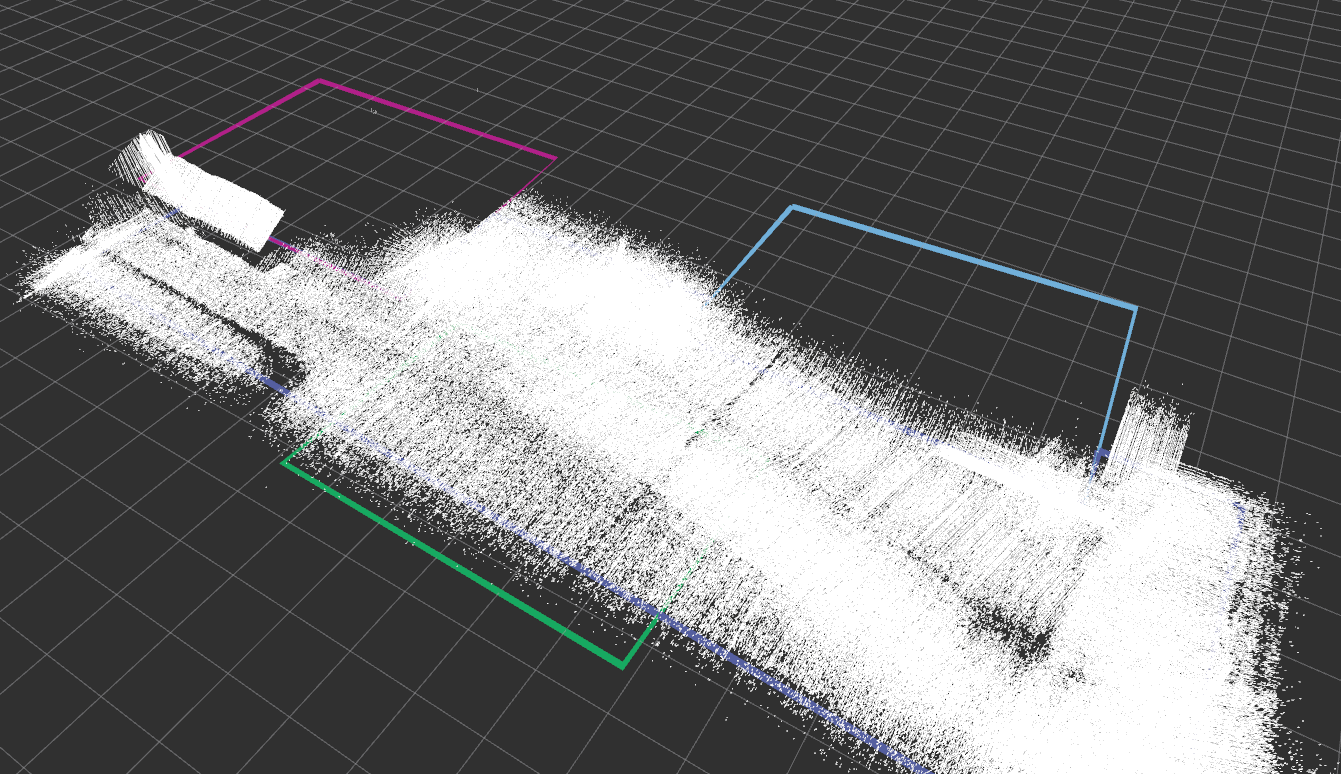}
\label{fig_pc_mission_d}
}

\subfloat[The combined point cloud from all missions]{
\includegraphics[width=0.48\linewidth]{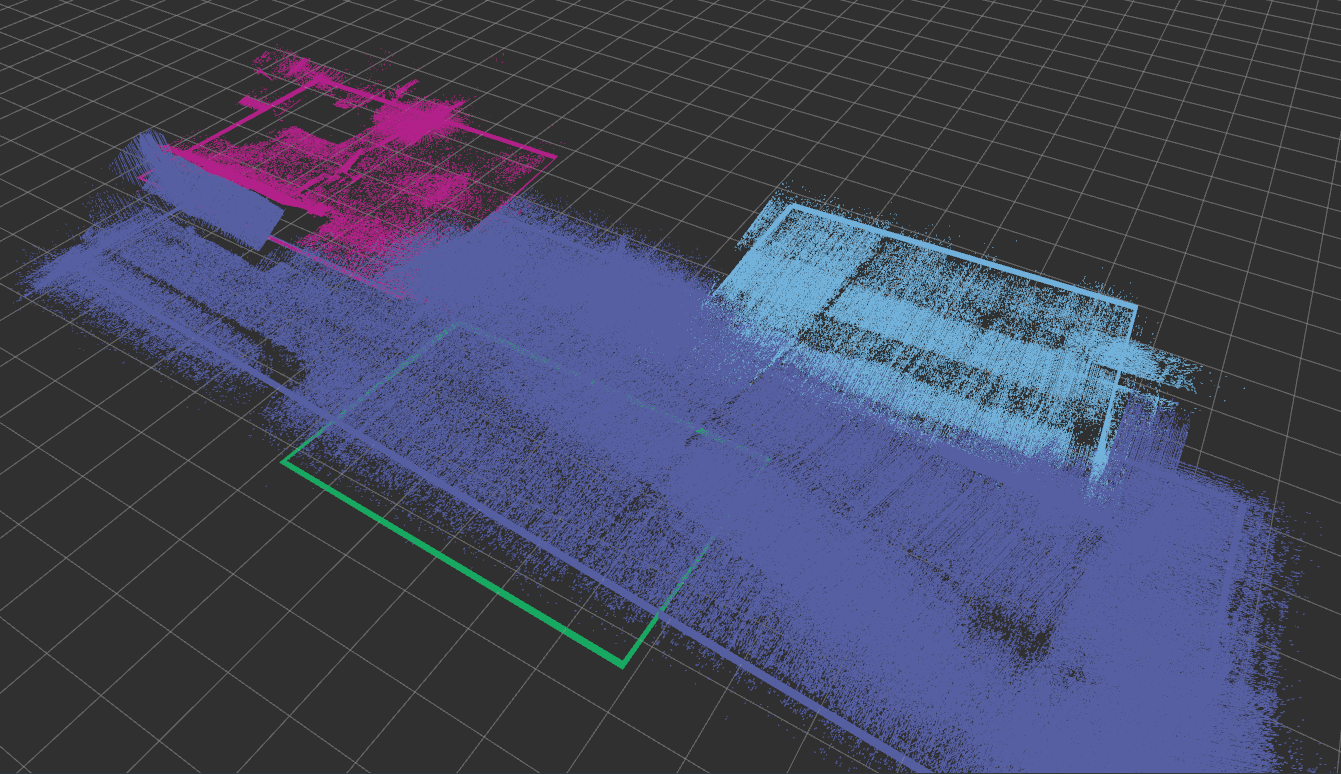}
\hspace{0.01pt}
\includegraphics[width=0.48\linewidth]{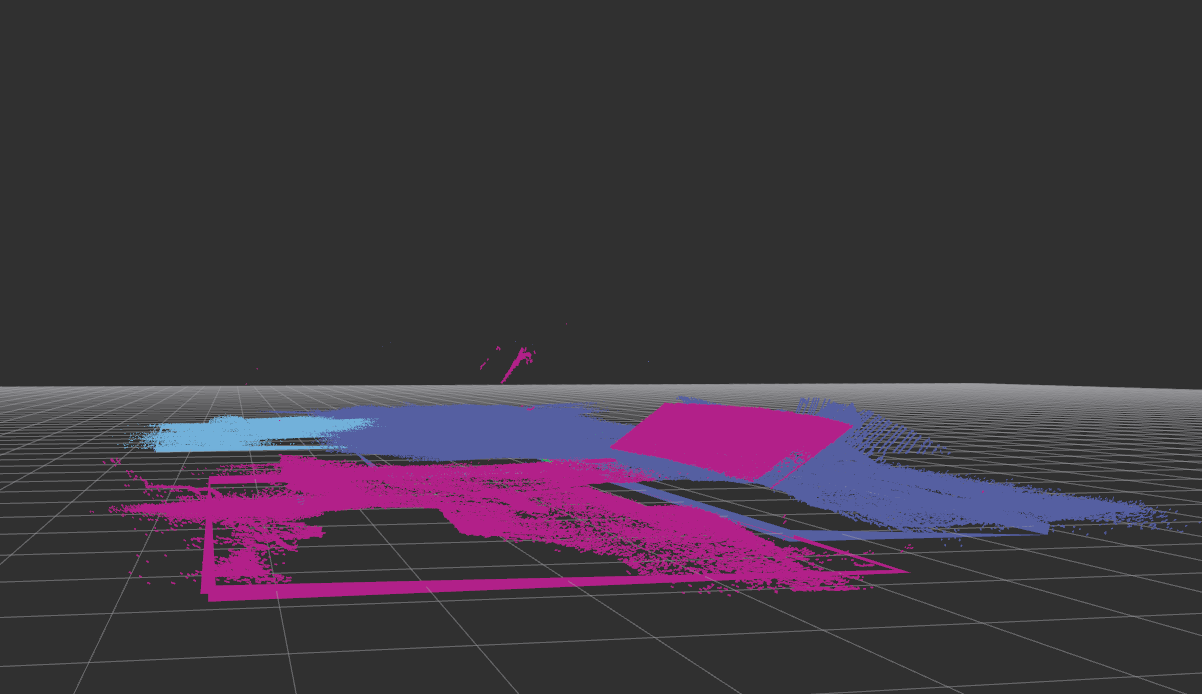}
\label{fig_pc_mission_e}
}
\caption{Resulting point cloud from flying the four missions.
The rectangles represent the four datasets $\mathcal{D}_A$ to $\mathcal{D}_D$.
The missions are run in sequence.
\Cref{fig_pc_mission_a,fig_pc_mission_b,fig_pc_mission_c,fig_pc_mission_d} show the results of the individual scan missions, while \Cref{fig_pc_mission_e} shows the point cloud generated by combining data from $\mathcal{D}_A$,  $\mathcal{D}_C$ and $\mathcal{D'}_D$.
}
\label{fig_pc_mission}
\end{figure}

\Cref{fig_pc_mission} shows the resulting point clouds for each mission as well as the final point cloud accessible to the ground operator $OP$. Here one can observe that point cloud data for region $\mathbf{B}$ is missing. This was due to the communication breakdown.

\subsection{Querying and  Query Capability}\label{sec:querying}

The HFKN Framework includes powerful query capability for human and robotic agents to acquire information about operational environments in different modalities. The resulting answers to queries can be used not only for situation awareness and decision making, but also to specify information gathering missions based on missing information. This was shown in the case study. We now consider query mechanisms~\cite{10.1145/3177850} in more detail and provide some additional examples.

Robotic or human agents, can query one agent individually using SPARQL, or several agents simultaneously using Federated Queries~\cite{fedqueries:2013}. From a mission perspective, a ground operator's queries play a special role where multi-modal interfaces can be used to advantage. In this case, rather than the human agent specifying SPARQL queries in detail, a graphical user interface has been developed that permits human agents to define queries graphically. For instance, given a map of an operational environment, a human operator can draw regions of interest (ROI) graphically, state what kind of information from that area they would like and press a query button. The interface then translates the graphical query into a SPARQL query which is directed to one or more agents on the team. The query reply can then be depicted both graphically and textually on the operator's screen. 

\begin{figure}[tb]
\centering
\includegraphics[width=1.0\linewidth]{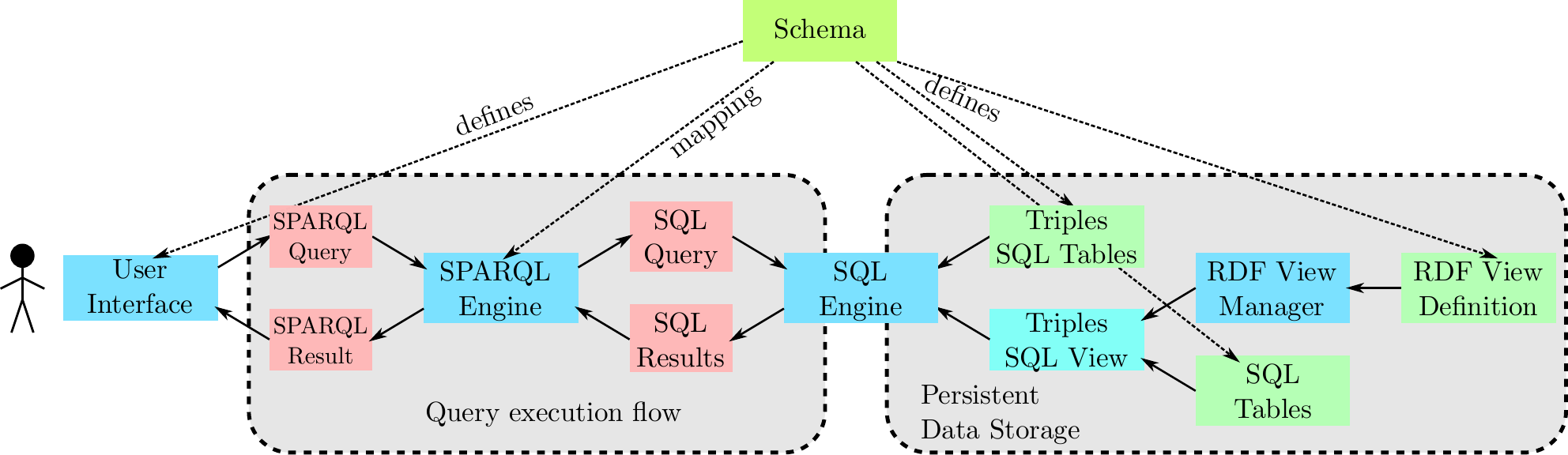}
\caption{Example of a SPARQL Query execution. Software components are shown in blue (i.e. User Interface, SPARQL Engine, SQL Engine and RDF View Manager). The data is shown in green (i.e. Triples SQL Tables, Triples SQL View, SQL Tables and RDF View Definition). The SQL and SPARQL Queries and their results are visualized in red.
The database schema defines the storage structure of RDF\OP Documents, SQL Tables, and the RDF View Definition used to map arbitrary SQL tables to a set of RDF Triples.
The schema also defines the structure of queries used to access the data and the conversion from SPARQL to SQL for the mapping between RDF Graph reference and actual SQL Tables and Views.
}
\label{fig_sparql_query_execution}
\end{figure}

The execution of a SPARQL Query from a ground operator perspective uses the following components (shown in \Cref{fig_sparql_query_execution}):
\begin{itemize}
  \item \textit{User Interface} - A user selects an area of interest and a data type of interest (e.g. list of buildings with their 3D models, potential victim locations with pictures, etc.). Based on that selection the \textit{User Interface} then automatically generates an appropriate \textit{SPARQL Query}.
  \item \textit{SPARQL Engine} - The SPARQL Engine converts the \textit{SPARQL Query} into a corresponding \textit{SQL Query}. After the query is executed it also converts the result back into a \textit{SPARQL Result}.
  \item \textit{SQL Engine} (in this case PostgreSQL) - The SQL Engine is responsible for execution of the SQL Query over the selected RDF Graphs (either stored as SQL tables or accessible through an SQL View).
  \item \textit{RDF View Manager} - The RDF View manager then generates an SQL View based on an RDF View definition, i.e. mapping any SQL Tables to an RDF Graph representation.
\end{itemize}

When a user selects an area of interest in the \textit{User Interface}, together with a data type of interest, a \textit{SPARQL Query} is generated and sent to the \textit{SPARQL Engine} which converts it into a \textit{SQL Query} over a virtual table with three fields (\textit{subject}, \textit{predicate}, \textit{object}).
This virtual table is the union of a set of RDF Graphs (represented as SQL Tables) and a set of RDF Views (mapping between any SQL Tables and an RDF Graph representation).
The RDF Views are generated by the \textit{RDF View Manager} based on an \textit{RDF View Definition} and an existing set of \textit{SQL Tables}.
Once the \textit{SQL Query} has been executed, the results are returned to the \textit{SPARQL Engine} which generates a \textit{SPARQL Result} which can be visualised in the \textit{User Interface}.

Schemas are used by the user interface to define the structure of allowable queries, and during execution by the SPARQL Engine to understand the mapping between RDF Graph URLs and the SQL Table or View, where the RDF data can be queried. For details and examples, see \Cref{sec:Appendix2}.

 In the case study, we showed how to collect information through specification of geographic regions or boundaries. This idea can be leveraged in many different ways. For example, suppose an operator or human rescuer is interested in knowledge about surrounding building structures and would like 3D models of existing building structures. The following query example shows how this can be done using the data from the case study missions.

In the case study, LIDAR sensor data and intermediate point cloud data were collected by several UAV agents and stored distributively in their respective \textit{SCModules}. Additionally, a human ground operator acquired some of this data through SPARQL queries and the use of the data exchange protocol. During this process, raw sensor data was stored in the agent's respective PostgreSQL databases in table form (see \Cref{sec:Appendix1}). Using the RDF View system, this data can be accessed using SPARQL as a high-level query language.

In the following, the prefixes before the URLs  will be used for defining the RDF View and SPARQL query used to retrieve 3D models of building structures in a geographical region:

\begin{itemize}
  \item \textit{askcore\_pointclouds}: \textit{http://askco.re/pointclouds\#}. This prefix is used for definitions related to point clouds. \textit{askcore\_pointclouds:patch} is the predicate that specifies the raw data of the point. \linebreak\textit{askcore\_pointclouds:intersection} is a function that computes the interesection of a point cloud with a geometric object. \textit{askcore\_pointclouds:union} computes the union of a set of point clouds.
  \item \textit{askcore\_types}: \textit{http://askco.re/types\#}. This prefix designates data types stored in the database, such as \textit{askcore\_pointclouds:pointclouds}.
  \item \textit{askcore\_graphs}: \textit{http://askco.re/graphs\#}. This prefix designates the RDF Graphs used in our system. \textit{askcore\_graphs:pointclouds\_view} is an RDF View which enables one to query point clouds stored in  SQL Tables using SPARQL queries.
  \item \textit{geo}: \textit{http://www.opengis.net/ont/geosparql\#}. This is the standard prefix for GeoSPARQL~\cite{battle2012enabling}, which enhances RDF and SPARQL with features specific to GIS.
  \item \textit{geof}: \textit{http://www.opengis.net/def/function/geosparql/}. This is the standard prefix for functions defined by GeoSPARQL.
\end{itemize}

An RDF View mapping the SQL table for point clouds to an RDF Triple representation using Sparqlify~\cite{STADLER-LDOW-2015} is shown in~\Cref{fig:sparql_create_view}. 

\AfterEndEnvironment{snugshade*}{\vspace{-2\FrameSep}}

\begin{listing}[h]
\begin{minted}{SPARQL}
Create View askcore_graphs:pointclouds_view As
  Construct {
    ?pc_uri a askcore_types:point_cloud ;
            askcore_pointclouds:patch ?patch ;
            geo:hasGeometry ?geometry .
  }
  With
    ?pc_uri = uri(%frame_base_uri, str(?id))
  From
    [[SELECT id, patch, ST_GeomFromEWKT(
            ST_AsEWKT(PC_EnvelopeGeometry(patch)))
            AS geometry FROM pointclouds]]
\end{minted}
  \caption{An RDF View mapping the SQL table for point clouds to an RDF Triple representation.}
  \label{fig:sparql_create_view}

\end{listing}

The SPARQL query presented in~\Cref{fig:sparql_get_pointcloud} is then used to retrieve 3D point clouds corresponding to the 2D footprints of buildings in the regions of interest from the case study.

\begin{listing}[h]
\begin{minted}{SPARQL}
SELECT ?building_uri ?building_2d_footprint 
(askcore_pointclouds:intersection(
                askcore_pointclouds:union(?point_cloud_3d),
?building_2d_footprint) AS ?building_3d_model) 
  FROM askcore_graphs:buildings
  FROM NAMED askcore_graphs:pointclouds_view
  WHERE {
    ?building_uri a <building> ;
                  geo:hasGeometry ?building_2d_footprint .
   GRAPH askcore_graphs:pointclouds_view {
    ?point_cloud_uri a askcore_types:point_cloud ;
                 geo:hasGeometry ?point_cloud_2d_footprint ;
                 askcore_pointclouds:patch ?point_cloud_3d .
  }
    FILTER(geof:sfIntersects(?building_2d_footprint,
           ?point_cloud_2d_footprint))
} GROUP BY ?building_uri ?point_cloud_uri %?building_2d_footprint
\end{minted}
  \caption{A SPARQL query used to retrieve 3D point clouds corresponding to 2D footprints of buildings.}
  \label{fig:sparql_get_pointcloud}

\end{listing}

\AfterEndEnvironment{snugshade*}{\vspace{\FrameSep}}

\begin{figure}[tb]
  \centering
  \includegraphics[width=0.98\linewidth]{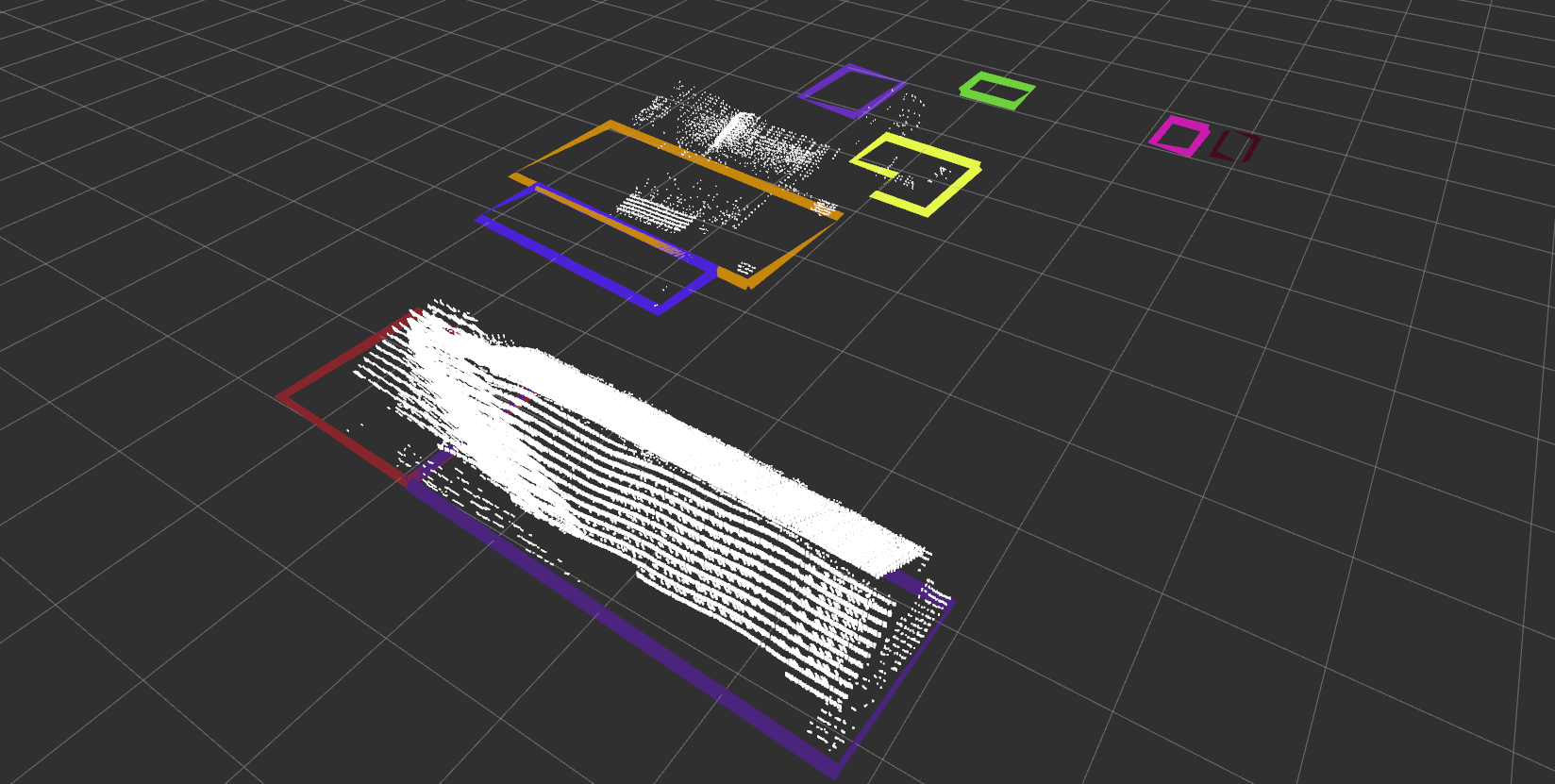}
  \caption{Results of querying the database for a point cloud representation of an existing set of buildings based on the point clouds acquired in \Cref{ssec:collab_exploration}. Note that some requested buildings have no point clouds in the team KDB repositories.}
  \label{fig:query_db_pc_sparql}
\end{figure}

The result of the execution of the SPARQL query is shown in \Cref{fig:query_db_pc_sparql}. For additional details about sensor data representations used and examples of other queries, see \Cref{sec:Appendix1,sec:Appendix2}.

\section{Related work}\label{sec:related_work}

The HFKN Framework envisioned is unique in its approach, but does take  inspiration  from some of the challenges and problems in related topic areas. In this section, we consider some of this work most relevant to the HFKN Framework.

\paragraph{Distributed Knowledge and Multi-Agent Systems}

Multi-Agent Systems (MAS)~\cite{book-MAS,book-MAS-Games} have direct relevance to the HFKN Framework since we view each robotic system as an agent with capabilities to plan and reason. In MAS, the topics of distributed and common knowledge are central to logical approaches to multi-agent problem solving and inference~\cite{book-MAS-Knowledge}. In the HFKN Framework, each agent has its own local knowledge in the form of RDF Documents/Graphs that are distributed across agents. The multi-agent system consisting of a team of agents has common knowledge in the form of the shared, synchronized RDF Documents/Graphs. Collections of RDF triples (RDF Documents) can be viewed logically~\cite{Hayes-RDF,Bruijn:2007, Bruijn:2010}.  This being the case, certain kinds of querying mechanisms, analogous to deductive databases~\cite{Greco:2016}, can be viewed as doing logical inference. SPARQL-DL~\cite{sirin2007sparql} is an interesting example of this.


Communication among agents is also an important topic in MAS. A common approach for communication in MAS is to use an implementation of the Agent Communication Language (ACL) such as FIPA ACL~\cite{POSLAD-TAAS-2007}. ACL contains mechanisms for agents to query other agents and transfer \textit{beliefs} between them. This provides an infrastructural mechanism for belief transfer but also requires a semantics for beliefs such as that used in the BDI approach~\cite{Bratman1987,Rao91}. It should be noted that the Delegation Framework and its integration with the HFKN Framework, does use ACL and speech acts for general communication between agents and that general communication mechanism carries over for use with SCModules.

One of the challenges with such multi-agent systems that is related to belief transfer is for agents to determine what agents to ask for certain information required during a mission. Sending a broadcast is not always practical, since for large numbers of agents, this approach could degrade the available bandwidth of the communication network. A potential solution involves using matchmaker agents~\cite{CHEN-ICIKM-2000} that hold information about which agents can answer questions. Our approach is much different because each agent's public/shared information is synchronized, so the collective knowledge from a team is accessible directly to each agent. Consequently, each agent can then internally query such information.  Raw sensor data is indirectly accessible through metadata and dataset transfer protocols, but an agent can still directly query a dataset's metadata to acquire useful knowledge.  

\paragraph{Distributed and Federated Database Systems}
The main challenge in Federated Database Systems (FDS)~\cite{SHETH-ACMCS-1990,BENT-CITA-2008} is to provide a unified query mechanism that hides data inconsistencies. This is often called the \textit{database integration} problem \cite{PARENT-CACM-1998}. The main challenges in Distributed Database systems~\cite{book-DDBMS,MOIZ-IJCA-2011,CORBETT-ACTCS-2013} is the consistency problem where schema definitions and data should be consistent and equal across the different database instances.

Traditional database technology provides common techniques for storing data and querying data. For databases running on a single computer and using smaller amounts of data, data consistency is less of an issue.
But this does not scale for large numbers of users and large amounts of data. This has led to the development of Distributed Database Management Systems (DDBMS)~\cite{book-DDBMS}.
Homogeneous DDBMS are systems where the schema definitions and data should be consistent and equal across the different database instances.
A common approach to solving this consistency problem is to use database replication~\cite{MOIZ-IJCA-2011}, where a master has  write permissions on a subset of the data, and when changes occur, they are propagated to the slaves. Homogeneous DDBMS also solve the problem of load balancing between servers.
In the case of Big Data, when a single computer cannot store all of its data in memory, it is necessary to use a Heterogeneous DDBMS approach, such as Spanner~\cite{CORBETT-ACTCS-2013} or Dynamo~\cite{decandia2007dynamo}. Here, the system controls which server stores which data depending on user needs and system requirements.
In \cite{CORBETT-ACTCS-2013}, the authors propose a dynamic system to lock tables so that any instance can write on a subset of the table schema. This approach is also capable of handling inaccuracies in the timestamping of transactions. In \cite{decandia2007dynamo} the data is replicated across multiple hosts. Dynamo trades-off consistency for availability, where data is guaranteed to be eventually consistent, that is all updates reach all replicas eventually.
Generally, these systems rely on a central server or policy for handling the spread and distribution of data.
Their goal is to optimize the efficient accessibility of data by end-users and deal with load balancing between servers.

In Homogeneous and Heterogeneous DDBMS, the control on the availability and writing of data is left to the system. In Federated Database Systems (FDS), each database instance is autonomous, in the sense that there is no assumption as to the individual schema and the availability and location of where specific data resides. What is of particular focus is the development of a common query mechanism across heterogeneous databases. Such systems have to deal with many types of data inconsistencies, such as naming of concepts, precision, schema alignment, etc.

FDS attempts to provide unified query mechanisms that hide data inconsistencies in order to deal with the \textit{database integration} problem \cite{PARENT-CACM-1998}. A solution to the distributed mapping of integrated answers to queries was proposed in \cite{SHETH-ACMCS-1990}, where the system is required to be static, and each database instance must remain in the federation.  A more dynamic approach was recently proposed in~\cite{BENT-CITA-2008}, where the FDS is implemented as a graph. When a query is executed, it is propagated along the graph and at each node the results are aggregated, correcting the inconsistencies incrementally.

The HFKN Framework shares some ideas from both DDMS and FDS technologies. The SQL database associated with each agent is not intended to be shared, globally consistent or replicated across agents. These are heterogeneous databases by design. On the other hand, the schema for representing sensor data of any kind is common to all agents in the system. Even at the RDF document/graph level intended to maintain semantic representations and meta-information about datasets describing collected sensor data, the HFKN framework does not require complete consistency. A form of \textit{weak consistency} is guaranteed as discussed in \Cref{ssec:synch_consistency}.


With FDS, the schema (i.e. structure of the data)  and concepts  (i.e. the meaning of the data) are the same for all agents. In the HFKN framework, each agent has full autonomous control of what kind of data it stores. Agents can join and leave the federation at any time. This is different  from  previous  work \cite{SHETH-ACMCS-1990} with FDS where  an  agent  can  leave  a federation only after obtaining a permission.  In our application scenarios, communication is assumed to be unreliable. Consequently, the approach proposed in this paper is designed to handle dynamic changes in the federation structure without any notice.

Like FDS, our approach does use a unified query mechanism, SPARQL (SPARQL Protocol and RDF Query Language)~\cite{SEABORNE-W3C-2008} for this collection of shared RDF documents/graphs, but also locally for internal agent queries of all its RDF documents, shared and unshared. SPARQL  was developed for querying information stored in RDF Graphs. 


\paragraph{Semantic Web and Robotics}
Our choice of representation for information and knowledge, RDF documents/graphs, is inspired by Tim Berners-Lee's vision of a Semantic Web~\cite{BERNERS-LEE-SA-2001,HENDLER-IS-2001,10.1145/3382097}, where web pages in the World Wide Web are annotated with representations of their semantic content in the form of collections of RDF triples (RDF documents) and ontologies and linked across the WWW. These information structures can be reasoned about using logical inference mechanisms such as those based on Description Logics~\cite{book-DLH}, in addition to related powerful ontology inference mechanisms such as OWL~\cite{mcguinness2004owl,book-DLH}. The modern equivalent of these ideas has resulted in standardization~\cite{rdf-standards} and the linked-data~\cite{book-LD} research area of which knowledge graphs are a prime example. Many additional tools and technologies have been developed since the original idea of the Semantic Web was proposed and these are used in the backbone of many companies' knowledge-intensive products.

More recently, there has been a trend toward leveraging Semantic Web ideas with robotics~\cite{socrob:2016,Distributed_KnowledgeBases_2015,NEUHAUS-SSN,LEMAIGNAN-IROS-2010,malec1,LIM-TSMC-2011}. Many existing ontologies are available and useful for describing robotic systems \cite{SCHLENOFF-ROSE-2013,CARBONERA-IROS-2013} and sensing for robotics \cite{NEUHAUS-SSN}.

Several frameworks using Semantic Web technologies have been implemented on robotics systems. The OpenRobots Ontology (ORO) \cite{LEMAIGNAN-IROS-2010} presents a processing framework leveraging Semantic Web technologies to enhance communication between robots and humans. ORO makes use of a \textit{common sense} ontology, an events system, a model of the cognitive state of other agents, and a long term memory. In \cite{LIM-TSMC-2011}, the authors present a method to connect low-level sensor data and high-level semantic information using RDF and an ontology for low-level information.

The HFKN Framework relies heavily on Semantic Web technologies. This choice is due in part to the conceptual fit between our ideas and these technologies and the wider availability of collective knowledge on the Internet provided in the form of ontologies and RDF Documents/Graphs, the latter often called \textit{knowledge graphs}. Due to the similarities, these knowledge sources can be leveraged, integrated, and used to best advantage by the collaborative agents in the scenarios we focus on.

One of the most mature frameworks combining Semantic Web technologies and robotics is \textit{KnowRob}~\cite{BEETZ-ICRA-2018, tenorth2015representations}, a knowledge processing framework for robotic agents. KnowRob supports reasoning over semantic information while taking into account planning processes. KnowRob needs to be used in conjunction with a knowledge base such as the Robot Scene Graph (RSG)~\cite{7353612, 6630614}.  RSG is a framework for representing 3D information about a robot's environment in a graph structure stored in the volatile memory of the robot. Work that combines RSG and KnowRob is presented in \cite{YAZDANI201980}. 

In the HFKN Framework,
an agent's SCModule stores information in its PostgreSQL database in the KDB. This architecture results in the information being stored on the hard drive whereas continually used information is cached in memory.
The benefit of this is that the information is saved long term and is persistent even when a particular robot agent goes offline. The information is not duplicated but shared across the modules of a robot. However, there is additional overhead when accessing the information stored in the PostgreSQL database when compared to the in-memory solution used with RSG. Still, efficient caching minimizes this as an issue. 

RSG has recently been extended with a synchronization mechanism in~\cite{7353612}, which allows the distribution of RSG nodes across a team of agents. Since some of the information stored in RSG can be large (e.g. point clouds), the authors have implemented a QoS solution that checks the network's quality and then downsamples the information relative to network congestion levels.

The HFKN framework deals with much more  general forms of synchronisation then those provided in RSG.
Additionally, only low bandwidth information is automatically exchanged between agents through synchronization. Large datasets (e.g. point clouds and images) are only transferred after an explicit request from an agent.
This results in a by-need philosophy, where agent \textit{SCModule}s transfer full resolution information only when needed. Note that there is a throttling mechanism that controls data transfer bandwidth when using the data exchange protocol. This is, in fact, a form of QoS mechanism. One could envision combining these ideas with the QoS techniques used in RSG, but we save such issues for future work.

Research in the area of Cloud Robotics~\cite{waibel11roboearth,7006734,bozcuoglu2018exchange} is also of interest, but less directly related to the specific challenges in this paper. Additionally, the work in both Cloud Robotics and Semantic Robots has focused primarily on single robotic systems rather than dynamic teams of robots as the HFKN framework does.

\paragraph{Applications with Mapping} 
The topic of Distributed Simultaneous Localisation and Mapping (SLAM) has some relation to our applications of the HFKN Framework.
One of the basic functionalities in any robotic platform is building a map of the environment. 
Many algorithms have been developed for a single robot \cite{BAILEY-IRAM-2006a,BAILEY-IRAM-2006b}, which deal with the problem of fusing information coming from multiple sensors in a robot.
Such algorithms have been extended to support fusion of sensor data from multiple robots. Earlier work involved considering a set of local maps connected in a global graph shared between robots \cite{VIDAL-RAS-2011}. This approach was superseded by introducing \textit{landmarks} directly in global graphs~\cite{CUNNINGHAM-ICRA-2012}.
However, such approaches require a central location for the fusion of the global map. A decentralized approach was proposed in~\cite{CIESLEWSKI-ICRA-2015}. The authors suggested using a version control system for the map, but this still required a lock when fusing part of the map distributed between platforms.

A practical application of building a common map by a group of robots is to solve the problem of exploration of a large environment~\cite{VINCENT-AMAI-2008}. Most of these approaches use either a centralized server or a global lock, while HFKN is a fully decentralized system. The lock is implicit in the synchronization algorithm.
The approaches considered here are specific to solving the distributed fusion problem, whereas HFKN is a much more general framework.

As mentioned earlier, the KDB is GIS (Geographic information systems)-like in concept since it stores low-level sensor data, intermediate information, and high-level semantic knowledge. Although the KDB architecture is not layered in the GIS sense, SPARQL queries about specific geographical regions return information about those regions at many different conceptual layers.

GIS systems can be used to store a wide range of vector information (points, lines, polygons, etc.  \cite{COX1997449}). They have also been extended to allow storage of 3D information such as point clouds~\cite{CURA-3GC-2016} or point clouds augmented with semantic knowledge~\cite{POUX-3GC-2017}. The latter is directly related to how the KDB can semantically label lower-level data geographically. 

As the state of an operational environment is continuously evolving, it is natural to use remote sensors to update GIS~\cite{congalton1991remote}.
In \cite{MEGURO-IROS-2006}, a centralized GIS server is used to share rescue information, where autonomous robots push observations to the GIS server and where rescuers can then access that information. As discussed previously, the HFKN Framework collects data and knowledge dynamically via agents yet stores it in a distributed manner. The HFKN Framework uses a weak form of replicated centralization in that each agent shares metadata about other agent's public data and knowledge. The synchronization process keeps this data and knowledge \textit{weakly} consistent across agents. There is no centralized server or source of global information.

\section{Conclusion}\label{sec:conclusions}

We have presented a general system architecture and framework (SymbiCloud HFKN Framework) for supporting dynamic, distributed situation awareness for heterogeneous teams of human/robotic agents.  The system builds upon previous work using a delegation framework for dynamic task allocation for collaborative human/robotic systems that is also part of the general system architecture. 

The major focus of the paper has been the description and specification of core system algorithms for supporting distributed data and knowledge collection and the storage, synchronization, aggregation and transfer of such data and knowledge among teams of collaborating agents. In this case, our teams have consisted of humans and UAVs.

Major effort has been placed on incorporating resiliency into the algorithms and the resulting engineered system to deal with difficult problems associated with unreliable communication among agents, out-of-range issues, agents entering and leaving mission environments, in addition to the heterogeneity of both the data collected and the systems used to collect such data.

A seamless unification of data, information and knowledge has been achieved through the use of  Semantic Web technologies as a basis for structuring and querying collected sensor data, meta-information about such sensor data, and post-processed knowledge in the form of RDF  (knowledge) graphs. The system can easily be extended with additional ontologies other than those described in the paper to provide richer semantic decision support.

Although the focus of this paper has been on the dynamic collection of  sensor data for constructing 3D maps and limited object representations such as people and building structures, the system is general in nature and any type of semantic information that can be modeled using RDF  (knowledge) Graphs could be collected, represented, processed and queried in a similar manner. 

Empirical experimentation in simulation has been performed showing the speed and scalability of the proposed system and algorithms. Additionally, limited field experimentation has been done using emergency rescue scenarios with teams of UAVs and humans to show the viability of the ideas in the engineered HFKN system. 

\addcontentsline{toc}{section}{References}

\bibliographystyle{splncs04}  
\bibliography{biblio}

\newpage 
\appendix
\gdef\thesection{\@Alph\c@section}%
\section{Representation of Data Structures in SCModules}\label{sec:Appendix1}
\setcounter{table}{0}
\setcounter{figure}{0}
\definecolor{orange}{rgb}{1,0.6,0.2}
\newcommand{\datatabcolor}{\cellcolor{orange!75}}
\ifdefined\forarxiv
\newcommand{\datatabresize}{\resizebox{0.8\width}{!} }
\else
\newcommand{\datatabresize}{\resizebox{0.7\width}{!} }
\fi

In \Cref{sec:querying}, \Cref{fig_sparql_query_execution}, we provided an overview of different ways to query an \textit{SCModule} using SPARQL. In this appendix, additional details are provided concerning some of the basics Schemas and RDF Views used in the HFKN Framework, together with a selected set of SQL table representations.



\paragraph{Triples table}
An RDF Graph consists of triples ($subject$, $predicate$, $object$) which are stored in a SQL table with three fields presented in \Cref{tab:data_rdf_triples}.
The SQL table fields correspond directly to the components of a triple (see also~\Cref{fig:rdf_triple}). The type of the $object$ field provides a means to efficiently store various data types such as integers, strings, lists, images, point clouds, etc.

\begin{table}[h]
\begin{center}
\datatabresize{
\begin{tabular}{ | c | c | p{8cm} | }
\hline
\multicolumn{3}{|c|}{\datatabcolor\textbf{RDF Triple}} \\
\hline
\textbf{Type}&\textbf{Name}&\textbf{Comments}\\
\hline
string&subject&Subject of the RDF Triple.\\
\hline
string&predicate&Predicate of the RDF Triple.\\
\hline
any&object&A complex type which allows to efficiently store any type of data, ranging from integers, strings, lists, point clouds, images, etc.\\
\hline
\end{tabular}
}
\caption{Fields of the \textit{RDF Triple} SQL table.} \label{tab:data_rdf_triples}
\end{center}
\end{table}

\paragraph{RDF Document}
\label{annex:rdf_document}
An RDF Document is stored using three tables:

\begin{itemize}
  \item A \textit{triples tables} corresponding to a version of the RDF Graph, which is available for answering queries or performing change operations (see \Cref{tab:data_rdf_triples}).
  \item A \textit{revision table} containing the metadata associated with each RDF Revision (see \Cref{tab:data_rdf_revision}).
  \item A \textit{delta table} containing the RDF deltas (see \Cref{tab:data_rdf_delta}).
\end{itemize}

The database also contains metadata about the RDF Document:

\begin{itemize}
  \item The revision hash of the current instance.
  \item The URI used to identify the RDF Graph.
\end{itemize}

\begin{table}[h]
\begin{center}
\datatabresize{
\begin{tabular}{ | c | c | p{8cm} | }
\hline
\multicolumn{3}{|c|}{\datatabcolor\textbf{RDF Revision}} \\
\hline
\textbf{Type}&\textbf{Name}&\textbf{Comments}\\
\hline
uint8[]&hash&The unique hash identifying the revision.\\
\hline
uint8[]&author&The UUID of the agent that created the revision.\\
\hline
uint8[]&signature&A cryptographic signature (using RSA) of the hash authentifying the author.\\
\hline
uint&timestamp&The time when the revision was created.\\
\hline
\end{tabular}
}
\caption{Fields of the \textit{RDF revision} SQL table.} \label{tab:data_rdf_revision}
\end{center}
\end{table}

\begin{table}[h]
\begin{center}
\datatabresize{
\begin{tabular}{ | c | c | p{8cm} | }
\hline
\multicolumn{3}{|c|}{\datatabcolor\textbf{RDF Delta}} \\
\hline
\textbf{Type}&\textbf{Name}&\textbf{Comments}\\
\hline
uint8[]&parent&The unique hash of the parent revision.\\
\hline
uint8[]&child&The unique hash of the child revision.\\
\hline
string&delta&A SPARQL Update representing the change between the parent revision and the child revision.\\
\hline
\end{tabular}
}
\caption{Fields of the \textit{RDF delta} SQL table.} \label{tab:data_rdf_delta}
\end{center}
\end{table}

\paragraph{RDF Delta}
\label{appendix:rdf_delta}

Deltas are encoded using a subset of SPARQL Update query\cite{SEABORNE-W3C-2008}.
Only \sparql|INSERT DATA| and \sparql|DELETE DATA| queries are allowed, because they are the only two types of query to unambiguously identify the inserted and removed triples.

RDF Blank Nodes\footnote{Blank nodes can be used to identify an element of an RDF Triple without providing an explicit URI} are encoded using the \textit{skolemization} process\cite{MALLEA-ISWC-2011}, i.e. blank node are replaced by an unique URI assigned by the KDB Manager.

An example of an RDF Delta with one triple being added and one removed is encoded as a SPARQL Update query in the following way:

\begin{minted}{SPARQL}
PREFIX ex: <http://example.org/>

INSERT DATA {
 ex:a ex:b ex:c
}
DELETE DATA {
 ex:d ex:e ex:f
}
\end{minted}

\paragraph{\textit{Position} and \textit{Transformation} tables}

\begin{figure}[tb]
\centering
\def\svgwidth{0.7\columnwidth}
\resizebox{0.5\linewidth}{!}{
\begingroup%
  \makeatletter%
  \providecommand\color[2][]{%
    \errmessage{(Inkscape) Color is used for the text in Inkscape, but the package 'color.sty' is not loaded}%
    \renewcommand\color[2][]{}%
  }%
  \providecommand\transparent[1]{%
    \errmessage{(Inkscape) Transparency is used (non-zero) for the text in Inkscape, but the package 'transparent.sty' is not loaded}%
    \renewcommand\transparent[1]{}%
  }%
  \providecommand\rotatebox[2]{#2}%
  \newcommand*\fsize{\dimexpr\f@size pt\relax}%
  \newcommand*\lineheight[1]{\fontsize{\fsize}{#1\fsize}\selectfont}%
  \ifx\svgwidth\undefined%
    \setlength{\unitlength}{168.00835797bp}%
    \ifx\svgscale\undefined%
      \relax%
    \else%
      \setlength{\unitlength}{\unitlength * \real{\svgscale}}%
    \fi%
  \else%
    \setlength{\unitlength}{\svgwidth}%
  \fi%
  \global\let\svgwidth\undefined%
  \global\let\svgscale\undefined%
  \makeatother%
  \begin{picture}(1,0.46206167)%
    \lineheight{1}%
    \setlength\tabcolsep{0pt}%
    \put(0,0){\includegraphics[width=\unitlength,page=1]{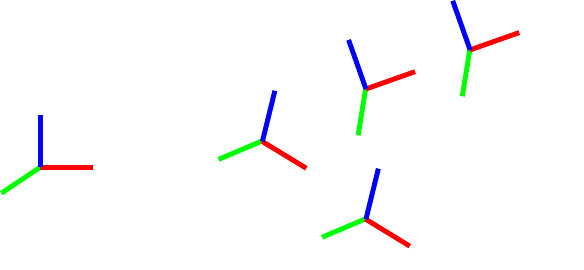}}%
    \put(0.32279571,0.13493717){\color[rgb]{0,0,0}\makebox(0,0)[lt]{\lineheight{1.25}\smash{\begin{tabular}[t]{l}\textit{base\_link}\end{tabular}}}}%
    \put(0.0469803,0.1197913){\color[rgb]{0,0,0}\makebox(0,0)[lt]{\lineheight{1.25}\smash{\begin{tabular}[t]{l}\textit{world}\end{tabular}}}}%
    \put(0.64165743,0.26088748){\color[rgb]{0,0,0}\makebox(0,0)[lt]{\lineheight{1.25}\smash{\begin{tabular}[t]{l}\textit{pan\_tilt}\end{tabular}}}}%
    \put(0.83536588,0.33661721){\color[rgb]{0,0,0}\makebox(0,0)[lt]{\lineheight{1.25}\smash{\begin{tabular}[t]{l}\textit{camera}\end{tabular}}}}%
    \put(0.59280839,0.00063218){\color[rgb]{0,0,0}\makebox(0,0)[lt]{\lineheight{1.25}\smash{\begin{tabular}[t]{l}\textit{IMU}\end{tabular}}}}%
    \put(0,0){\includegraphics[width=\unitlength,page=2]{tf.pdf}}%
  \end{picture}%
\endgroup%
}

\caption{Example of a transformation tree for a single agent. The origin of the world is represented by the frame \textit{world} and the root frame for the agent is \textit{body\_link}.}
\label{fig:tf}
\end{figure}

Internally our robots use the ROS TF library~\cite{6556373} for representing coordinate frames of all the agents.
TF allows for keeping track of all transformations between different components of a robot (and between robots) represented as a tree.
Each node in the tree corresponds to a frame and includes the rotation and translation between that frame and its parent (see \Cref{fig:tf}).
The root of the tree is usually the origin of the world.

The definition of the origin of the world is agent-specific and can be different for all agents. Therefore in \textit{SCModule}, we cut the tree at the main reference frame of each agent and record the corresponding geoposes as different entries in the Position table. This allows for maintaining a global coordinate frame for all agents in the system.
For example, in \Cref{fig:tf}, only the \textit{base\_link} and all its children (\textit{pan\_tilt}, \textit{camera}, \textit{IMU}) are saved in the Transformation table (top of \Cref{tab:tf_geopose}) and the geopose (latitude, longitude, altitude and orientation) of the \textit{base\_link} is saved in the Position table (bottom of \Cref{tab:tf_geopose}).

\begin{table}[h]
\begin{center}
\datatabresize{
\begin{tabular}{ | c | c | p{8cm} | }
\hline
\multicolumn{3}{|c|}{\datatabcolor\textbf{Transformation}} \\
\hline
\textbf{Type}&\textbf{Name}&\textbf{Comments}\\
\hline
string&agentUri&URI of the agent.\\
\hline
string&franeUri&URI of the frame.\\
\hline
string&parentUri&URI of the parent frame.\\
\hline
uint&timestamp&Data acquisition time.\\
\hline
float[3]&translation&Translation between the parent and current frame.\\
\hline
float[4]&rotation&A quaternion representing the rotation between the parent and current frame.\\
\hline
\hline
\multicolumn{3}{|c|}{\datatabcolor\textbf{Position}} \\
\hline
\textbf{Type}&\textbf{Name}&\textbf{Comments}\\
\hline
string&agentUri&URI of the agent.\\
\hline
float&latitude&Latitude of the agent in the WSG84~\cite{decker1986world}.\\
\hline
float&longitude&Longitude of the agent in the WSG84.\\
\hline
float&altitude&Altitude of the agent using the EGM96~\cite{williamson1998development} gravity model.\\
\hline
float[4]&rotation&A quaternion representing the orientation of the robot, such as x-axis, y-axis and x-axis are pointing respectively North, East and up.\\
\hline
\end{tabular}
}
\caption{Fields of the \textit{Transformation} and \textit{Rotation} SQL tables.} \label{tab:tf_geopose}
\end{center}
\end{table}

\paragraph{Images}

Image data is stored in two tables: \textit{CameraFrame} and \textit{CameraInfo} presented in \Cref{tab:data_image}. The former contains basic information about the parameters of an image such as the time of acquisition, its size, encoding type, etc. The most bandwidth-intensive part is the data (i.e. pixel values). The latter contains calibration information of the camera sensor, such as the matrix of camera intrinsic parameters K as well as the lens distortion parameters D.

\begin{table}[h]
\begin{center}
\datatabresize{
\begin{tabular}{ | c | c | l | }
\hline
\multicolumn{3}{|c|}{\datatabcolor\textbf{CameraFrame}} \\
\hline
\textbf{Type}&\textbf{Name}&\textbf{Comments}\\
\hline
string&sensorUri&URI of the sensor.\\
\hline
string&datasetUri&URI of the dataset.\\
\hline
uint&timestamp&Data acquisition time.\\
\hline
string&frameId&Transform frame for the image.\\
\hline
uint&width&Width of the image in pixels.\\
\hline
uint&height&Height of the image in pixels.\\
\hline
string&encoding&Encoding the image data.\\
\hline
uint&step&Data step of the image data.\\
\hline
string&compression&jpeg or raw - indicates if the image is compressed.\\
\hline
uint8[ ]&data&Image data.\\
\hline
\hline
\multicolumn{3}{|c|}{\datatabcolor\textbf{CameraInfo}} \\
\hline
\textbf{Type}&\textbf{Name}&\textbf{Comments}\\
\hline
uint&width&Width of the image in pixels.\\
\hline
uint&height&Height of the image in pixels.\\
\hline
string&distortionModel&Name of distortion model.\\
\hline
float[5]&D&2 radial and 3 tangential lens distortion parameters.\\
\hline
float[9]&K&Intrinsic parameters, 3x3 row-major matrix.\\
\hline
\end{tabular}
}
\caption{Fields of the \textit{CameraFrame} and \textit{CameraInfo} SQL tables.} \label{tab:data_image}
\end{center}
\end{table}




\paragraph{Point clouds}

\begin{figure}[tb]
  \centering
  \includegraphics[width=0.5\linewidth]{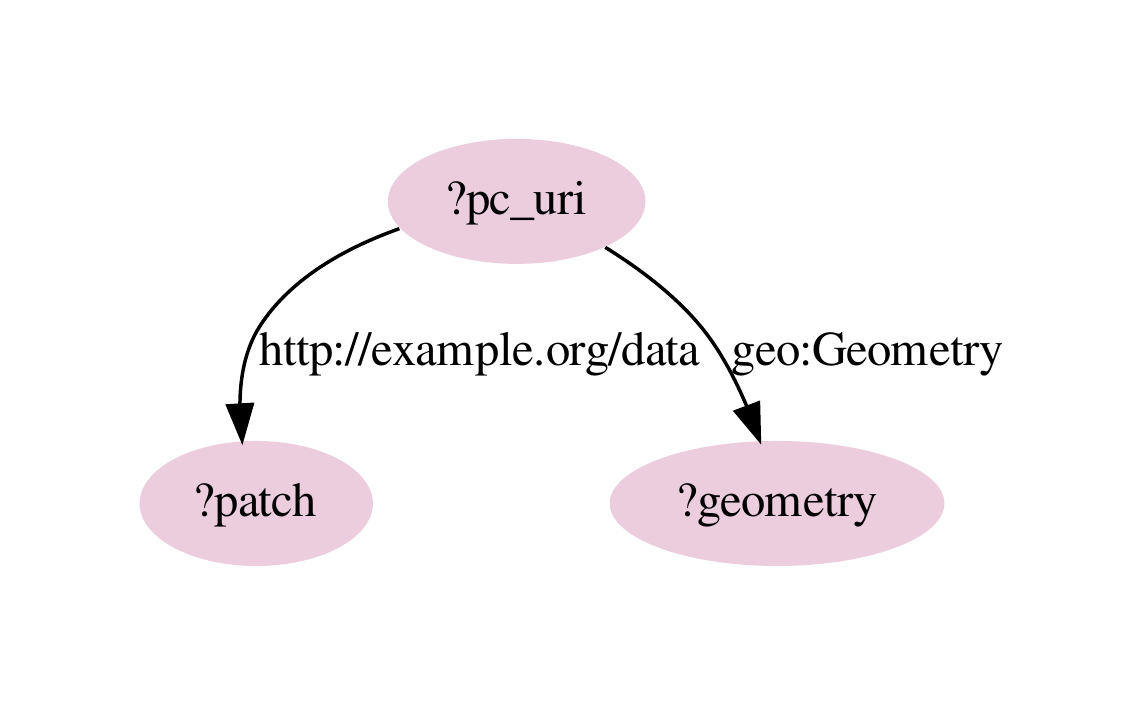}
  \caption{The RDF View (\textit{http://example.org/point\_clouds}) definition for point clouds.}
  \label{fig:point_cloud_views}
\end{figure}

Sensor data representing point clouds is stored in one table presented in \Cref{tab:data_point_cloud}.
Point cloud data typically consists of many measurements, and storing them as individual records in a database is inefficient. Instead, we use data representation proposed in~\cite{ramsey2013lidar}.
The base type \textit{PcPoint} is used to encode a single point LIDAR measurement.
A collection of those points is then grouped in a complex type \textit{PcPatch} which is stored as a record in the database reducing the total number of entries.
The RDF View defining the mapping between the table and corresponding RDF Triples used for querying of point cloud data is shown in \Cref{fig:point_cloud_views}.

\begin{table}[h]
\begin{center}
\datatabresize{
\begin{tabular}{ | c | c | p{8cm} | }
\hline
\multicolumn{3}{|c|}{\datatabcolor\textbf{PointCloud}} \\
\hline
\textbf{Type}&\textbf{Name}&\textbf{Comments}\\
\hline
PcPatch&patch&A complex type that can represent arbitrary point cloud data in PostgreSQL~\cite{ramsey2013lidar}.\\
\hline
\end{tabular}
}
\caption{Fields of the \textit{PointCloud} SQL table.} \label{tab:data_point_cloud}
\end{center}
\end{table}

\paragraph{LIDAR data}

\begin{figure}[tb]
  \centering
  \includegraphics[width=\linewidth]{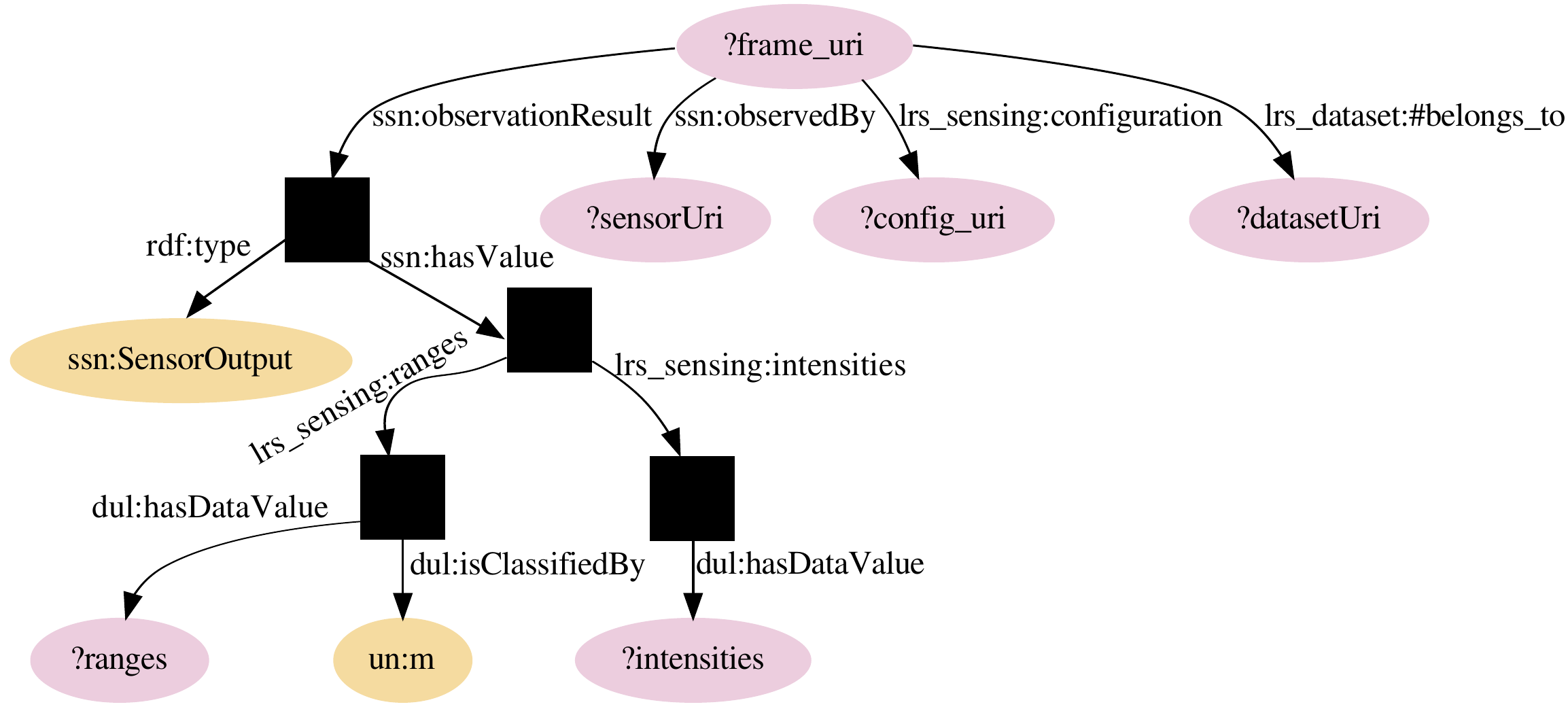}
  \includegraphics[width=\linewidth]{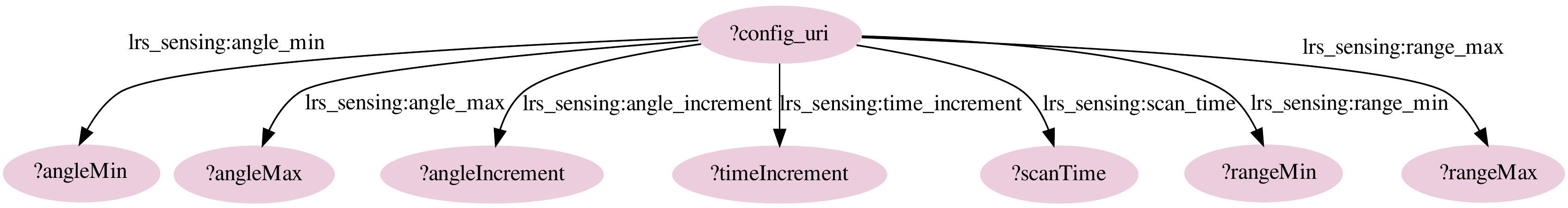}
  \caption{The RDF View definition for LIDAR Frames (\textit{http://example.org/lidar\_frames}, top) and the associated configuration of the sensor (\textit{http://example.org/lidar\_config}, bottom)}
  \label{fig:lidar_views}
\end{figure}

LIDAR data is stored in two tables: \textit{LidarFrame} and \textit{LidarConfig} presented in \Cref{tab:data_lidar}. The former contains information about the time of acquisition, a pointer to the sensor's configuration, and the actual LIDAR readings in the form of ranges and return intensities. The latter contains information about the sensor configuration used during the data collection. Most importantly, angular and range resolutions are specified. The RDF View definitions for the mapping between the SQL Tables and a set of triples that can be queried with SPARQL are shown in \Cref{fig:lidar_views}.


\begin{table}[h]
\begin{center}
\datatabresize{
\begin{tabular}{ | c | c | l | }
\hline
 \multicolumn{3}{|c|}{\datatabcolor\textbf{LidarFrame}} \\
\hline
\textbf{Type}&\textbf{Name}&\textbf{Comments}\\
\hline
string&sensorUri&URI of the sensor.\\
\hline
string&datasetUri&URI of the dataset.\\
\hline
string&frameId&Transform frame for the LIDAR data. \\
\hline
uint&timestamp&Data acquisition time.\\
\hline
LidarConfig&config&A pointer to the sensor configuration.\\
\hline
float&ranges[]&Range readings of the sensor.\\
\hline
float&intensities[]&Intensity readings of the sensor.\\
\hline
\hline
\multicolumn{3}{|c|}{\datatabcolor\textbf{LidarConfig}} \\
\hline
\textbf{Type}&\textbf{Name}&\textbf{Comments}\\
\hline
float&angleMin&Angle for the first measurement.\\
\hline
float&angleMax&Angle for the last measurement.\\
\hline
float&angleIncrement&Angle between two measurements.\\
\hline
float&timeIncrement&Time between two consecutive measurements.\\
\hline
float&rangeMin&Minimum measurement distance.\\
\hline
float&rangeMax&Maximum measurement distance.\\
\hline
\end{tabular}
}
\caption{Fields of the \textit{LidarFrame} and \textit{LidarConfig} SQL tables.} \label{tab:data_lidar}
\end{center}
\end{table}





\clearpage

\section{Additional SPARQL Queries of Interest}\label{sec:Appendix2}

Based on the specifications in \Cref{sec:Appendix1}, this appendix provides additional examples of sensor-related SPARQL queries and their use in acquiring object data about building structures.

\paragraph{Query for the range data of a LIDAR sensor}

The following query returns a set of ranges from the specified dataset, i.e. \sparql|?range| from the dataset \sparql|ex:dataset1|:

\begin{minted}{SPARQL}
SELECT ?ranges FROM <lidar> WHERE {
  ?frameUri lrs_dataset:belongs_to ex:dataset1 ; 
    ssn:observationResult [
          a ssn:SensorOutput ;
          ssn:hasValue [
            lrs_sensing:ranges [ dul:hasDataValue ?ranges ;
            dul:isClassifiedBy un:m ]   ;
          ] ;
      ] ;
}
\end{minted}

The first part of the query selects the \sparql|?frameUri| that belongs to the dataset \sparql|ex:dataset1|:

\begin{minted}{SPARQL}
  ?frameUri lrs_dataset:belongs_to ex:dataset1 ; 
\end{minted}

The second part of the query retrieves the \sparql|?ranges| data:

\begin{minted}{SPARQL}
    ssn:observationResult [
          ...
      ] ;
\end{minted}

\paragraph{Query for acquiring a list of buildings and their 3D models}

This is an example of a query to access a list of buildings \sparql|?building_name| and their corresponding 3D models \sparql|?building_3d_model|:

\begin{minted}{SPARQL}
SELECT ?building_name, ex:filter(ex:combine(?point_cloud_3d),
       ?building_2d_footprint) AS ?building_3d_model WHERE {
    ?building_uri a ex:building ;
                  geo:Geometry ?building_2d_footprint ;
                  foaf:name ?building_name .
    ?point_cloud_uri a lrs_sensing:point_cloud ;
                 geo:Geometry ?point_cloud_2d_footprint ;
                 ex:data ?point_cloud_3d .
    FILTER(geo:intersects(?building_2d_footprint,
           ?point_cloud_2d_footprint))
  } GROUP BY ?building_name
\end{minted}

The first part of the query lists all buildings. Their 2D footprints are stored in \sparql|?building_2d_footprint| and their names in \sparql|?building_name|:

\begin{minted}{SPARQL}
  ?building_uri a ex:building ;
                  geo:Geometry ?building_2d_footprint ;
                  foaf:name ?building_name .
\end{minted}

\ifdefined\forarxiv
The second part of the query lists all the point clouds \sparql|?point_cloud_3d| whose footprints\linebreak \sparql|?point_cloud_2d_footprint| were obtained by projecting all the points on a 2D plane intersecting with a specified building footprint \sparql|?building_2d_footprint|:
\else
The second part of the query lists all the point clouds \sparql|?point_cloud_3d| whose footprints \sparql|?point_cloud_2d_footprint| were obtained by projecting all the points on a 2D plane intersecting with a specified building footprint\linebreak \sparql|?building_2d_footprint|:
\fi

\begin{minted}{SPARQL}
  ?point_cloud_uri a lrs_sensing:point_cloud ;
                geo:Geometry ?point_cloud_2d_footprint ;
                ex:data ?point_cloud_3d .
    FILTER(geo:intersects(?building_2d_footprint,
           ?point_cloud_2d_footprint))
\end{minted}

Finally, the point clouds are combined to give a single result for each building with:

\begin{minted}{SPARQL}
  ex:filter(ex:combine(?point_cloud_3d), ?building_2d_footprint)
    AS ?building_3d_model
\end{minted}

The \sparql|ex:combine| function takes a set of 3D point clouds and combines them into a single one.
The \sparql|ex:filter| function takes a point cloud and a 2D footprint and output a point cloud for which each point is inside the footprint.

\end{document}